\documentstyle[11pt,aaspp4,flushrt]{article}
\received{1999 February 12} \accepted{~~}

\input{psfig.tex}
\font\lbf cmbx12 at 14truept
\font\lrm cmr12 at 14truept

\begin{document}

\psfig{figure=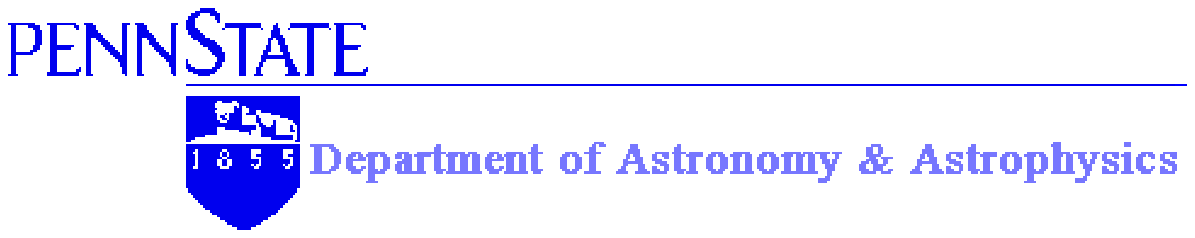,width=8in,rheight=1.3in}

\title{\lbf An X-ray Spectral Survey of Radio-Loud AGN With ASCA} 
\author{\lrm Rita M. Sambruna\footnote{Department of Astronomy and
Astrophysics, The Pennsylvania State University, 525 Davey Lab, State
College, PA 16802, e-mail: {\tt rms@astro.psu.edu, mce@astro.psu.edu}.},
Michael Eracleous$^1$, \& Richard F. Mushotzky
\footnote{NASA/GSFC, Code 662, Greenbelt, MD 20771,
e-mail: {\tt mushotzky@lheavx.gsfc.nasa.gov}.}}

\author{To appear in the {\it Astrophysical Journal}}

\begin{abstract}

We present a uniform and systematic analysis of the 0.6--10 keV X-ray
spectra of radio-loud active galactic nuclei (AGN) observed by {\it
ASCA}. The sample, which is not statistically complete, includes 10
Broad Line Radio Galaxies (BLRGs), 5 radio-loud Quasars (QSRs), 9
Narrow Line Radio Galaxies (NLRGs), and 10 Radio Galaxies (RGs) of
mixed FR~I and FR~II types. For several sources the {\it ASCA} data
are presented here for the first time. The exposure times of the
observations and the fluxes of the objects vary over a wide range; as
a result, so does the signal-to-noise ratio of the individual X-ray
spectra.  At soft X-rays, about 50\% of NLRGs and 100\% of RGs exhibit
a thermal plasma emission component, with a bimodal distribution of
temperatures and luminosities. This indicates that the emission arises
in hot gas either in a surrounding cluster or loose group or in a hot
corona, consistent with previous {\it ROSAT} and optical results.  At
energies above 2 keV, a hard power-law component (photon index,
$\Gamma \sim 1.7-1.8$) is detected in 90\% of cases. The power-law
photon indices and luminosities in BLRGs, QSRs, and NLRGs are similar.
This is consistent with simple orientation-based unification schemes
for lobe-dominated radio-loud sources in which BLRGs, QSRs, and NLRGs
harbor the same type of central engine.  Moreover, excess cold
absorption in the range $10^{21}$--$10^{24}$~cm$^{-2}$ is detected in
most (but not all) NLRGs, consistent with absorption by an obscuring
torus, as postulated by unification scenarios. The {\it ASCA} data
provide initial evidence that the immediate gaseous environment of the
X-ray source of BLRGs may be different than in Seyfert 1s: absorption
edges of ionized oxygen, common in the latter, are detected in only
one BLRG. Instead we detect large columns of {\it cold} gas in a
fraction ($\sim$ 44--60\%) of BLRGs and QSRs, comparable to the
columns detected in NLRGs, which is puzzling. This difference hints at
different physical and/or geometrical properties of the medium around
the X-ray source in radio-loud AGN compared to their radio-quiet
counterparts, that can be explored further with future X-ray
observations. For the full sample, the nuclear X-ray luminosity is
correlated with the luminosity of the [{\sc O\,iii}] emission line,
the FIR emission at 12~$\mu$m, and the lobe radio power at 5~GHz. The
Fe K$\alpha$ line is detected in 50\% of BLRGs and in one QSR, with a
large range of intrinsic widths and equivalent widths. In the handful
of NLRGs where it is detected, the line is generally
unresolved. Comparing the average power-law photon indices of the
various classes of radio-loud AGN to their radio-quiet counterparts
from the literature, we find only a weak indication that the {\it
ASCA} 2--10~keV spectra of BLRGs are flatter than those of Seyfert 1s
of comparable X-ray luminosity. This result is at odds with evidence
from samples studied by other authors suggesting that radio-loud AGN
have flatter spectra than radio-quiet ones. Rather, it supports the
idea that a beamed synchrotron self-Compton component related to the
radio source (jet) is responsible for the flatter slopes in those
radio-loud AGN. We argue that because of the way those samples were
constructed, beamed X-ray emission from the radio jets probably
contributed to the observed X-ray spectra.

The sample studied here includes 6 Weak Line Radio Galaxies (WLRGs),
powerful radio galaxies characterized by [O{\sc\,iii}]~$\lambda\lambda
4569,5007$ lines of unusually low luminosity and by unusually high
[O{\sc\,ii}]/[O{\sc\,iii}] line ratios. The {\it ASCA} spectra of
WLRGs can be generally decomposed into a soft thermal component with
$kT\sim 1$ keV, plus a hard component, described either by a flat
($\langle \Gamma \rangle=1.5$) absorbed power law, or a very hot ($kT
\sim 100$ keV) thermal bremmstrahlung model. Their intrinsic
luminosities are in the range $L_{\rm 2-10~keV}\sim
10^{40}$--$10^{42}$~erg~s$^{-1}$, two orders of magnitude lower than
in other sources in our sample. If the hard X-ray emission is
attributed to a low-luminosity AGN, an interesting possibility is that
WLRGs represent an extreme population of radio galaxies in which the
central black hole is accreting at a rate well below the Eddington
rate.

\end{abstract}

\noindent {\underline{\em Subject Headings:}} 
Galaxies:active -- X-rays:galaxies -- Radiation mechanisms:non-thermal
-- Radio galaxies:individual. 
 
\section{Introduction} 

The X-ray spectra of Active Galactic Nuclei (AGN) are a diagnostic of
the structure of their accretion flows. Recent studies of X-ray-bright
Seyfert galaxies with {\it ROSAT}, {\it GINGA}, and {\it ASCA} have
shown their 0.1--100 keV spectra to be complex. The intrinsic X-ray
continuum is typically described by a power law model with a mean
photon index of $\langle\Gamma\rangle \approx$ 1.9, albeit with an
appreciable dispersion around this value. A soft excess is sometimes
detected below 1~keV and modelled as thermal emission from an
accretion disk. At higher energies, an Fe K$\alpha$ emission line is
detected between rest-frame energies of 6 and 7 keV in several sources
(Nandra \& Pounds 1994; Mushotzky et al. 1995; Nandra et al. 1997a).
The line is strong (EW $\sim$ 250 eV) and broad, with a tail extending
to the red (Tanaka et al. 1995; Nandra et al. 1997a). The line profile
is consistent with gravitational and Doppler broadening in a black
hole accretion disk (Fabian et al. 1989), although other explanations
could be possible (e.g., Weaver \& Yaqoob 1998). At higher energies,
{\it GINGA} observations have shown that many Seyferts exhibit a
Compton reflection bump peaking around 20--30 keV (Nandra \& Pounds
1994), now confirmed by {\it RXTE} observations (Lee et al. 1998;
Weaver, Krolik, \& Pier 1998).  This spectral feature has been
interpreted as the result of reprocessing of the primary continuum by
cold matter around the X-ray source, presumably the accretion disk
(Lightman \& White 1988; Guilbert \& Rees 1988; George \& Fabian
1991). The X-rays also probe the immediate environment around the
X-ray source of an AGN. At low energies, absorption features from
ionized gas are detected with {\it ASCA} in about 50\% of Seyfert~1s
(Reynolds 1997; George et al. 1998).  The inferred column densities of
this ``warm absorber'' are in the range $N_{\rm W} \sim
10^{21}$--$10^{24}$~cm$^{-2}$.

In contrast to Seyferts and other radio-quiet AGN, the hard X-ray
spectra of radio-loud AGN are not known in much detail. Pre-{\it ASCA}
studies of heterogenous samples of radio-loud AGN with {\it EXOSAT},
{\it Einstein}, and {\it GINGA} have shown that these objects are
systematically more X-ray luminous than their radio-quiet counterparts
of similar optical luminosity.  In addition, they have flatter
continua than Seyferts with photon indices differing by $\Delta\Gamma
\sim 0.5$ (Wilkes \& Elvis 1987; Lawson et al. 1992; Shastri et
al. 1993; Lawson \& Turner 1997).  However, these studies were plagued
by the low sensitivity of the instruments, especially at low energies
where the diffuse thermal emission from the host galaxy or cluster can
make a substantial contribution (Worrall et al. 1994). Observations
with high spectral resolution and sensitivity over a wide energy band
are the key to disentangling the various spectral features and
measuring the intrinsic X-ray properties of radio-loud AGN more
reliably.

The high spectral resolution of the {\it ASCA} SIS at both low and
high X-ray energies, combined with its broad-band (0.6--10 keV)
sensitivity, are well suited to studying the X-ray properties of
radio-loud AGN. Indeed, results for a few individual bright sources,
mainly broad-line radio galaxies, show that some of these objects have
complex spectra, with a broad Fe line and intrinsic power-law photon
indices of $\Gamma\sim 1.7-2.0$ (Eracleous, Halpern, \& Livio 1996;
Grandi et al. 1997; Sambruna et al. 1998), although others do not
(Eracleous \& Halpern 1998; Reynolds et al. 1998). The hard spectral
response of {\it ASCA} is essential for separating spectrally the soft
thermal component from the hard nuclear power law component and better
studying the latter.

In order to characterize the X-ray properties of radio-loud AGNs we
have carried out a systematic systematic analysis of archival {\it
ASCA} data for a sample of such objects. For several of these objects
for which the data analysis is presented here for the first time.
Other goals of this work include testing unified schemes and exploring
the origin of the radio-loud/radio-quiet AGN dichotomy. The sample is
made up of various classes of radio-loud AGN, which can mainly be
divided into ``type~1'' objects (those with both broad and narrow
optical emission lines) and ``type 2'' (those with only narrow optical
lines). According to the simple orientation-based unification schemes,
type~1 and type~2 sources contain similar active nuclei seen at
different angles with respect to the axis of a thick, obscuring torus
(e.g., Urry \& Padovani 1995). The X-ray properties provide a test of
these ideas: because of the high penetrating power of the hard X-rays,
it is possible to measure the intrinsic luminosity and spectral shape
of the central engine in type~2 objects up to line-of-sight column
densities of $\approx 10^{24}$ cm$^{-2}$. In principle, a test of the
models should also be provided by the Fe line, which is expected to
have higher Equivalent Widths (EWs) in type 2 AGN.

Another use of the X-ray observations is to constrain ideas for the
origin of the radio-loud/radio-quiet AGN dichotomy. Indeed, different
X-ray properties between the two classes are predicted in the context
of models which attribute the dichotomy to a different structure of
the accretion disk or to the intrinsic properties of the black hole.
We thus compare the average X-ray spectral properties of our sample of
radio-loud AGN with those of samples of radio-quiet AGN available in
the literature.  Such a comparison, carried out by Wo\'zniak et
al. (1998) using {\it ASCA} as well as other, non-simultaneous data,
has already yielded interesting results. These authors found that
Broad Line Radio Galaxies, the alleged radio-loud counterparts of
Seyfert 1s, have systematically flatter slopes, narrower and weaker Fe
K$\alpha$ lines, and weak or absent Compton reflection components at
$\gtrsim$ 10 keV. 

In addition, we use the spectroscopic capabilities of {\it ASCA} to
characterize the medium surrounding radio galaxies. Recent results
{\it ROSAT} observations (e.g., Worrall \& Birkinshaw 1994; Crawford
et al. 1999) established that most radio sources are embedded in
diffuse, hot haloes with core radii up to several hundreds of kpc. The
{\it ASCA} sensitivity allows us to detect the thermal component and
measure with accuracy its spectral parameters, complementing {\it ROSAT}. 

The paper is organized as follows. The sample selection criteria and
definitions are given in \S2, together with a description of the {\it
ASCA} observations. Data analysis and results of the spectral fits are
reported in \S3 and \S4, respectively. In \S5 we present the average
nuclear properties of the various subclasses, while in \S6 we examine
correlations of the nuclear X-ray luminosity with the luminosity at
other wavelengths. The summary and discussion of the results are given
in \S7 and \S8, with the conclusions following in \S9. The Appendix
provides a full discussion of the results for the individual sources,
with particular emphasis on the sources for which the {\it ASCA} data
are presented here for the first time. Throughout this paper we assume
a Hubble constant of $H_0=75~{\rm km~s^{-1}~Mpc^{-1}}$ and a
deceleration parameter of $q_0=0.5$.

\section{The Sample and the Data} 

Our sample includes all the radio-loud AGN observed by {\it ASCA} for
which data were available in the public archive up to 1998 September.
We adopted the quantitative criteria for radio-loudness of Kellermann
et al. (1989, 1994), namely objects with either a 5~GHz radio power of
$P_{\rm \,5\,GHz} > 10^{25}~{\rm W~Hz^{-1}}$ or with rest-frame
{5~GHz-to-4400~\AA} flux-density ratios of ${\cal R}_{\rm ro} >
10$. In addition, we included an analysis of the {\it ASCA} data of
the high-redshift radio galaxy 4C~+41.17 and the low-power radio
galaxy IC~310, whose X-ray observations have never been published.
The former object is a star-forming radio galaxy at $z=3.798$ (see Dey
et al. 1997 and references therein); its {\it ASCA} data are affected
by abnormally large positional uncertainties (see Appendix). IC~310 is
a ``head-tail'' radio source, which does not satisfy either of the
strict criteria listed above ($P_{\rm \,5\,GHz} = 1.5\times
10^{23}~{\rm W~Hz^{-1}}$, ${\cal R}_{\rm ro} > 4$), although it is
traditionally regarded as a radio galaxy (Simon 1979; Miley 1980;
Owen, Ledlow, \& Keel 1996). Neither of these objects is included in
the statistical tests that we carry out in later sections of the
paper; their properties are discussed in the Appendix. Blazars and
high-redshift ($z > 1$) radio-loud quasars were excluded from the
sample since their observed X-ray emission is affected by beaming. For
a review of their {\it ASCA} data, see Cappi et al. (1997), Reeves et
al. (1997), Kubo et al. (1998), Vignali et al. (1999).  Our selection
criteria gave a total of 39 objects (including IC~310 and 4C~+41.17),
which are listed in Table~1.

The objects in the final sample were divided into four subclasses
according to their optical spectroscopic properties, such as the
presence of broad, permitted lines in their optical spectra and the
luminosity of the [O{\sc\,iii}]~$\lambda5007$ line. Objects with
broad, permitted optical emission lines (typically Balmer and/or
Mg{\sc\,ii}~$\lambda2800$ lines) were classified either as Quasars or
Broad Line Radio Galaxies (hereafter QSRs and BLRGs, respectively).
Objects without broad emission lines were classified as Narrow Line
Radio Galaxies or Radio Galaxies (hereafter NLRGs and RGs,
respectively). The distinction between QSRs and BLRGs was based on the
[O{\sc\,iii}] luminosity, by adopting $L_{\rm
[O\sc\,iii]}>10^{43.5}~{\rm erg~s^{-1}}$ as the condition for
classifying an object as a QSR. Similarly NLRGs were separated from
RGs by requiring that the former subclass have $L_{\rm [O\,\sc
iii]}>10^{41}~{\rm erg~s^{-1}}$. The [O{\sc\,iii}] luminosity criteria
we have adopted are admittedly arbitrary.  However, they ensure that
objects traditionally regarded as QSRs fall in the appropriate
category, and similarly for RGs. The giant double-lobed radio source
4C +74.26 was tentatively classified as a QSR based on its optical
luminosity and the equivalent width of its [O{\sc\,iii}] line
(Brinkmann et al. 1998). This turned out to be a reasonable
classification {\it a posteriori} when we found the X-ray luminosity
of this object to be comparable to that of QSRs. Similarly,
PKS~2251+11 (also known as PG~2251+11) was classified as a BLRG based
on its [O{\sc\,iii}] luminosity, a classification which is consistent
with its X-ray properties as described in later sections.  This
classification scheme gives 10 BLRGs, 12 NLRGs, 6 QSRs, and 11 RGs.
The sample is by no means statistically complete or unbiased; indeed,
since the targets were originally selected as part of independent
programs by different observers, they may reflect a bias toward the
brightest sources of each type.

All the BLRGs, NLRGs, and QSRs of Table~1 turn out to be powerful
radio sources that satisfy both of the above criteria for radio
loudness.  This is not the case for the RGs, however. Although all of
the RGs have ${\cal R}_{\rm ro} > 10$, only three of them have $P_{\rm
\,5\,GHz} > 10^{25}~{\rm W~Hz^{-1}}$: 3C~28, PKS~0625--53, and
3C~353. Most of the other objects are intermediate-power FR~I radio
galaxies (Fanaroff \& Riley 1974) with $10^{25}~{\rm W~Hz^{-1}} >
P_{\rm \,5\,GHz} > 10^{22}~{\rm W~Hz^{-1}}$. Included among the RGs
are 6 Weak-Line Radio Galaxies (WLRGs; identified as such in Tables
1--4), of either FR~I or FR~II morphology. These were recently
identified in sensitive spectroscopic surveys in optical as
moderately-powerful radio galaxies with low ionization state of their
narrow-line gas, which manifests itself as a high value of the
[O{\sc\,ii}]/[O{\sc\, iii}] emission-line ratio, in the range of 1--10
(Tadhunter et al. 1998). WLRGs have [O{\sc\,iii}] luminosities an
order of magnitude or more lower than those of other radio galaxies of
similar radio power (Tadhunter et al. 1998; Laing et al. 1994). These
sources are thus characterized by small ratios of the [O{\sc\,iii}] to
radio power.

In Table 1 we list the basic properties of each object: redshift
(column~2), coordinates (column~3), other common names (column~4),
Galactic column density from H{\sc\,i}~21~cm maps (column~5), ratio of
the core radio power to the lobe radio power $R$ (column~6), and
luminosities at other wavelengths (columns~7--10) with relevant
references (column~11). The Galactic column densities were taken from
Elvis, Lockman, \& Wilkes (1989), Murphy et al. (1996), and Stark et
al. (1992), and have a formal 90\% uncertainty of $\pm 1 \times
10^{20}$ cm$^{-2}$ or better.  The luminosities given are the
rest-frame monochromatic luminosities ($\nu L_{\nu}$) at 5~GHz and in
the IRAS 12~$\mu$m and 60~$\mu$m bands, and the integrated [O{\sc\,
iii}]~$\lambda5007$ emission-line luminosity.  The [O{\sc\,iii}]
luminosity was corrected for foreground absorption in the Galaxy
\footnote{The correction was computed according to the Seaton (1979)
law, using the Galactic H{\sc\,i} column density listed in Table~1
and $N_{\rm H}/E(B-V)=5\times 10^{21}~{\rm cm^{-2}~mag^{-1}}$ (Savage
\& Mathis 1979).} and in the case of Centaurus~A we also adopted the
correction for extinction in the host galaxy (Storchi-Bergmann et
al. 1997).  The monochromatic luminosities were transformed to the
rest frame of the source using the measured spectral index (the
5-to-2.7~GHz or 5-to-1.4~GHz spectral index for the radio band, the
12-to-25~$\mu$m spectral index for the 12~$\mu$m band and the
25-to-60~$\mu$m spectral index for the 60~$\mu$m band).

Table~2 summarizes the {\it ASCA} observations, with the observation
date (column 2), the image designation number in the {\it ASCA}
archive (column 3), the effective exposure time after data screening
(column 4), and the net source count rate recorded by the SIS0
(0.6--10 keV) and GIS2 (0.7--10 keV) detectors, or the corresponding
3$\sigma$ upper limit (columns 5 and 6).  As Table~2 shows, 100\% of
the BLRGs, 75\% of the NLRGs, 83\% of the QSRs, and 100\% of the RGs
were detected. The signal-to-noise ratio in the SIS spans a broad
range of 3--300.  A few sources, including 3C~303, 3C~390.3, 3C~219,
and Cygnus~A, were observed repeatedly by {\it ASCA}. Observations of
the same target at different epochs are indicated with sequential
numbers in Table~2 and in all tables thereafter.  In the case of
3C~303, which was observed at two epochs about 8 months apart, no flux
variations were found between the two datasets, which were thus
coadded for spectral analysis.  The data for several of these sources
were published by other authors, as described in the Appendix. To
ensure uniformity in the spectral analysis we re-analyzed the data
according to our own data screening criteria.

\section{Data Screening and Analysis}

The sources listed in Table~1 were observed by {\it ASCA} (Tanaka,
Inoue, \& Holt 1994) in a variety of SIS modes, including
\verb+FAINT+, \verb+BRIGHT+, and \verb+BRIGHT2+. In order to apply
standard data analysis methods, the \verb+FAINT+ mode was converted
into \verb+BRIGHT2+, applying the corrections for echo effect and dark
frame error which takes into account a more precise calibration at the
lower energies. If the SIS data were taken in mixed \verb+FAINT+ and
\verb+BRIGHT+ mode, they were converted to \verb+BRIGHT+ mode in order
to have a better signal-to-noise ratio, except for the brighter cases,
where the \verb+BRIGHT2+ mode was used instead for a more detailed
spectral analysis. The GIS data were all taken in PH mode. The SIS
data were taken in 1CCD mode, except for Cyg~A and IC~310 (4CCD),
Fornax~A, Centaurus~A, and 4C~+41.17 (2CCD).

In order to perform a systematic analysis it is important
that the extraction criteria are as uniform as possible, so that the
systematic uncertainties are uniform as well. The screening
criteria for both SIS and GIS included the rejection of data taken
during passages over the South Atlantic Anomaly and with geomagnetic
cutoff rigidity lower than 6 GeV/c.  We retained data accumulated for
Bright Earth angles greater than 20$^{\circ}$ and Elevation Angles
greater than 10$^{\circ}$ for the SIS and $5^{\circ}$ for the GIS.
Only data corresponding to SIS grades 0, 2, 3, and 4 were retained in
the spectral analysis (in general, inclusion of grade 6 did not
improve the quality of the data at the higher energies). We checked
each data set individually to ensure  that the above data selection
criteria gave the best signal-to-noise ratio. These criteria are
similar to those used by most authors  when analyzing {\it ASCA}
observations of Seyfert galaxies and radio-quiet quasars. As a result
we can readily compare our results with theirs. 

Source spectra and light curves were extracted from circular regions
centered on the source position with a radius of 4$^{\prime}$ in the
case of the SIS and 6$^{\prime}$ in the case of the GIS, which has a
larger intrinsic point spread function. In the case of the GIS we
evaluated the background in circular regions of radius 4$^{\prime}$,
each located about 12$^{\prime}$ away from the target where the source
counts are negligible, and avoiding other sources in the field. In a
few radio galaxies the {\it ASCA} data are dominated by the diffuse
emission on a cluster scale (Table 3); in these cases we used a larger
extraction cell ($\sim 8^{\prime}$ at $z=0.05$) in order to collect
most of the flux. In the case of the SIS, the background was measured
in a circular region of radius 2$^{\prime}$ located 6$^{\prime}$ away
from the target on the same chip, or from blank-sky observations, when
needed. For the fainter sources in Table~2, we checked that our
results did not depend on the background by measuring the latter at
different locations on the detector field of view, when possible.  The
light curves were visually inspected to search for intervals of data
dropout, which were eliminated from the final analysis, and for source
variability (see \S3.1). Since, in general, no temporal variations
were detected (see \S3.1), in all cases spectra were produced by
accumulating photons over the entire duration of the observation. We
extracted SIS and GIS spectra only when 100 or more counts were
detected (Table 2), a condition met by 34 sources (10 BLRGs, 9 NLRGs,
5 QSRs, and 10 RGs). 

The {\it ASCA} spectra were fitted using the \verb+XSPEC+ (v.10)
software package. The spectra were rebinned so that each bin included
at least 20 counts in order for the $\chi^2$ test to be safely
valid. Joint fits to the data from the 4 detectors were performed in
each case, leaving the normalizations as independent parameters to
account for uncalibrated differences between the effective areas.  The
GIS response matrices from the latest release of 1995 March were used
for the GIS spectra, while for the SIS spectra we used the response
matrices generated by the \verb+sisrmg+ program (v.1.1, 1997 April),
including the latest calibration files.  The \verb+ascaarf+ program
(v.2.72, 1997 Mar 14) was used to generate the ancillary response
functions (a convolution of the telescope effective area curve, the
source point-spread function, and the detector transmission
function). The SIS and GIS spectra were fitted in the energy ranges
0.6--10~keV and 0.7--10 keV, respectively, where the spectral response
for each instrument is best known.

\subsection{Timing Analysis}

The SIS and GIS light curves were visually inspected for time
variability on timescales up to the length of the observation. In
practice, this analysis was limited to type~1 AGN (BLRGs
and QSRs), whose light curves have the highest signal-to-noise ratio
(Table 2). In the case of NLRGs and RGs, an additional complication is
the difficulty in separating the nuclear emission from the thermal
emission of the host galaxy or cluster.

No significant short-term flux variability was present in the data,
with the exception of 3C~120, where flux changes with amplitude $\sim$
20\% are present toward the end of the {\it ASCA} observation (Grandi
et al. 1997), as confirmed by a $\chi^2$ test.  In the remaining BLRGs
and QSRs there is no evidence for significant flux variability on the
short timescales of the {\it ASCA} observations ($<$ 10 hours). For
3C~382, there is an indication of variability ($P_{\chi^2} \sim
0.002$) when the SIS light curves are finely binned (bin size 128 s),
which is not confirmed, however, when the same light curves are binned
more coarsely. The lack of flux variability of BLRGs on shorter
timescales is in contrast to the large-amplitude variations observed
in Seyfert 1s with {\it ASCA} on timescales as short as a few hours
(Nandra et al. 1997b).  

\section{Model Fits to the X-Ray Spectra} 

\subsection{Methodology}

The goal of our spectral analysis is to find models that provide a
good description of the 0.6--10 keV spectra of all the sources in our
sample. This includes models for a possible Fe line around 6--7~keV,
which is present in most Seyferts (Nandra et al. 1997a; Turner et al.
1997a) and in several BLRGs (Eracleous et al. 1995; Grandi et
al. 1997; Sambruna et al. 1998). The properties of the line profile
are important diagnostics of the physical conditions at the heart of
the emission region. In practice, however, because of the limited {\it
ASCA} pass-band and possible continuum spectral complexity, the line
profile depends sensitively on the determination of the underlying
continuum shape. Conversely, the determination of the continuum shape
can be affected by the presence of an Fe line.  We thus adopted the
following fitting procedure. To search for the spectral model which
best fits the continuum, we first examined the data from the four
detectors jointly in the energy range 0.6--10 keV excluding the
rest-frame energies 5--7 keV, a procedure which is now common (e.g.,
Nandra et al. 1997a). Once the best-fit continuum model was found, we
included the 5--7 keV energy range, added a Gaussian emission line to
the model, and fitted it to the data to determine the parameters of
both the continuum and the line. We attempted fits with several
continuum models, as described below. To decide whether the addition
of one or more free parameters improves the fit significantly (and thus
discriminate among continuum models), the F-test was used, adopting as
a minimum threshold for improvement a probability of $P_{\rm
F}\gtrsim95$\%. To assess the significance of the detection of an Fe
line we conservatively assumed that it contributed three free
parameters to the model even when the line energy and/or the width
were kept fixed in the fit (since an {\it a priori} choice was made
for this parameter).

The best-fitting models and their parameters are reported in Tables 3
and 4. Table 3 gives the continuum parameters and Table 4 gives the Fe
line parameters; they both refer to the same best-fitting models.  All
the quoted uncertainties correspond to the 90\% (1.6$\sigma$)
confidence level for one parameter of interest ($\chi^2=\chi^2_{\rm
min} + 2.7$).  Figure~1 shows the best-fitting {\it continuum} models
(upper panels) and the corresponding post-fit residuals (lower
panels), expressed as the ratio of the model to the data. In the cases
where an Fe line was required to fit the spectrum, the line profile
model was excluded from the model plotted in Figure~1 to make the line
appear in the ratio spectrum.  A variety of models are needed to
explain the {\it ASCA} continua of radio galaxies. It is apparent from
Figure~1 that the Fe line is present in several sources.

\subsection{Suite of Models for the X-Ray Continuum and Synopsis of
Results}

Here we describe the spectral models which were fitted to the {\it
ASCA} continua according to the procedure outlined above. In all
cases, a column density fixed to the Galactic value was included,
affecting all spectral components. The photoelectric absorption cross
sections were taken from Morrison \& McCammon (1983). In all models,
the normalization factor(s) of the model were treated as free
parameters. The continuum fit results are reported in Table 3 where we
list the best-fit model and corresponding parameters (column 2), the
reduced $\chi^2_{\rm r}$ and degrees of freedom (column 3), the
observed X-ray flux integrated in 0.5--2 keV for the spectral
component at low energies (column 4) and in 2--10 keV (column 5) for
the component at high energies. The corresponding intrinsic
(absorption-corrected and in the source rest-frame) luminosities are
listed in columns 6 and 7, respectively.

\noindent{\it 1) Absorbed Power Law.} Parameters: photon index
$\Gamma$ and column density $N_{\rm H}$ (cm$^{-2}$). The latter was
first fixed to the Galactic value from Table 1, and then left free to
vary to search for possible excess absorption.

\noindent{\it 2) Power Law + Absorption Edge.} Power law parameters:
photon index $\Gamma$, column density $N_{\rm H}$ (cm$^{-2}$). Edge
parameters: energy $E_{\rm e}$ (keV), optical depth $\tau_{\rm e}$. An
absorption edge is detected only in 1/34 of the sources, namely in the
longest of the three observations of 3C~390.3. For the remaining BLRGs
and QSRs, we report the 90\% upper limits to the optical depth of the
{O{\sc\,vii}} (0.739 keV) K-edge in the Appendix in all cases where
the latter is redshifted within the SIS sensitivity range.

\noindent{\it 3) Raymond-Smith Thermal Plasma.} Thermal emission from
a hot, diffuse gas, including emission lines of several elements.
Parameters: temperature $kT$ (keV), elemental abundance $Y$. This
model is an adequate representation of the X-ray spectra of 3/34 of
the sources, yielding temperatures and metallicities typical of
clusters of galaxies.

\noindent{\it 4) Absorbed Power Law + Raymond-Smith Plasma.}  Power
law parameters: photon index $\Gamma$, column density $N_{\rm H}$
(cm$^{-2}$). Raymond-Smith component parameters: temperature $kT$
(keV), elemental abundance $Y$. This model is an adequate
representation of the data for 14/34 of the sources.  For the RG
3C~28, the detection of the power-law component is ambiguous (see
Appendix) and this source was not included in the statistical tests.

\noindent{\it 5) Partial Covering/Scattering.} The partial covering
model parametrizes the physical situation of a ``patchy'' absorber,
where the nuclear emission leaks through an absorbing medium covering
only a fraction of the source (e.g., Holt et al. 1980). It can also
describe the case of radiation scattered into the line of sight by a
uniformly covering medium. Parameters: photon index $\Gamma$, column
density $N_{\rm H}$ (cm$^{-2}$), covering fraction $f_{\rm c}$ (the
fraction of the source which is covered by the medium, or the fraction
of the primary continuum flux scattered into the line of sight). This
model provides a good fit to the data of 3/34 of the sources.

\noindent{\it 6) Dual Absorber.} This model represents the physical
situation of a nuclear power-law continuum which is seen through two
different layers of gas, one uniformly covering the source and one
partially covering the source. Parameters: photon index $\Gamma$,
uniform column density $N_{\rm H,1}$ (cm$^{-2}$), partially covering
column density $N_{\rm H,2}$ (cm$^{-2}$) and corresponding covering
fraction $f_{\rm c,2}$. The spectrum of only 1/34 of the sources, the
BLRG 3C~445, requires such a description.

\noindent{\it 7) Power Law + Thermal Bremmstrahlung.} The model
consists of a power law of photon index $\Gamma$ plus a thermal
bremmstrahlung of temperature $kT$ (keV), parameterizing excess flux
at the soft energies in one BLRG (3C~382). This model also provides an
acceptable (and equivalent to the power law) fit to the hard X-ray
continuum of WLRGs, with temperatures $kT \sim 100$ keV. 

\noindent{\it 8) Double Power Law (sum of two power law models)}. This
model parameterizes a curved (upward or downward) continuum. The
parameters are the power-law photon indices, $\Gamma_1$ and $\Gamma_2$
for the lower- and higher-energy components, respectively, and the
column density $N_{\rm H}$ (cm$^{-2}$), acting on both power laws.
This model describes the {\it ASCA} spectrum of the QSR, 4C~+74.26,
with an upward curved continuum.

\noindent{\it 9) Power Law + Reflection.} This model parameterizes
Compton scattering of the primary continuum by cold, dense matter in
the immediate vicinity of the X-ray source (Lightman \& White 1988;
Guilbert \& Rees 1988).  We used the model \verb+pexrav+ (Magzdiarz \&
Zdziarski 1995) in \verb+XSPEC+ to parameterize reflection for the
BLRGs and NLRGs for which a significant Fe line was detected (Table
4). The free parameters are as follows: photon index of the primary
power law $\Gamma$, power law normalization $N$, reflected component
normalization $A$. In Table 3 we give the reflection normalization $A$
($A$=1 for an isotropic illuminating source).  The inclination angle
was fixed to the value from independent measurements from the
literature, when available (see, for example, the compilation of
Eracleous \& Halpern 1998), or fixed at 30$^{\circ}$ (nearly face-on)
for BLRGs, and 90$^{\circ}$ (edge-on) for NLRGs. The power law plus
reflection model gives an acceptable (both from a statistical and
physical point of view) fit to the data of a single BLRG, 3C~390.3,
during the longest of the three observations. 

\subsection{Models for the Fe Line} 

Having determined the model that best-fits the continuum, we next
attempted spectral fits to the full 0.6--10 keV datasets adding back
the data at 5--7 keV rest-frame energies and including a Gaussian
component to parameterize the Fe K$\alpha$ line.  As illustrated in
Figure~1, evidence for an Fe line around 6 keV is present in several
sources.

We first fitted the data with an unresolved-line model, fixing the
Gaussian energy dispersion to $\sigma=0.05$~keV. This is comparable to
the SIS resolution during the first few years of the {\it ASCA}
mission; however, the SIS resolution has been degrading with time,
reaching $\sigma=0.2$~keV during AO3. In the cases of spectra with low
signal-to-noise ratio, the line energy was also fixed at the
rest-frame energy of the Fe K$\alpha$ emission line, 6.4 keV.  The
addition of the Gaussian component led to an improved fit at the
$P_{\rm F} \gtrsim 95$\% confidence level in 6 BLRGs, 1 QSR, 3 NLRGs,
and 1 RG (with a marginal detection at a $P_{\rm F} \gtrsim 90$\%
confidence level in 3 additional RGs). Next, the fit was repeated
allowing the line to be resolved. This gave an improved fit in all
BLRGs and the QSR where the line was detected, with inferred FWHM
$\gtrsim 20,000$ km s$^{-1}$, but not in the NLRGs and the RGs, where
the line is unresolved.  The results of the narrow- and broad-line
fits are reported in Table 4, where we list the fitted line energy
(column 2), Gaussian energy dispersion $\sigma_{\rm rest}$ (column 3),
corresponding FWHM (column 4), and EW (column 5); all quantities have
been transformed to the source rest-frame. In the cases where the Fe
line is not detected, the 90\% upper limit on the EW for a narrow
($\sigma=0.05$ keV) line at a rest energy of 6.4~keV is reported.

While the signal-to-noise ratio of the NLRGs and RGs spectra is, in
general, relatively low, an additional difficulty in detecting an Fe
line in these objects is the presence of a thermal plasma
contribution.  This component contributes to the Fe line emission, as
shown, for example, in the spectra of PKS~2152--69 and Cygnus~A in
Figure~1.  It is worth noting that in several NLRGs and RGs the
residuals plotted in Figure~1 (SIS0 and SIS1) do show some excess flux
around 6 keV at the $\sim 1-2\sigma$ level which suggests a line. It
will be important to observe these weak sources with more sensitive
instruments such as {\it XMM} and {\it Astro-E} to confirm the Fe line
and study its profile.

To refine our analysis of the Fe-line profiles, we fitted the {\it
ASCA} data of the brightest BLRGs and the single QSR exhibiting a
broad line with an accretion disk profile (Fabian et al. 1989;
\verb+diskline+ in \verb+XSPEC+). The model assumes that the line is
produced in inner parts of the disk inclined at an angle $i$, between
the two radii $R_{\rm in}$ and $R_{\rm out}$ (normalized to the
gravitational radius), with an emissivity law $\epsilon \propto
R^{-\beta}$. Following Nandra et al. (1997a), we fixed $R_{\rm in}=6$
and $R_{\rm out}=1000$ (average parameters for Seyfert 1s). The fits
are rather insensitive to the values of the radii. The value of
$\beta$ was allowed to vary while the inclination angle was
constrained by independent information on the radio morphology
(superluminal speeds set an upper limit while a lower limit is given
by the observed size of the large-scale, double-lobed radio source;
Eracleous \& Halpern 1998; Brinkmann et al. 1998). We find that a
disk-like profile model provides a significantly better description
only in the case of the long observation of 3C~390.3, where
$\Delta\chi^2$=15 with respect to the simple Gaussian profile (for the
same number of degrees of freedom). The model parameters are: $E_{\rm
rest}=6.43 \pm 0.06$ keV, EW=$132^{+40}_{-48}$ eV, $i \sim 33^{\circ}$
(pegged at the upper limit allowed by the radio morphology),
$\chi^2_{\rm r}=1.03$ (for 1158 degrees of freedom). Interestingly,
this BLRG is well-known for the double-peaked profile of its Balmer
lines, which has also been interpreted as due to emission from an
accretion disk (Eracleous \& Halpern 1994).

\section{Distribution of Nuclear X-ray Properties}

Figures 2a, 2b, and 2c show the distribution of the intrinsic spectral
properties for the various classes of radio-loud AGN in our sample:
the 2--10~keV photon index $\Gamma$, the 2--10~keV intrinsic
(absorption-corrected) luminosity in the source rest-frame, and the
excess column density $N_{\rm H}$. The latter is defined as the
difference between the fitted X-ray column of the power law component
and the Galactic column in the direction to the source.  Because of
reports that the column densities measured by the {\it ASCA} SIS are
overestimated by as much as $3\times10^{20}$ cm$^{-2}$ (Dotani et
al. 1996), we corrected the column densities in Figure~2c by
subtracting the above amount.  Only the sources where excess column
densities were measured are plotted in Figure~2c.  The average value
and standard deviation of each of the quantities plotted in Figure~2
is reported in Table~5. For the photon index the weighted average is
also listed for comparison with reports in the literature on other AGN
samples. The weighted averages were computed using inverse variance
weights; as such they are dominated by the objects with the highest
signal-to-noise ratios.

\noindent{\it Photon Index} (Figure 2a): All classes of radio-loud AGN
have similar distributions of photon indices in the range 1--2.2, with
mean values $\langle \Gamma \rangle=1.7-1.8$. When the RG class is
divided into WLRGs and remaining sources, the WLRGs have much flatter
spectra on average than any of the other classes, with $\langle \Gamma
\rangle=1.5$. The distributions and corresponding mean values for the
various classes are not formally different according to the
Kolmogorov-Smirnov and Student t-tests.

\noindent{\it X-ray Luminosity} (Figure 2b): The distributions of
intrinsic X-ray luminosity of BLRGs, QSRs, and NLRGs overlap over the
total range $10^{42.5}$--$10^{45.0}$~erg~s$^{-1}$. The luminosity
distribution of RGs is broad, with most (3/5) of the WLRGs populating
the low-luminosity tail at $\lesssim 10^{41}$~erg~s$^{-1}$, more than
two orders of magnitude lower than the other classes. The
distributions of BLRGs, QSRs, and NLRGs are only marginally different:
a KS test yields a probability that the distributions are different of
94.7\% for BLRGs {\it vs.} QSRs, 79.8\% for BLRGs {\it vs.} NLRGs, and
of 96.7\% for QSRs {\it vs.} NLRGs. The distribution of X-ray
luminosities of RGs differs from the distributions of all the other
classes at $\gtrsim$ 99.7\% confidence. 

\noindent{\it X-ray column density} (Figure 2c): NLRGs and RGs have
large intrinsic columns obscuring the source of the power-law
component, with $N_{\rm H} \sim 10^{21}$--$10^{24}~{\rm cm}^{-2}$.  A
somewhat unexpected result is that similarly high excess columns are
also observed in a fraction of BLRGs and QSRs, as apparent from Table
3.  Table 5 lists the average column densities of the various classes;
for BLRGs and QSRs we listed the mean values with and without 3C~445
and 3C~234, respectively, since in these two sources the absorbing
medium appears to have a complex structure (see Table 3).  The BLRG
3C~111 is located behind a molecular cloud in the Galaxy (Bania,
Marscher, \& Barvainis 1991), which could be responsible for the
excess X-ray absorption observed (see discussion in Reynolds et
al. 1998); this source was thus excluded from Figure 2c and the
average column density in Table 5. The fraction of BLRGs and QSRs in
Figure 2c where soft X-ray absorption is observed is 6/9, or 67\%, and
3/5, or 60\%, respectively, for which $\Delta N_{\rm H} \gtrsim 1
\times 10^{20}$ cm$^{-2}$. Excess columns larger than $\Delta
N_{\rm H} \gtrsim 10^{21}$ cm$^{-2}$ are observed in 4/9 (44\%) BLRGs
and in 3/5 (60\%) QSRs. 

\noindent{\it Trends with X-ray luminosity:} There are no apparent
trends when the X-ray photon index and fitted column density of each
individual object are plotted against the X-ray luminosity.  However,
from the {\it average} values of the photon index (weighted average,
dominated by the brightest/best observed sources) and of the 2--10 keV
luminosity in Table 5 there is an apparent trend of flattening slope
with increasing luminosity going from RGs to QSRs (Fig.~2d). A Kendall
non-parametric correlation test gives only a probability of $\sim$
95\% that a correlation is present. WLRGs stick out by having lower
X-ray luminosities and much flatter spectral slopes than the remaining
RGs and all NLRGs. 

\section{Multiwavelength Luminosity Correlations}

In this section we investigate correlations between the {\it
intrinsic} nuclear X-ray luminosity $L_{\rm 2-10~keV}$ (corrected for
absorption and converted to rest-frame energies; Table 3) and the
power at other wavelengths (from Table 1). To estimate the
significance of correlations, we used the survival analysis package
\verb+ASURV+ (Isobe, Feigelson, \& Nelson 1986) which allows us to
perform correlation analysis with censored data and take proper
account of upper limits. We used the Kendall non-parametric test and
calculated the correlation coefficient $\tau_{12}$ and corresponding
probability $P_{\tau}$ that the correlation arises by chance. A small
value of $P_{\tau}$ indicates a significant correlation.

In examing luminosity-luminosity correlations for inhomogeneous
samples like ours, it is customary to estimate the correlation
probability once the common redshift dependence of both variables has
been removed (partial correlation analysis). However, Feigelson \&
Berg (1984) have argued that, in the presence of censored data, there
is no need to remove the redshift dependence once the non-detections
are properly weighted into the correlation test.  The reason is that
measuring a luminosity $L_1$ at a wavelength $\lambda_1$ does not
affect {\it a priori} the distribution of luminosities $L_2$ at a
second wavelength $\lambda_2$ if $L_1$ and $L_2$ are not intrinsically
correlated.  Moreover, since the fluxes at a given wavelength were
generally measured in a uniform way (i.e., same instrument and
sensitivity), the addition of upper limits can only weaken any
intrinsic correlation between two luminosities.  There is, however,
the possibility of spurious correlations arising as a result of a
common dependence on a third luminosity. To evaluate the correlation
between two luminosities removing their common dependence on a third
luminosity, we used the partial correlation test for censored data
developed by Akritas \& Siebert (1996), where upper/lower limits are
properly weighted. The software was kindly provided by M. Akritas.
From the partial Kendall coefficient, $\tau_{12,3}$, and the variance,
the probability $P_{12,3}$ of erroneously rejecting the null
hypothesis (i.e., the probability of no correlation) can be estimated
(Akritas \& Siebert 1996). A small value of $P_{12,3}$ indicates a
significant correlation.  The results of the correlation analysis are
presented in Table 6. The format of the table is as follows: column 1
gives the total number of datapoints (we combine together the various
classes); columns 2 and 3 list the independent variables and number of
upper limits, while columns 4 and 5 list the dependent variables and
number of upper limits. Columns 6 and 7 list the Kendall correlation
coefficient, $\tau_{12}$, and corresponding probability, $P_{\tau}$,
while columns 8--12 give the results of the partial correlation
analysis, namely the third variable and number of corresponding upper
limits, the partial Kendall coefficient $\tau_{12,3}$ and the square
root of the variance, and the partial correlation probability
$P_{12,3}$.  We emphasize that the luminosities used to perform the
correlation tests in Table 6 are intrinsic, {\it rest-frame}
values. Also, the QSR 3C~234 was excluded from every correlation test,
since its true X-ray luminosity is uncertain, and so were 3C~28 and
IC~310 (see Appendix).

It has been long known that the extended radio luminosity of radio
galaxies is strongly correlated to the luminosity of the narrow
emission lines (Rawlings et al. 1989; Rawlings \& Saunders 1991). We
also find a correlation between the 5~GHz lobe radio power and
$L_{[{\rm O\,\sc iii}]}$, with $\tau_{12}$=0.57 and $P_{\tau}=1 \times
10^{-4}$. Thus the lobe radio luminosity was included in the partial
correlation analysis.
 
From Table 6 it is apparent that, after removing the various common
dependences on other variables, the nuclear X-ray luminosity is
primarily correlated with the [{\sc O\,iii}] and 12~$\mu$m
luminosities. These are the two most significant correlations, with a
chance probability of $P_{12,3} \lesssim 0.1$\%. There is also a
weaker correlation between $L_{\rm 2-10~keV}$ and the lobe radio
power, with $P_{12,3} \lesssim$ 3\%. These correlations are shown in
Figure 3a, 3b, and 3c, respectively. The correlations between $L_{\rm
2-10~keV}$ and the core radio and 60~$\mu$m luminosities appear to be
the results of a common dependence on $L_{[{\rm O\,\sc iii}]}$. We
also verified that these correlations survive when the redshift
dependence is removed. Our results hold when the subclass of WLRGs
(which appears to have different properties; Tadhunter et al. 1998,
Laing et al. 1994) is removed.

We note that the correlation between the X-ray and [{\sc O\,iii}]
luminosities, shown in Figure 3a, holds over several orders of
magnitude. The outlier at high [{\sc O\,iii}] luminosity is the NLRG
3C~321 where a starburst could be present (see Appendix). We performed
a linear regression analysis to determine the slope of the correlation
between $L_{[\rm O\,{\sc iii}]}$ and $L_{\rm 2-10~keV}$, excluding
3C~321, the upper limits in Figure~3a, and WLRGs.  We find $L_{[{\rm
O\,\sc iii}]} \propto L_{\rm 2-10~keV}^{1.2}$ or, more precisely,
$$
\log L_{[{\rm O\,\sc iii}]}=(1.2\pm0.2)\; \log \; L_{\rm 2-10~keV}-(11\pm 8)
$$
when NLRGs and RGs are included, and 
$$
\log L_{[{\rm O\,\sc iii}]}=(1.1\pm0.2)\; \log \; L_{\rm 2-10~keV}-(6\pm22)\;
$$
when NLRGs and RGs are excluded (i.e., for BLRGs and QSRs only).  We
also derive an empirical relation between the X-ray and 12~$\mu$m
luminosities, and between the X-ray and radio lobe luminosities, using
only the firm detections and excluding the WLRGs, of the forms:
$L_{\rm 12~\mu m} \propto L_{\rm 2-10~keV}^{0.8}$ or, more precisely,
$$
\log L_{\rm 12~\mu m}=(0.8\pm0.1)\; \log \; L_{\rm 2-10~keV}+(8\pm5)\;, 
$$
and $L_{\rm lobe} \propto L_{\rm 2-10~keV}^{0.9}$ or, more precisely,
$$
\log L_{\rm lobe}=(0.9\pm0.2)\; \log \; L_{\rm 2-10~keV}+(2\pm7)\; .
$$
The error bars are based on the dispersion of the data points about the
best fit. These relationships are plotted in Figures 3a-c (dotted
lines). We conclude that there is a strong link between the nuclear
X-ray power and the power of the [{\sc O\,iii}] emission line, the FIR
luminosity at 12~$\mu$m, and the lobe radio power for the sample of
BLRGs, QSRs, NLRGs, and RGs. This suggests that the same process is
ultimately responsible for powering the emission in all of these bands
supporting and extending the conclusions of earlier studies (e.g.,
Rawlings \& Saunders 1991). The correlation between the nuclear X-ray
luminosity and the lobe radio power is particularly interesting and
somewhat difficult to understand. It is unlikely that there is a
direct causal connection between the two luminosities because the
emitting regions (the AGN and the radio lobes) are separated by
distances of order 100~kpc. We therefore take this correlation as an
indication that the level of activity has been fairly steady over a
time interval corresponding to the light-travel time between the AGN
and the radio lobes, which is of order $10^5$ to $10^6$ years. WLRGs
have underluminous [O{\sc\,iii}] emission lines (Fig.~3a) while being
very powerful in the radio (Fig.~3c), consistent with their defining
properties.

The origin of the non-linear relationship between the [O\,{\sc iii}]
and X-ray luminosities is not clear. Wilkes et al. (1994) find a
non-linear correlation (slope 1.41) between the optical and X-ray
luminosity for a large sample of radio-loud and radio-quiet
quasars. We note that, using only the BLRGs and QSRs of our sample
(for which the optical luminosity is dominated by the nuclear emission
with negligible starlight contamination), we find a correlation with
$L_{\rm opt} \propto L_{\rm 2-10~keV}^{1.03}$. For the same
sub-sample, $L_{[{\rm O\,\sc iii}]} \propto L_{\rm 2-10~keV}^{1.05}$,
and a partial correlation analysis shows the latter is primary, with
P$_{12,3} <$ 0.04\%. Although based on a handful of sources, these
results indicate that the [{\sc O\,iii}] luminosity is linearly and
primarily correlated with the nuclear X-ray luminosity in type-1
radio-loud AGN. The addition of NLRGs and RGs results in a steepening
of the correlation. This could be due simply to a non-linear response
of the gas in the NLRs to the photoionizing continuum.


\section{Summary of Observational Results} 

We studied the X-ray spectral characteristics of a sample of
radio-loud AGN observed with {\it ASCA}, including 10 BLRGs, 5 QSRs, 9
NLRGs, and 10 RGs, the latter including 6 WLRGs. The principal
observational results are:

\begin{itemize}

\item At low X-ray energies, a thermal emission component is present
in the spectra of 67\% of NLRGs and 100\% of RGs, with temperatures in
the range 0.7--6 keV. No such thermal component is detected in any of
the BLRGs or QSRs. 

\item A power-law component is detected above 2 keV in 94\% sources,
including 5/6 WLRGs.

\item The intrinsic 2--10 keV luminosity distribution of the power
law component is similar in BLRGs, QSRs, and NLRGs in the range
$10^{42}$--$10^{45}$ erg s$^{-1}$. The distribution for RGs shows a
low-luminosity tail, populated by the WLRGs, which have $L_{\rm
2-10~keV} \sim 10^{40}$--$10^{42}$~erg~s$^{-1}$.

\item The distribution of the 2--10 keV photon index is similar in
BLRGs, QSRs, and NLRGs, with an average photon index $\langle \Gamma
\rangle=1.7-1.8$.

\item In WLRGs the hard X-rays can be described equally well by either
an absorbed power-law model with $\langle \Gamma \rangle=1.5$ (with 
individual slopes as flat as $\Gamma=1.3$), or a hot bremmstrahlung
model with $kT \sim 100$ keV. 

\item Absorption of the hard power-law component by cold matter is
detected in 71\% of NLRGs and 89\% RGs, with $N_{\rm H} \sim
10^{21-24}$ cm$^{-2}$.  Similar column densities are also measured in
44\% of BLRGs and 60\% of QSRs.

\item No ionized absorption between 0.6--1.0 keV is detected in the
BLRGs and QSRs (with the exception of 3C~390.3). 

\item The Fe K$\alpha$ line is detected in 67\% of BLRGs, 20\% of
QSRs, 33\% of NLRGs, and 30\% of RGs. In BLRGs and QSRs the line is
broad, with ${\rm FWHM} \gtrsim$ 20,000 km s$^{-1}$. In the BLRG
3C~390.3, the Fe K$\alpha$ line profile is consistent with emission
from a disk. In NLRGs and RGs the Fe line is unresolved. Some of these
detections should be regarded with caution because of the poor
signal-to-noise ratio of the spectra (e.g., 3C~321).

\item The nuclear 2--10 keV luminosity is strongly correlated with the
luminosity of the [O{\sc\,iii}] emission line, with a linear trend
($L_{[{\rm O\,\sc iii}]} \propto L_{\rm 2-10~keV}^{1.05}$) for BLRGs
and QSRs, and non-linearly ($L_{[{\rm O\,\sc iii}]} \propto L_{\rm
2-10~keV}^{1.23}$) when NLRGs and RGs are added. The nuclear X-ray
luminosity is also primarily correlated with the FIR emission at
12~$\mu$m ($L_{\rm 12~\mu m} \propto L_{\rm 2-10~keV}^{0.83}$), and
weakly with the lobe radio power ($L_{\rm lobe} \propto L_{\rm
2-10~keV}^{0.92}$).

\end{itemize}

\section{Discussion} 

\subsection{The Soft X-ray Thermal Component}

In about 67\% of NLRGs and 100\% of RGs a thermal component, best
described by a Raymond-Smith plasma model, is detected with {\it ASCA}
at low X-ray energies. This confirms previous results from {\it ROSAT}
and {\it Einstein} that radio galaxies are embedded in a hot, diffuse
medium (Fabbiano 1989; Worrall et al. 1994; Trussoni et al.  1997;
Crawford et al. 1999), as well as more recent results obtained with
{\it SAX} (Padovani et al. 1999). We detect the thermal emission
component over a large range of redshifts ($z=0.007-0.46$) and
intrinsic radio powers ($L_{\rm
lobe}=10^{40.8}$--$10^{44.5}$~erg~s$^{-1}$). While {\it ASCA} does not
have a high enough angular resolution to study the morphology of the
thermal component in detail, its high sensitivity allows us to
determine its spectral properties.

The temperatures and luminosities determined from the {\it ASCA}
spectra have a bimodal distribution. This is best illustrated in
Figure 4, where the measured temperature is plotted against the X-ray
luminosity. For comparison with published values for ``normal''
clusters (see below), we plot the intrinsic 0.1--2.4 keV luminosity of
the thermal components. In four objects (PKS~0131--36, Centaurus~A,
Fornax~A, and 3C~321), it is possible that the thermal component is
related to a starburst or is of nuclear origin (see below and
Appendix). These objects are marked with different symbols (stars) in
Figure~4. It is apparent from Figure 4 that the sources make up two
different groups in the $L-kT$ plane. Several sources are grouped into
a region of low luminosity and temperature ($L_{\rm 0.1-2.4~keV} \sim
10^{40-43}$ erg s$^{-1}$, $kT \sim 1$ keV), close to the values
measured in poor groups (Mulchaey \& Zabludoff 1998) and/or in the hot
coronae of elliptical galaxies (e.g., Fabbiano 1989). The remaining
sources (4C~+55.16, 3C~295, Cygnus~A, 3C~28, IC~310, and PKS~0625--53)
have higher temperatures and luminosities, $L_{\rm 0.1-2.4~keV}
\gtrsim 10^{43}$ erg s$^{-1}$ and temperatures $kT \sim 3-6$ keV,
suggestive of larger-scale emission such as from the intergalactic gas
in clusters of galaxies.  Indeed, for these sources (except for
PKS~0625--53) there is independent evidence from optical data for
cluster associations (see Appendix).  In general, our results support
the findings of Zirbel (1997) that radio galaxies on average inhabit
regions of moderate to low galaxy density.  Unfortunately, our sample
is insufficient to test the conclusion of Zirbel (1997) that FR~I
radio galaxies are in richer environments than FR~IIs.

It is interesting to compare the X-ray spectral properties of the
clusters hosting radio sources to those of clusters without radio
sources. The latter are known to exhibit a tight correlation in the
$L-kT$ plane which evolves slowly with redshift (Mushotzky \& Scharf
1997). Recently, Markevitch (1998) parametrized the relation between
the temperature and luminosity of a number of clusters at $z \sim$
0.05, both with and without the cooling flow region. We show the
resulting relations in Figure~4 in the form of dashed and dotted lines
respectively, and compare it to our measured values for the sources in
the upper-right diagrams which are known to reside in optical galaxy
dense environments. Although the number of objects in our plot is
small, it is apparent that on average the host clusters of radio
sources do not differ significantly from normal clusters with or
without cooling flows. The main outlier is PKS~$0625-53$, which
appears to be underluminous for its temperature [we double-checked the
{\it ASCA} flux of the cluster with that measured by {\it ROSAT} in a
similar energy range (Brinkmann, Siebert, \& Boller 1994), and found
good agreement].  Moreover, the metallicities of the host clusters
inferred from the X-ray spectra are very similar to those of
``normal'' clusters, with an average value $\langle Y \rangle = 0.48$
and dispersion 0.20. These results suggest that the presence of the
radio source does not affect the large-scale properties of its
environment, a conclusion reached independently by Zirbel (1997).

The role of the diffuse hot gas in confining the radio lobes and jets
has been long discussed in the literature (Birkinshaw \& Worrall 1992;
Worrall, Birkinshaw, \& Cameron 1995). For example, Morganti et
al. (1988) studied a sample of radio galaxies in the radio and X-rays
with the {\it Einstein} HRI and found a correlation between the radio
size and the cluster deprojected central density, with smaller sources
inhabiting denser environments. They also find a general balance
between the thermal gas pressure and the radio lobe pressure,
indicating that the gas is responsible for confining the lobes. These
results confirmed the earlier claims of Feigelson \& Berg (1985), who
found a tight correlation between the radio lobe power and the
0.3--4.5 keV luminosities and concluded that the radio emission is
enhanced in denser (and thus more luminous) clusters due to inhibited
adiabatic expansion losses.  However, Miller et al. (1985), examining
a sample of radio galaxies in radio and X-rays, reached opposite
conclusions. These authors showed that the external medium does not
have enough pressure to confine the radio lobes which are continuously
expanding into the gas. It has also been suggested that the external
medium is fueling the central active nucleus through a cooling flow,
and that the core radio emission depends on the amount of gas (and
thus X-ray emission) in the vicinity of the AGN (Fabbiano, Gioia, \&
Trinchieri 1989). Indeed, evidence for a cooling flow has been claimed
for at least three radio galaxies in our sample, 3C~270 (Worrall \&
Birkinshaw 1994), NGC~6251 (Birkinshaw \& Worrall 1992), and 4C~+55.16
(Iwasawa et al. 1999a) on the basis of {\it ROSAT} and {\it ASCA}
data.  In the case of the first two radio galaxies, Worrall \&
Birkinshaw (1994) and Birkinshaw \& Worrall (1992) suggest
specifically that the gas in the cooling flow is feeding the active
nucleus.

If the amount of gas around the AGN is directly related to the
activity in the lobes and/or in the core of the radio galaxy, one
would naively expect to find a correlation between the gas X-ray
luminosity and the radio lobe and core luminosities. We thus looked
for possible correlations between these quantities in those radio
galaxies in which thermal-plasma emission was detected with {\it
ASCA}. We found only weak correlations between the X-ray luminosity of
the thermal components, $L_{\rm X,th}$ (Table 3) and the radio
luminosities of both the lobes ($L_{\rm r,lobe}$) and the cores
($L_{\rm r,core}$): the Kendall correlation coefficient and
corresponding chance probability are $\tau_{12}$=0.463 and $P_{\tau}
\sim$ 5\% in the case of the $L_{\rm X,th}$--$L_{\rm r,lobe}$
correlation, and $\tau_{12}$=0.428, $P_{\tau} \sim$ 7\% for the
$L_{\rm X,th}$--$L_{\rm r,core}$ correlation.  X-ray observations with
higher spatial resolution (e.g., with {\it AXAF}) will undoubtedly
shed more light on the role of the ambient gas in confining the radio
structure and in triggering the nuclear activity.


\subsection{IR Emission from Radio-Loud AGN} 

As shown in Table 1, several radio galaxies are powerful Far-IR (FIR)
emitters. The origin of the FIR emission in radio galaxies is still a
matter of debate. One possibility is thermal emission by dust in the
ISM of the galaxy which is heated by UV light from the AGN (e.g.,
Impey \& Gregorini 1993; Antonucci, Barvainis, \& Alloin 1990). This
possibility is also supported by the detection of
CO($J=1\rightarrow0$) line emission in several radio galaxies, with
large inferred H$_2$ masses (Mazzarella et al. 1993). Another
possibility is that the IR flux is the result of beamed non-thermal
emission from the synchrotron jet (Hoekstra, Barthel, \& Hes 1997).
An attractive feature of this scenario is that it seems to account for
the higher 60~$\mu$m luminosities of quasars compared to NLRGs (Hes,
Barthel, \& Hoekstra 1995; Heckman et al. 1994). While this hypothesis
can hold for the most distant quasars (as in the samples of Heckman et
al. and Hes et al.), we note that the sources in our sample are by
selection at low redshifts and are all lobe-dominated ($\log R <0$),
making the contribution of a beamed component to the FIR emission
negligible.  A third possibility is production of FIR flux in an
advection-dominated accretion flow (ADAF), which will be discussed
below (\S8.4). 

On the other hand, there is abundant observational evidence that dust
is a constituent of the interstellar medium (ISM) of elliptical
galaxies (Knapp et al. 1989; Bregman 1998; Goudfrooij 1998). It is
introduced into the ISM by winds from evolved stars and planetary
nebulae (Tsai \& Mathews 1995) or externally through galaxy-galaxy
mergers (e.g., van Dokkum \& Franx 1995). Obvious locations for the
dust are ``cold'' environments where the dust grains are shielded from
sputtering by the hot radiation of stars and/or the active nucleus,
for example in the commonly-observed dark lanes (e.g., de Koff et
al. 1996). Another possible location is a flat, large disk formed by
the cooling ISM in rotating ellipticals (Brighenti \& Mathews 1997).
An unresolved question is the origin of the radiation that heats the
dust.  The most likely possibilities include young, hot stars in the
galaxy or an active nucleus. The strong correlation we find between
the X-ray nuclear luminosity and the 12~$\mu$m luminosity (\S6)
bolsters the idea that the nucleus is primarily responsible for
heating the cold interstellar dust that emits in the FIR band. This is
also supported by previous studies at other wavelengths which also
established links between the FIR emission and other nuclear-related
activity. For example, Impey \& Gregorini (1993) find a correlation
between the FIR emission and the core radio emission, with more
powerful radio sources having warmer FIR colors. Golombek, Miley, \&
Neugebauer (1988) report that the IRAS fluxes of radio galaxies
correlate with the presence of strong emission lines in their optical
spectra. Interestingly, {\it ASCA} detects large amounts of cold gas
in several radio galaxies with column densities around
$10^{21}$--$10^{23}$~cm$^{-2}$, suggesting the presence of dust in
these objects. In particular, strong excess column densities are
detected in the BLRG 3C~445 and the QSRs 3C~234 and 3C~109. These
objects are among the most powerful FIR emitters in our sample and are
known for exhibiting unusually steep IR-to-optical spectra (Elvis et
al. 1984).

Impey \& Gregorini (1993) find indirect evidence that most radio
galaxies harbor large amounts of cold gas and suggest that
star-formation is occurring in these objects at a rate comparable to
that in spirals (see also Mazzarella et al. 1993). Since the 60~$\mu$m
emission is most sensitive to thermal emission from a starburst (e.g.,
Spinoglio et al. 1995), we have collected in Table 1 the IRAS
60~$\mu$m fluxes of the radio galaxies in our sample. We first note
that only a few objects are detected at 60~$\mu$m, which {\it per se}
is already an indication that a starburst contribution is unlikely in
radio galaxies as a class.  To help identify objects where star
formation could be occurring, we plot in Figure 5 the ratios of the
60~$\mu$m to 12~$\mu$m luminosity versus the nuclear hard X-ray
(power-law) luminosity. We include only those objects which were
clearly detected at 60~$\mu$m. Objects with a starburst contribution
in the FIR should be characterized by excess emission at 60~$\mu$m at
a given nuclear X-ray luminosity.  From Figure 5 it is apparent that
large IR excesses are present only in the cases of Cygnus~A,
Centaurus~A, and the WLRGs PKS~0131--36 and Fornax~A, where the
60-to-12~$\mu$m flux ratios are a factor $\gtrsim$ 2 larger than in
``pure'' AGN (BLRGs and QSRs). Thus we conclude that a starburst
contribution to the observed FIR flux is negligible {\it in general}
in our sample. Indeed, the large radio fluxes locate our sources above
the range in which the FIR/radio correlation for starburst galaxies
holds (Condon, Frayer, \& Broderick 1991).  Nevertheless, a starburst
contribution could be important in some individual cases in our
sample. It is very interesting that two such sources are WLRGs (see
discussion below in \S8.4).

In summary, we conclude that the most likely explanation for the FIR
emission observed in several radio galaxies in our sample is thermal
reradiation from dust in the galaxy ISM illuminated by the active
nucleus. This is supported by the strong correlation between $L_{\rm
2-10~keV}$ and $L_{\rm 12\,\mu m}$ which adds to previous evidence for
a link between the FIR emission and the nuclear activity. A starburst
contribution could be present only in a few individual radio galaxies,
including two WLRGs.

\subsection{Constraints on Unification Schemes for Radio-Loud AGN}

Current unification scenarios for radio-loud AGN postulate that the
differences between classes are merely the result of a different
orientation of the observer relative to the symmetry axis of the
system (Urry \& Padovani 1995 and references therein). An essential
ingredient of these orientation-based scenarios is the large obscuring
torus, which hides the nucleus and the broad emission-line region in
type 2 objects (NLRGs and RGs) but not in type 1 objects (BLRGs or
QSRs).  Observational support for this view comes from optical
spectropolarimetric studies (e.g., Antonucci \& Barvainis 1990), which
generally show the presence of broad emission lines in polarized light
in type 2 objects.

X-ray observations can provide an independent test of these ideas
since the hard (2--10 keV) X-rays can penetrate through the large
absorbing column of the torus. In principle one could get a direct
view of the nucleus at $\gtrsim 10$~keV for column densities up to
approximately $10^{24}$~cm$^{-2}$, above which the torus becomes
optically thick even to these energies. In this context one would
expect NLRGs to look just like BLRGs or QSRs viewed from the side,
i.e., to have a similar hard X-ray continuum shape and luminosity and
heavy absorption. Bearing in mind that our sample is of modest size
and certainly not statistically complete, our results are
qualitatively in agreement with these predictions. The continuum slope
distributions of the three classes are very similar, spanning the same
range of values (see Figure 2a).  The nuclear power law emission in
NLRGs (and RGs) is generally heavily absorbed, with measured column
densities $N_{\rm H} \sim 10^{21}$--$10^{24}$~cm$^{-2}$.  The
distribution of the intrinsic (absorption-corrected) 2--10 keV nuclear
luminosities of NLRGs is similar to that of BLRGs and QSRs, suggesting
that these sources contain nuclei with similar levels of activity, in
general.

Additional support for unification scenarios comes from the strong
correlation between the 2--10 keV luminosity and the [{\sc O\,iii}]
emission-line luminosity (Fig.~3a). Remarkably, this correlation holds
for more than four decades in luminosity from QSRs to RGs.  Besides
supporting unification scenarios {\it qualitatively}, this correlation
also suggests that the narrow emission-line regions of radio-loud AGN
are photoionized by the nuclear emission, similarly to radio-quiet AGN
(Mulchaey et al. 1994).
It is worth noting, however, that the relation between the X-ray and
[{\sc O\,iii}] luminosities is non-linear, $L_{[{\rm O\,\sc iii}]}
\propto L_{\rm 2-10~keV}^{1.2}$, when NLRGs and RGs are
included. A possible explanation for this effect includes  
a higher ionization state of the narrow-line gas with X-ray
luminosity. Our results confirm earlier claims that the luminosity of
the forbidden oxygen lines scales non-linearly with the non-thermal
(optical) continuum, based on a mixed sample of radio-loud and
radio-quiet sources (e.g., Yee 1980; Rawlings et al. 1989).

However, our detection with {\it ASCA} of a population of
underluminous radio sources is not easily to reconcile with a simple
orientation-based scheme. As Figure~2b shows, we find that a sub-class
of RGs, the WLRGs, host nuclei which are underluminous in the hard
X-rays by almost two orders of magnitude than the remaining objects.
This indicates that another possible free parameter of unified
schemes is the nuclear luminosity.  A similar suggestion was also made
by Hill, Goodrich, \& De Poy (1996) based on the consideration that it
is difficult to explain the higher UV, FIR, and [{\sc O\,iii}]
luminosities of quasars with respect to NLRGs based on obscuration
effects alone. Similarly, WLRGs have underluminous [{\sc O\,iii}]
emission lines for their X-ray luminosities (Fig.~3a). We speculate
that this is due to their weaker far-UV ionizing continua (as
discussed below, \S8.5). Thus, the {\it ASCA} data strengthen the case
that the unified schemes in their simplest form are insufficient to
account for the entire observational picture. 

In principle, the profile of the Fe line affords another test of
unification scenarios.  Studies of Seyfert galaxies with {\it GINGA}
and {\it ASCA} have shown that in Seyfert 1s the Fe line is broad and
asymmetric to the red, suggesting an origin in an accretion disk
around the central black hole (Nandra et al. 1997a; but see also the
critique of Weaver \& Yaqoob 1998). In Seyfert 2s, on the other hand,
the Fe line typically has a very large EW (Smith \& Done 1997; Turner
et al. 1997a) and its profile consists of a narrow core, which is the
signature of the outer obscuring torus (Weaver \& Reynolds 1998),
superposed on a broad base. Unfortunately, the {\it ASCA} spectra of
radio-loud AGN set only loose constraints on the Fe line, in practice,
because of their poor signal-to-noise ratio. The poor signal-to-noise
ratio is a particularly serious handicap in NLRGs and RGs because
their spectra are contaminated by a contribution from a
thermally-emitting plasma. We were able to detect the Fe line with
high confidence in two type~2 objects, Cygnus~A and Centaurus~A (see
Table 4), and only marginally in the WLRGs PKS~0131--36 and 3C~270,
and for NGC~6251. In all cases the line is unresolved, with an EW in
the range 100--700 eV. The measured EWs are consistent with
fluorescence in cold material with column densities
$10^{22}$--$10^{23}$~cm$^{-2}$ (Leahy \& Creighton 1993).
Thus the {\it ASCA} data for the Fe line would tend to support
unification schemes in their simple form, but do not allow us to reach
more specific conclusions.

Even in BLRGs the study of the Fe K$\alpha$ line profile is plagued by
practical difficulties, the most important one being the dependence of
the profile on the description of the underlying continuum in the {\it
ASCA} band. This is exemplified by the case of the BLRG 3C~382, where
different authors fit the continuum with different models and find
different line profiles (see discussion in the Appendix). Thus we
cannot confirm the result of Wo\'zniak et al. (1998) that BLRGs as a
class have narrow and weak Fe lines. We measure instead a large range
of EWs in BLRGs which could well be a result of this practical
problem.  Significant progress can be made only with broad-band X-ray
spectra with a high spectral resolution at the Fe line region. Such
spectra can be obtained though simultaneous {\it RXTE} + {\it ASCA}
observations (e.g., Weaver, Krolik, \& Pier 1998). Future instruments,
namely {\it XMM} and {\it Astro-E}, can also provide data that are
useful in this respect.

\subsection{Soft X-ray absorption in radio-loud AGN}

An interesting result of our work is the detection with {\it ASCA} of
soft X-ray excess cold absorption in a fraction (44--60\%) of BLRGs
and QSRs, with excess column densities $\Delta N_{\rm H} \gtrsim
10^{21}$ cm$^{-2}$ (Figure 2c and Table 5). The presence of excess
X-ray absorption is also confirmed by published {\it ROSAT} and {\it
SAX} data of several of these sources (3C~120, Grandi et al. 1997;
3C~390.3, Eracleous et al. 1995, Grandi et al. 1999; 3C~109, Allen \&
Fabian 1992; 4C~+74.26, Brinkmann et al. 1998). In the following
subsections we discuss two possible scenarios for the origin of the
excess X-ray absorption, including a local origin in the Galaxy and
an intrinsic absorber to the source. 

\subsubsection{Origin in the Galaxy} 

A first possibility is that the excess column density is local to the
Galaxy, and is due to absorption by molecular gas and/or dust, to
which radio surveys at 21 cm are insensitive.  A local origin for the
excess absorption is entirely possible in the case of the BLRG 3C~111,
which is located behind a dark cloud in the Galaxy (Bania, Marscher, \&
Barvainis 1991). From the intensity of the $^{12}$CO emission in the
direction of 3C~111, the latter authors estimate that the molecular
hydrogen column density in the direction to this source is $N_{\rm
H_2} \approx 9 \times 10^{21}$ cm$^{-2}$. The total Galactic column in
the direction of 3C~111 is $N_{\rm H}^{\rm Gal} \approx 1 \times
10^{22}$ cm$^{-2}$, enough to account for the excess X-ray column
measured by {\it ASCA} (Table 3; see also the discussion in Reynolds et
al. 1998). We conclude that the excess X-ray column in 3C~111 could
have an origin in the Galaxy, and exclude this source in the following
discussion. 

We checked for CO emission in the direction of all of the BLRGs and
QSRs with highly-absorbed X-ray spectra by examining published surveys
(D\'esert, Bazell, \& Boulanger 1988; Liszt \& Wilson 1993; Liszt
1994). Only 3C~120 was observed, but no CO emission was detected in
the direction to this source (Liszt \& Wilson 1993). Of the remaining
sources, only 4C~+74.26 is at a relatively low Galactic latitude
($b=19.\!\!^{\circ}5$), such that the probability of being associated
with a molecular cloud could be non-negligible. The other sources are
located at high latitudes, $|b| \gtrsim 27^{\circ}$, where the average
density of the molecular gas is low, $\langle N_{\rm H_2} \rangle =
7.3 \times 10^{19}$ cm$^{-2}$ (Blitz, Bazell, \& D\'esert 1990). We thus
conclude that molecular gas is unlikely to be responsible for the high
absorbing columns in most of our BLRGs and QSRs, with the possible
exception of 4C~+74.26.

The total column density of gas {\it and} dust in the Galaxy can also
be traced by means of its emission in the infrared band, in particular
at 100~$\mu$m. The {\it IRAS} survey showed that the Galactic IR
emission consists of several discrete ``cirrus'' features, associated
with dusty clouds with column densities $\sim 10^{20}$ cm$^{-2}$
superposed onto a more diffuse component (Low et al. 1984), with
larger IR fluxes corresponding to larger column densities. From the
available {\it IRAS} 100~$\mu$m maps and adopting a conversion factor
from IR flux to column density of $8.5 \times
10^{-21}$~mJy~sr$^{-1}$/cm$^{-2}$ (Rowan-Robinson et al. 1991), we
derive total column densities in good agreement with the Galactic
columns in Table 1 for all the X-ray absorbed BLRGs and QSRs, showing
that cirrus obscuration is negligible.

In summary, on the basis of the available CO and IR data, we conclude
that contributions from the molecular gas and dust in the Galaxy to
the observed X-ray excess absorption in BLRGs and QSRs is unlikely,
with the exception of 3C~111 and, possibly, 4C~+74.26. In the
remaining sources, the X-ray absorber is extragalactic in origin,
and is most plausibly associated with the AGN itself.

\subsubsection{Intrinsic X-ray absorption} 

Two locations are possible for an intrinsic absorber: the host galaxy
or a surrounding cluster, and the AGN central engine itself.  A
location of the X-ray absorber in the host galaxy (e.g., a cooling
flow) and/or in the larger-scale environment of the radio source
cannot be ruled out {\it a priori}. We note, however, that all the
BLRGs and QSRs from our sample with highly absorbed spectra are found
to inhabit relatively poor optical clusters (see list of Zirbel 1997),
which is common for FR~II radio galaxies. Moreover, there are no
reports of cooling flows around these objects. Instead, the presence
of intrinsic reddening in several cases (3C~445, 3C~109, 3C~234)
supports a nuclear origin for the X-ray absorber, or at for least part
of it. In addition, the latter was observed to vary significantly in
at least one BLRG, 3C~390.3 (Grandi et al. 1999) \footnote{In 3C~390.3
the {\it ASCA} data suggest the presence of an additional component of
warmer absorbing gas. In fact, we detect an absorption edge at $\sim$
0.7 keV (which we interpret as the K-edge of O{\sc\,vii}), with column
densities $N_{\rm H,gas} \approx 8 \times 10^{20}$ cm$^{-2}$ (see
Appendix), similar to the column of cold gas. The presence of this
additional warm absorber could complicate absorption variability
studies in this source.}, supporting a nuclear location. Based on the
above evidence we consider it unlikely that the absorber is associated
with the host galaxy or its environment. 

The {\it ASCA} data provide initial constraints on the physical state
of the X-ray absorber in most sources.  A cold gas is clearly
indicated in the cases of 3C~390.3, 3C~445, 3C~109, and 3C~234,
because of the shape of the spectra at soft X-rays (Table 3 and Figure
1). This is also supported by {\it ROSAT} PSPC and {\it SAX}
observations of these objects (Allen \& Fabian 1992; Sambruna et
al. 1998; Grandi et al. 1999). In 3C~445 and 3C~234, there seems to be
reprocessing of the soft X-ray flux (modelled as a dual
absorber/partial covering; see also Appendix). The presence of cold
gas in these BLRGs and QSRs is at odds with the view that the line of
sight to the nucleus in type~1 objects should be free of cold
material. 

In the case of 3C~303, 4C~+74.26, and PKS~2155+11 the nature of the
absorber detected by {\it ASCA} is less clear. Although the data are
well fitted by a cold absorber (Table 3), a warmer gas can not be
excluded. In the latter case, the most prominent features would be
absorption edges of highly-ionized oxygen (O{\sc\,vii/viii} at
rest-frame energies 0.7--0.8~keV). These would be redshifted to
0.5--0.6~keV, at the cutoff of the SIS sensitivity, mimicking ``cold''
absorption. Future broad-band, high-resolution X-ray observations of
these sources will provide better constraints.

Our BLRGs and QSRs with excess X-ray absorption are not isolated
cases. X-ray column densities larger than Galactic were also measured
with the {\it ROSAT} PSPC in the BLRGs 3C~332 ($z$=0.151), 3C~287.1
($z$=0.216), and Arp~102B ($z$=0.0244) (Crawford \& Fabian 1995;
Halpern 1997). Interestingly, soft X-ray absorption is also detected
with {\it ASCA} and {\it ROSAT} in about 30\% of high-redshift
(radio-loud) QSRs up to $z \sim 3$ (Elvis et al. 1994, 1998; Cappi et
al. 1997; Fiore et al. 1998), with intrinsic column densities $\sim
10^{22}$ cm$^{-2}$. These columns were not observed in radio-quiet
QSOs of similar redshift, suggesting the excess absorption is found
preferentially in radio-loud objects.  Based on the statistical
association of X-ray and optical absorption, Elvis et al. (1998)
propose that the soft X-ray absorption is due to a photoionized gas in
the central engine of the quasar. This interpretation is supported in
a few sources of intermediate redshift ($0.4 \lesssim z \lesssim 1$)
by X-ray spectra of high signal-to-noise ratio, which allow detailed
modelling (Fiore et al. 1993; Mathur et al. 1994). Absorption in a
dusty nuclear outflow has also been suggested (Mathur 1994).

Assuming that the absorber is intrinsic in all the BLRGs and QSRs of
our sample, we determined the intrinsic (rest-frame) column densities
by fitting an absorption model to their {\it ASCA} spectra. The model
consisted of the best-fit continuum model in Table 3, plus local
absorption ($z=0$, column fixed to the Galactic value of Table~1) and
intrinsic absorption in the rest-frame of the AGN ($z_{abs}=z_{\rm
source}$, column treated as a free parameter).  The best-fitting
values of the intrinsic column, $N_{\rm H, intr}$, are given in
Table~7 together with the absorption-corrected 2--10 keV
luminosity. 3C~120 and Pictor~A are not included, since $N_{\rm
H, intr}$ is $\lesssim 3 \times 10^{20}$ cm$^{-2}$, consistent
with the SIS systematic uncertainties (Dotani et al. 1996). The
sources reported in Table 7 are the BLRGs and QSRs of our sample where
{\it bona fide} excess X-ray absorption is found from the {\it ASCA}
data.

Figure~6 shows the plot of $N_{\rm H, intr}$ versus the X-ray
luminosity for the BLRGs and QSRs of Table~7 (filled dots and
triangles, respectively). Also plotted for comparison are the high-$z$
radio-loud quasars from the samples of Cappi et al. (1997) and Elvis
et al. (1994), where large intrinsic columns were detected with {\it
ROSAT} and {\it ASCA} (open triangles). The three open dots represent
the BLRGs Arp~102B, 3C~332, and 3C~287.1, where excess X-ray columns
were measured with the {\it ROSAT} PSPC (Halpern 1997; Crawford \&
Fabian 1995); the asterisks are the QSRs 3C~351 and 3C~212, where an
ionized absorber is thought to be present (Fiore et al. 1993; Mathur
1994). It is noteworthy that Arp~102B and 3C~332 also feature
metastable Fe{\,\sc ii} absorption lines in the near UV which could
originate in the same medium that is responsible for the X-ray
absorption (Halpern 1997). It is also very interesting that a few of
the heavily X-ray-absorbed BLRGs and QSRs of our sample, including
3C~445, 3C~234, and 3C~109, exhibit strong polarization of the optical
continuum and/or broad lines (Antonucci 1984; Goodrich \& Cohen 1992;
Corbett et al. 1998; Kay et al. 1999), confirming the presence of
large amounts of gas along very close to their central engines.

Figure~6 shows no obvious trend between the absorbing column density
and the X-ray luminosity. On the contrary, it appears that radio-loud
AGN exhibit similar excess X-ray column densities over more than four
orders of magnitude in luminosity. This observation contradicts the
``receding torus'' scenario of Hill et al. (1996) in which the
absorbing medium subtends a smaller solid angle to the primary X-ray
source with increasing luminosity.


In conclusion, we detected with {\it ASCA} excess X-ray absorption in
a fraction of BLRGs and QSRs, also confirmed in a few cases by
independent observations with other X-ray instruments. A cold
absorber, most likely of nuclear origin, is the favored explanation in
most cases, although a higher-ionization gas cannot be ruled out in
the more distant sources of our sample based on the {\it ASCA} data
alone. Interestingly, the excess X-ray columns are similar to those
observed in higher-$z$ radio-loud quasars over more than 4 decades in
X-ray nuclear luminosity. This stresses the importance of the X-ray
absorber as a fundamental constituent of radio-loud AGN. Future
higher-resolution X-ray observations are necessary to clarify the
physical and dynamical conditions of the medium, and constrain its
location.

\subsection{The Low-Luminosity AGN in Weak Line Radio Galaxies} 

Perhaps one of the most interesting results of our study is the
detection of WLRGs in hard X-rays.  As discussed above, these objects
represent a relatively recent discovery, sticking out in the sample of
Tadhunter et al. (1998) as powerful radio-galaxies with underluminous
[{\sc O\,iii}] emission lines and line ratios indicative of a low
ionization state of the narrow-line region.  Radio galaxies with
similar properties were also discussed by Laing et al. (1994) who
identified a class of ``low-excitation'' radio sources and noted that
they formed a separate group with distinct optical properties.
Tadhunter et al. (1998) advance the hypothesis that the most likely
explanation for the properties of WLRGs is that they harbor a low
luminosity AGN. The {\it ASCA} data presented here give us the first
view of these systems at hard X-rays.  We find that in five out of the
six WLRGs observed, the X-ray spectrum can be decomposed into a hard
X-ray component plus a soft thermal component with $kT \sim 1$
keV. The hard component can be described by either a flat power law,
with $\langle \Gamma \rangle=1.5$ (and individual slopes as flat as
$\Gamma=1.3$). The intrinsic luminosity of the hard component is
$L_{\rm 2-10~keV} \sim 10^{40}-10^{42}$~erg~s$^{-1}$, two orders of
magnitude fainter than in the other radio sources.

The [{\sc O\,ii}]/[{\sc O\,iii}] line ratios of WLRGs locate them in
the LINER/H\,{\sc ii}-region part of the diagnostic line-ratio
diagrams of Filippenko (1996), raising the possibility that WLRGs
represent the radio-loud analogs of LINERs. In fact, Fornax~A is
included in the LINER sample studied by Fabbiano (1996) on the basis
of its optical spectroscopic properties presented by Baum, Heckman, \&
van Breugel (1992). The similarity between LINERs and WLRGs is
reinforced by the ubiquitous soft X-ray thermal component, which could
be related to a starburst in one (PKS~0131--36) and possibly two
(Fornax~A) of the WLRGs.  As in WLRGs, the {\it ASCA} spectra of
LINERs and other low-luminosity AGNs are also made up of a soft,
thermal component plus a hard power law.  However, the spectra of
(radio-quiet) LINERs are considerably steeper than those of WLRGs of
similar luminosities. The photon indices in LINERs are $\Gamma \sim$
1.7--1.8 (Serlemitsos, Ptak, \& Yaqoob 1996), compared to $\Gamma
\sim$ 1.5 in WLRGs. We note that the X-ray spectra of WLRGs are also
much flatter than those of all other radio-loud AGN in our sample,
suggesting a fundamental difference.

There is considerable debate as to whether the hard X-ray emission of
LINERs originates in a low-luminosity active nucleus or whether
alternative mechanisms (e.g., the collective emission from a large
number of X-ray binaries at near-Eddington luminosities) are more
viable.  Similar arguments hold for WLRGs.  In the case of WLRGs,
however, the presence of kpc-sized radio jets and lobes as well as
compact radio cores makes an ironclad case for the presence of an AGN.
The real question is whether the observed X-ray flux comes from the
AGN or from some other process. While only future high-resolution {\it
AXAF} observations will settle the issue definitively, we can still
make a few comments here. Available {\it ROSAT} PSPC and HRI images do
not show evidence for discrete X-ray binaries or other point-like
sources in Fornax~A (Fabbiano 1996; Kim, Fabbiano, \& Mackie 1998), in
3C~270 (Worrall \& Birkinshaw 1994), or in PKS~$0131-36$ (archival
{\it ROSAT} HRI data). In 3C~270, the {\it ROSAT} PSPC data show
explicitly the presence of a spatially unresolved component, with a
power law spectrum, which is attributed to an AGN (Worrall \&
Birkinshaw 1994).  There is marginal evidence for an Fe line in the
{\it ASCA} spectra of two WLRGs, with EWs $\sim$ 250 eV. If confirmed
this would provide support in favor of an AGN origin of the bulk of
the X-ray luminosity. 

An interesting possibility is that the accretion flows in the active
nuclei of WLRGs are advection dominated (hereafter ADAFs; Ichimaru
1977; Rees et al. 1982; Narayan \& Yi 1994, 1995; see Narayan,
Mahadevan, \& Quataert 1998 for a recent review). ADAFs are possible
if the accretion rate is very low relative to the Eddington rate,
i.e., when $\dot m \lesssim \alpha^2 \dot m_{\rm Edd}$ (where $\alpha$
is the Shakura-Sunyaev viscosity parameter). At these low densities,
the Coulomb collisions between the ions and the electrons are an
inefficient means of equilibration, and the plasma settles into a two
temperature phase, with the ions being much hotter than the electrons
($T_{\rm i} \sim 10^{11}$ K, $T_{\rm e}\sim 10^9$ K). The ions take
most of the potential energy and advect it to the black hole, while
the relativistic electrons radiate a small fraction of it. The
emission is by the synchrotron, inverse Compton, and bremmstrahlung
mechanisms, with the output in the radio, optical, and hard X-rays,
respectively. The X-rays in particular provide very sensitive
constraints on ADAF models, since the slope of the X-ray spectrum is a
sensitive function of the accretion rate (see, for example, Figure~6
of Narayan et al. 1998). At low densities the dominant X-ray emission
mechanism is bremmstrahlung with $h\nu_{\rm B,max} \sim kT_{\rm e}
\sim 100$ keV (Narayan et al. 1998), while when the density increases
inverse Compton losses are more important and the X-ray spectrum is a
steeper power law.  ADAFs have been succesful in accounting for the
relatively flat X-ray spectra and low X-ray luminosities of NGC~4258
(Lasota et al. 1996), M87 (Reynolds et al.  1996), and M60 (Di Matteo
\& Fabian 1997). These models are particularly attractive for
radio-loud AGN since some of the gas in an ADAF is unbound and may
escape in an outflow, potentially forming a jet (Blandford \& Begelman
1999; see also Narayan \& Yi 1994, 1995 for ADAFs and relativistic
jets). 

The alternative possibility that the flat X-ray slopes of WLRGs are due
to the beamed contribution from the inner parts of the radio jet is
ruled out by the lack of correlation between their nuclear X-ray
luminosities and the core radio emission. In fact, all WLRGs in the
present sample are lobe-dominated (Table 1), assuring that the jet
emission is beamed away from the line of sight. 

Are ADAFs consistent with the observed low luminosities and flat X-ray
slopes of WLRGs? We study the case of 3C~270 (hosted by the giant
elliptical NGC~4261) since for this WLRG there is an accurate
dynamical estimate of the central mass from recent {\it HST}
observations (Ferrarese, Ford, \& Jaffe 1996): $M_{\rm BH}=(4.9\pm
1.0) \times 10^8$ $M_{\odot}$ (a factor 10 lower than in M87). This
implies an Eddington luminosity $L_{\rm Edd} \sim (5-8) \times
10^{46}$ erg s$^{-1}$. The radio-to-X-ray spectral energy distribution
of 3C~270 is shown in Figure~7 (using data from Table 1, the {\it
ASCA} data, optical nuclear fluxes from Ferrarese et al. 1996, and an
UV upper limit from the {\it HST} observations of Zirbel \& Baum
1998).  Also plotted for comparison is the spectral distribution of
M87, where an ADAF is thought to be present (Reynolds et al. 1996),
and for which a recent analysis of the {\it ASCA} data has given a
flat X-ray slope, $\Gamma=1.40^{+0.37}_{-0.46}$ (Allen, Di Matteo, \&
Fabian 1999), similar to what we measure in 3C~270 and other
WLRGs. Integration of the flux of 3C~270 from radio to X-rays gives a
bolometric luminosity of $L_{\rm Bol} \sim 2 \times 10^{43}$ erg
s$^{-1}$; this probably overestimates the true nuclear nuclear
luminosity, because it includes the IR datapoints, where contribution
from the nuclear dust could be present. Thus, $L_{\rm Bol}/L_{\rm Edd}
\lesssim (2-4) \times 10^{-4}$, placing 3C~270 in the ADAF
regime. Note the lack of UV continuum emission, too shallow to be
entirely due to dust absorption. We speculate that the lack of a
strong far-UV ionizing continuum is the cause of the observed
underluminous [{\sc O\,iii}] emission lines.

It is worth noting that not all WLRGs have abnormally flat X-ray
continua even though they all have low X-ray luminosities. For
example, in 3C~353 the measured 2--10 keV photon index is 1.9, very
similar to the values found in Seyfert galaxies.
Interestingly, 3C~353 is a powerful FR~II radio galaxy, while 3C~270
is a lower-power FR~I radio galaxy\footnote{Of the 23 WLRGs in
Tadhunter et al. (1998), 61\% are of the FR~I type while 40\% have
FR~II morphologies.}. 
It is tempting to speculate that WLRGs represent that segment of
the population of radio-loud AGN in which the accretion rate is so low
that an ADAF is inevitable. The exact shape of the X-ray spectrum
would depend on the exact value of the accretion rate, hence a range
of spectral slopes would be possible. 

Also interesting is the tentative detection with {\it ASCA} of an Fe
line in two WLRGs, 3C~270 and PKS~0131--36 (Table 4). Future
observations of these and other WLRGs at higher-sensitivity (e.g.,
with {\it XMM}) will allow us to study the line profile and energy,
discriminating among possible scenarios for its origin. Particularly
intriguing is the possibility that the Fe line originate from the
coronal gas in the outer parts of the ADAF (Narayan \& Raymond
1999). In this case, soft X-ray lines would be expected as well, and
from their relative strength the size and dynamical conditions of the
ADAF can be diagnosed (Narayan \& Raymond 1999).

In the context of ADAF models a large energy output in the IR/FIR band
is expected (e.g., Narayan et al. 1998). Since the FIR flux originates
in the thin disk exterior to the ADAF, it would be expected to vary on
short timescales. FIR flux variability in WLRGs could thus provide a
diagnostic tool to discriminate between an AGN (ADAF) and a
dust/starburst origin for their FIR emission.


\subsection{The Radio-Loud/Radio-Quiet AGN Dichotomy} 

Various samples of radio-quiet AGN have been studied with {\it ASCA}
by different authors recently, allowing a systematic comparison with
our sample of radio-loud objects. The main advantage of comparing
samples of radio-loud and radio-quiet AGN observed with the same
instrument is that the systematic effects are the same and hence are
unlikely to affect the results. In particular, compilations of {\it
ASCA} results for Seyfert 1s, Seyfert 2s, and QSOs have been presented
by Nandra et al. (1997a), Turner et al. (1997a), and Reeves et
al. (1997), respectively. We extracted from these studies subgroups of
objects in a range of intrinsic (rest-frame) 2--10 keV luminosities
matching the range of the corresponding subclass of radio-loud
objects. We also used the collection of LINERs from Serlemitsos et
al. (1996) for comparison with the WLRGs. This gave 7 Seyfert 1s, 13
Seyfert 2s, 7 QSOs, and 4 LINERs. Their average photon indices and
luminosities are reported in Table 5.

The distribution of spectral indices of Seyfert~1s, Seyfert~2s, and
QSOs are compared to those of their radio-loud counterparts in Figure
2a (dotted histograms). A Kolmogorov-Smirnov test gives a probability
of only $P_{\rm KS} \sim$ 94\% that the distribution for Seyfert 1s
and BLRGs are different, while their average photon indices, $\langle
\Gamma_{\rm BLRGs} \rangle=1.6$ and $\langle \Gamma_{\rm Sy~1s}
\rangle=1.9$, differ at 93\% according to a Student t-test.  The
distribution of photon indices for the QSR/QSO, NLRG/Sy~2, WLRG/LINER
samples are not demonstrably different. The average values for the
last pair differ at the 95\% confidence level, with WLRGs having
flatter spectra than LINERs.

We thus conclude that {\it ASCA} does not provide evidence that
radio-loud AGN have flatter X-ray spectra than their radio-quiet
counterparts.  The observed difference in photon indices between BLRGs
and Seyfert~1s (the only case where the two distributions are
significantly different) is $\Delta\Gamma \sim$ 0.3. This is less than
the $\Delta\Gamma \sim$ 0.5 found in samples observed with {\it
Einstein} (Wilkes \& Elvis 1987) and {\it EXOSAT} (Lawson et
al. 1992). One difference from earlier results is that we have matched
the radio-loud and radio-quiet AGN samples in X-ray luminosity, which
was not always possible in previous studies. In so doing, we took into
account possible intrinsic trends with luminosity and we compared
spectral slopes by subclass. More importantly, our sample also differs
from previous ones in including only lobe-dominated radio sources in
which beaming is unlikely to be important. The effect of beaming was
more important in samples presented in the literature (Wilkes \& Elvis
1987; Shastri et al. 1993; Lawson \& Turner 1997), especially when
high-redshift objects were included. Our results thus show that the
X-ray spectral slopes of radio-loud and radio-quiet AGN are
intrinsically similar, confirming that a beamed component from the
inner regions of the radio jet is responsible for the flatter slopes
previously observed.


A leading model for the production of the X-ray continuum in Seyfert
galaxies is Comptonization of thermal photons from the disk (Haardt \&
Maraschi 1993), which can produce photon indices in the large range
1.4--2.0. This model can thus accomodate the observed spectral indices
of radio-loud AGN of $\Gamma \sim 1.7$ and there is no need for
special geometry or physical processes for them {\it based on their
{\it ASCA} spectral slopes alone.}  Besides, a few Seyferts also
exhibit flatter than average spectral indices, which are comparable to
those of WLRGs. Examples include Mrk~841 (George et al. 1993) and
1H~$0419-577$ (Turner et al. 1999). 

Perhaps the most intriguing observational evidence so far in favor of
differences in the accretion flows of radio-loud and radio-quiet AGN
is provided by high-energy observations.  Wo\'zniak et al. (1998)
studied archival {\it GINGA} and {\it CGRO}/OSSE spectra of BLRGs and
reported that these objects have weaker Compton reflection components
than Seyferts (Nandra \& Pounds 1994; Weaver et al. 1998). These
findings are also confirmed by recent observations with {\it RXTE}
(Eracleous, Sambruna, \& Mushotzky 1999, Rothschild et al. 1999) and
{\it SAX} (Grandi et al. 1999). The simplest interpretation is that
the dense, cold reprocessor in radio-loud objects subtends a smaller
solid angle to the central X-ray source. This would be the case, for
example, if the inner disk were of a vertically extended,
quasi-spherical structure such as an ion torus (Rees et al. 1982), or
an ADAF (see the estimate of Chen \& Halpern 1989).  The lower photon
indices measured by {\it ASCA} for radio-loud AGN fit well into this
picture, since flat X-ray continua are expected from ADAFs.  Also, the
non-linear variability trends of the X-ray flux, such as observed for
3C~390.3 (Leighly \& O'Brien 1997), find a natural explanation in
ADAFs.

If the cold reprocessor in BLRGs is indeed at larger distances from
the central black hole than in radio-quiet AGN, then one expects also
to observe narrower Fe K$\alpha$ lines in BLRGs. Indeed Wo\'zniak et
al. (1998) claimed to have observed such an effect, although we can
neither confirm nor disprove their results.  Figure 8 shows a plot of
the Fe line EWs, or their 90\% upper limits, versus intrinsic
continuum X-ray luminosity for the BLRG and QSR sub-classes of
radio-loud AGN that we have studied (updated from Eracleous \& Halpern
1998). The shaded area is the $EW-L_{\rm X}$ $1\sigma$ relationship
(the X-ray Baldwin effect) for radio-quiet AGN found by Nandra et
al. (1997c).  Radio-loud AGN have a large dispersion of EWs around the
shaded area. We suggest that uncertainties in determining the
continuum in the vicinity of the Fe line limit one's ability to
measure the properties of the line profile. 

Another result of our work is the edence for different gas
environments in radio-loud AGN. Recent systematic studies of Seyfert
1s with {\it ASCA} (Reynolds 1997; George et al. 1998) show that a
large fraction (13/18, or 72\%) of the objects studied exhibit ionized
absorption, which manifests itself as absorption edges in the observed
energy range 0.6--0.8 keV. Thanks to the sensitivity of {\it ASCA} at
energies $\lesssim$ 1 keV it is possible to set interesting limits to
the amount of ionized gas along the line of sight in the BLRGs of our
sample.  We find that only one BLRG (3C~390.3) exhibits an absorption
edge at $\sim$ 0.78 keV (which we thus interpret as the O{\sc\,vii}
edge), with optical depth $\tau\sim 0.3$. The latter corresponds to a
column density of ionized gas $N_{\rm H,gas} \sim 8 \times 10^{20}$
cm$^{-2}$, at the lower end of the distribution for Seyfert 1s of
similar X-ray luminosity (Reynolds 1997; George et al. 1998).  Thus,
radio-loud AGN appear to have smaller amounts of ionized gas around
their nuclear X-ray sources than Seyferts. Instead we have detected
large columns of cold gas in BLRGs and QSRs (Fig.~6), which are
uncommon in Seyfert galaxies. This result stresses a fundamental
difference in the circumnuclear environment of radio-loud and
radio-quiet AGNs, which could be important for the formation and
collimation of jets, and which needs to be further explored by future
{\it AXAF} and {\it XMM} higher-resolution observations. For example,
the absorbing medium can have the form of a cold accretion-disk wind,
which is invoked in models for jet collimation (e.g., Blandford \&
Payne 1982; K\"onigl \& Kartje 1994). It is interesting to note that
radio-loud quasars exhibit narrow UV absorption lines much more
frequently than radio-quiet quasars (Foltz et al. 1986; Anderson et al
1987), which may be an additional manifestation of the X-ray absorber.


\section{Conclusions} 

We have studied the X-ray spectral properties of a sample of
radio-loud AGN observed with {\it ASCA}. The sample includes 10 BLRGs,
9 NLRGs, 5 QSRs, and 10 RGs observed by {\it ASCA} up until 1998
September, with enough detected counts to allow spectral analysis. 

We find that the soft X-ray emission of about 50\% of the sources is
dominated by a thermal component, most likely associated to the hot,
diffuse gas detected in {\it ROSAT} images. The temperatures and
luminosities determined from the {\it ASCA} spectra have a bimodal
distribution, with some sources being dominated by cluster-type
emission while others having parameters typical of loose groups or hot
coronae. There is no difference in the temperature and luminosity
between clusters hosting radio galaxies and those without, suggesting
that the presence of an active nucleus is not closely connected to the
larger-scale properties of the environment. Future higher-resolution
observations with {\it AXAF} will elucidate the role of this external
gas in confining the radio jets and for fueling the central engine.

At hard X-rays, a power-law component is detected in most sources,
including NLRGs and RGs where it is heavily absorbed. The measured
2--10 keV intrinsic luminosities and photon indices have similar
distributions in BLRGs, QSRs, and NLRGs, with $L_{\rm 2-10~keV} \sim
10^{42}$--$10^{45}$~erg~s$^{-1}$ and $\langle \Gamma_{\rm 2-10~keV}
\rangle=$1.7--1.8. This is in agreement with simple orientation-based
unification scenarios, which postulate that all subclasses of AGN have
fairly similar central engines. The presence of significant amounts of
cold gas with column densities $N_{\rm H}\sim
10^{21}$--$10^{24}$~cm$^{-2}$ in NLRGs and RGs supports the presence
of an obscuring torus in these objects. However, significant excess
column densities are also detected in a fraction of BLRGs (4/9, or
44\%) and QSRs (3/5, or 60\%), in contrast to Seyfert~1
galaxies. Another pronounced difference between BLRGs and QSRs and
their radio-quiet counterparts is the absence of warm absorbers in the
radio-loud objects. In particular, absorption edges from
highly-ionized species, a hallmark of the X-ray spectra of Seyfert 1s,
are detected in only one BLRG with a column density that falls at the
lower end of the distribution among Seyferts of similar intrinsic
luminosity. These observational results suggest a systematic
difference in the conditions in the gas surrounding the X-ray source
in radio-loud and radio-quiet AGN, that needs to be studied with
future higher-resolution observations. 

The X-ray luminosity of the power-law spectral component is primarily
correlated with the luminosity of the [{\sc O\,iii}] emission line, in
a linear way for the sub-sample of BLRGs and QSRs ($L_{[{\sc O\,
iii}]} \propto L_{2-10~keV}^{1.05}$); the relationship steepens when
NLRGs and RGs are added ($L_{[{\sc O\,iii}]} \propto
L_{2-10~keV}^{1.23}$). The X-ray nuclear luminosity is also primarily
correlated with the FIR power at 12~$\mu$m ($L_{\rm 12~\mu m} \propto
L_{\rm 2-10~keV}^{0.83}$), and with the lobe radio power ($L_{\rm
lobe} \propto L_{\rm 2-10~keV}^{0.92}$), indicating that the same
basic physical process is responsible for turning on the nucleus and
feeding the distant radio lobes.

The Fe~K$\alpha$ line is detected in 50\% BLRGs and 20\% QSRs with a
large dispersion of EWs and velocity widths. The line is generally
broad with $FWHM \gtrsim 20,000$ km s$^{-1}$. In the few NLRGs and RGs
where the line is detected, it is narrow and unresolved, with EWs
consistent with an origin through fluorescence in cold material, such
as an obscuring torus.

Particularly intriguing is the detection in hard X-rays of the
subclass of WLRGs. The {\it ASCA} spectra of 5/6 WLRGs in our sample
are decomposed into a hard power law with $\langle \Gamma \rangle=1.5$
(with individual slopes as flat as $\Gamma=1.3$) and $L_{\rm 2-10~keV}
\sim 10^{40}$--$10^{42}$~erg~s$^{-1}$ and a Raymond-Smith thermal
plasma model with $kT \sim 1$ keV.  Based on their optical and X-ray
properties, WLRGs could be the radio-loud counterparts of LINERs,
although with considerably flatted hard X-ray spectra. If a
low-luminosity AGN is present in WLRGs (which seems to be the case in
most of the sources discussed here), we suggest that the latter is
fuelled at a rate well below the Eddington rate. Starburst activity
could also be present at soft X-rays, and, viewed from this
perspective, the true nature of some WLRGs may be different to
ascertain, as for LINERs.  Future {\it AXAF} and {\it XMM}
observations may clarify the picture.


We compared the X-ray spectral slopes of radio-loud AGN to those of
their radio-quiet counterparts in matched ranges of intrinsic
luminosity. We find only weak indication (94\% confidence) that BLRGs
have flatter X-ray spectra than Seyfert 1s by $\Delta \Gamma \sim
0.3$, while the slopes of the remaining subclasses (NLRGs/Seyfert 2s,
QSRs/QSOs) are not demonstrably different. This shows that the X-ray
continuum slopes of radio-loud and radio-quiet AGN are intrinsically
similar, confirming that a beamed component from the inner regions of
a relativistic jet is responsible for the flatter slopes observed in
radio-loud sources in previous works. 


\acknowledgements
 
RMS acknowledges support from NASA contract NAS--38252 and from an NRC
Fellowship at the Laboratory for High Energy Astrophysics at Goddard
where this research was started. During the early parts of this work,
ME was based at the University of California, Berkeley, and was
supported by a Hubble Fellowship (grant HF-01068.01-94A from the Space
Telescope Science Institute, which is operated for NASA by the
Association of Universities for Research in Astronomy, Inc., under
contract NAS~5-26255). We are grateful to Michael Akritas for kindly
sharing his code for partial correlation analysis of censored data,
and Eric Feigelson for interesting conversations about
astrostatistics. We also thank Tahir Yaqoob for all his help with
several questions about the {\it ASCA} data analysis. The referee,
Belinda Wilkes, provided constructive criticism and many comments
which improved the presentation of the paper. This research made use
of the {\it ASCA} data archive at HEASARC, Goddard Space Flight
Center, and of the NASA/IPAC Extragalactic Database (NED) which is
operated by the Jet Propulsion Laboratory, California Institute of
Technology, under contract with the National Aeronautics and Space
Administration.

\newpage
\appendix{\bf {Appendix: Comments on individual sources}}

\noindent{\bf 3C~111:} The {\it ASCA} data were previously studied by
Reynolds et al. (1998).  Our results are in agreement with theirs. In
particular, we confirm the detection of the Fe line
($\Delta\chi^2$=10.6 over the single absorbed power law for 3
additional parameters) with an EW $\sim$ 100 eV. We fixed the line
energy and width in Table 4 to their best-fitting values (E=6.4 keV,
$\sigma=0.50$ keV) to derive a better estimate of the EW. Because the
source is located behind a dark molecular clouds in the Galaxy (Bania
et al. 1991), it is unclear whether the excess X-ray absorption
detected with {\it ASCA} is local or intrinsic to the source (see
discussion by Reynolds et al. 1998).  No ionized absorption is
detected, with a 90\% confidence upper limit to the O{\sc\,vii}
K-edge of $\tau_e < 1.0$.

\noindent{\bf 3C~120:} Our results for the continuum are in full
agreement with previous analysis by Grandi et al. (1997). Our
best-fitting model for the Fe line, however, gives a larger width and
EW than found by those authors ($\sigma=0.89^{+0.17}_{-0.15}$ keV,
EW=509$^{+67}_{-85}$ eV), with larger uncertainties on the fitted
parameters. A possible cause of the discrepancy is that we used the
SIS data in \verb+BRIGHT2+ mode, while Grandi et al. (1997) used the
data in \verb+BRIGHT+ mode, and fixed the continuum slope during the
fits.  There is no evidence for a warm absorber; the optical depth of
the O{\sc\,vii} K-edge is unconstrained in the fit.

\noindent{\bf Pictor~A:} The {\it ASCA} data were analyzed by
Eracleous \& Halpern (1998). Our results are in full agreement with
theirs. The {\it ASCA} column density, $N_{\rm H} \sim 8 \times
10^{20}$ cm$^{-2}$, and photon index, $\Gamma \sim 1.8$, are
completely consistent with previous measurements with {\it EXOSAT} in
a slightly higher ($F_{\rm 2-10~keV} \sim 1.7 \times 10^{-11}$ erg
cm$^{-2}$ s$^{-1}$) intensity state (Singh, Rao, \& Vahia 1990). The
latter authors also report the detection of an Fe emission line
between 6 and 8 keV with a large EW $\sim$ 1.3 keV, which is not
confirmed with {\it ASCA}. Recent {\it SAX} observations of Pictor~A
showed a thermal-plasma component with $kT \sim 0.6$ keV at low
energies, contributing $\sim$ 5\% of the power law flux at 0.1--2.4
keV (Padovani et al. 1999). Adding a thermal component to the {\it
ASCA} data (both fixing the $N_{\rm H}$ and leaving it free) does not
lead to an improved fit ($\Delta\chi^2$=2 for 3 additional
parameters).  We do not detect absorption edges, with an (interesting)
90\% confidence upper limit to the optical depth of the K-edge of 
O{\sc\,vii} of $\tau_e < 0.09$. 

\noindent{\bf 3C~303:} The best-fitting model for the {\it ASCA}
continuum is a power law absorbed by a column of cold gas in excess to
the Galactic value.  A formally equivalent fit is obtained with a
power law, absorbed by the Galactic column of cold gas, plus and
absorption edge, with the following parameters:
$\Gamma=1.85^{+0.08}_{-0.04}$, edge rest-frame energy $E_{\rm
e}=0.685^{+0.051}_{-0.373}$ keV and optical depth
$\tau=0.98^{+9.02}_{-0.34}$, $\chi^2_{\rm r}/dofs$=1.01/365. The edge
energy falls well below the SIS sensitivity cutoff, suggesting that
the edge model is only a parameterization of excess cold absorption. A
{\it ROSAT} spectrum from the all-sky survey ($\sim$ 500 counts) is
fitted with a single power law with $\Gamma=2.07 \pm 0.56$ and
N$_H=2.6 \pm 1.6$ cm$^{-2}$, consistent with Galactic (Almudena Prieto
1996); however, the fit is not acceptable ($\chi^2_{\rm
r}/dofs$=1.49/26).  Future broader-band observations of 3C~303 are
necessary to better study the low-energy absorption in this BLRG. The
only previous pointed X-ray observations of 3C~303 were performed with
the {\it Einstein} IPC, which measured a flux in 0.5--3 keV of $F_{\rm
0.5-3~keV} \sim 1.2 \times 10^{-12}$ erg cm$^{-2}$ s$^{-1}$ (Fabbiano
et al. 1984), similar to {\it ASCA}, but no spectral information is
reported.

\noindent{\bf 3C~382:} The {\it ASCA} data were previously analyzed by
Reynolds (1997) and Wo\'zniak et al. (1998). The {\it ASCA} continuum
was modelled in different ways, which resulted in different
constraints on the Fe line profile.  Reynolds (1997) fits the {\it
ASCA} continuum with a single power law with photon index $\Gamma \sim
1.9$ plus two absorption edges at $\sim$ 0.6 and 0.7 keV (representing
O{\sc\,vii} and O{\sc\,viii}) with optical depths $\tau \sim$ 0.2. He
finds a strong (EW $\sim$ 1 keV) and broad ($\sigma \sim 2$ keV) Fe
K$\alpha$ line. Wo\'zniak et al. (1998) parameterize the continuum in
terms of a broken power law model with a break at 4 keV and
$\Gamma_{\rm hard} \sim 1.7$, $\Gamma_{\rm soft} \sim 2$, and find a
very weak (EW $\sim$ 27 eV) and unresolved ($\sigma \equiv 0$ keV) Fe
line. We fit the {\it ASCA} continuum with a single power law ($\Gamma
\sim 1.2$) plus a soft excess described by a thermal bremmstrahlung
($kT \sim 2$ keV), and find a moderately strong (EW $\sim$ 760 eV) and
broad ($\sigma \sim 1.4$ keV) Fe line (see Tables 3 and 4). This
illustrates the difficulty of modeling the Fe line in the relatively
narrow {\it ASCA} band. In support of our parameterization of the
data, we note that a soft excess in 3C~382 at energies $\lesssim$ 2
keV was found in previous {\it EXOSAT} (Barr \& Giommi 1992) and {\it
GINGA} (Kaastra, Kunieda, \& Awaki 1991) observations. The flat
continuum slope is not unprecedented: depending on the
parameterization of the soft excess the range of photon indices
measured from the {\it GINGA} spectrum is 1.03--1.51. The power law
slope also appears to vary significantly on timescales of days to
years with the X-ray flux (Barr \& Giommi 1992; Kaastra et al. 1991).
The {\it ASCA} observations correspond to an intermediate state. In
this model, the 90\% confidence upper limit to the K-edge of {\sc O\,
vii} is $\tau_e < 0.2$.

\noindent{\bf 3C~390.3:} The {\it ASCA} observations of this BLRG were
previously analyzed by Eracleous et al. (1995) and Leighly et
al. (1997). Eracleous et al. fit the 1993 data with an absorbed power
law plus reflection plus an Fe line; our results for the column
density, photon index, and line parameters are completely consistent
with those of Eracleous et al., who fixed the reflection strength to
0.5. This value is consistent with the one we find when this parameter
is left free to vary (Table 3). We also report the first detection of
a significant ($\Delta\chi^2$=17 for 2 additional degrees of freedom)
absorption edge at a rest-energy $E_{\rm e} \sim$ 0.78 keV with $\tau
\sim 0.3$; if the feature is identified with the absorption edge of
O{\sc\,vii} ($E=0.739$~eV), a column density of $N_{\rm H,warm} \sim 8
\times 10^{20}$ cm$^{-2}$ is implied for the ionized gas, similar to
the column density of cold gas.  We also find that the Fe K$\alpha$
line profile is better described by a disk-line model than by a
symmetric Gaussian (\S4.3); the disk-line model fitted parameters are
$E_{\rm rest}=6.43 \pm 0.06$ keV, EW=$132^{+40}_{-48}$ eV, inclination
$i \sim 33^{\circ}$ (pegged at the radio upper limit), $\chi^2_{\rm
r}/dofs=1.03/1158$.  Our results for the remaining 1995 {\it ASCA}
observations of 3C~390.3 are consistent with Leighly et
al. (1997). 3C~390.3 was recently observed with {\it SAX} in 0.1--100
keV (Grandi et al. 1999). The SAX data are consistent with a $\Gamma
\sim 1.8$ powerlaw continuum, plus a reflection component with
strength $R \equiv N_{refl}/N_{plaw} \sim 1$, plus a narrow ($\sigma
\sim 73$ eV) Fe line with EW $\sim$ 140 eV. At soft X-rays, the SAX
spectrum is consistent with cold absorption in excess to Galactic,
N$_H \sim 10^{21}$ cm$^{-2}$ (Grandi et al. 1999).  The latter authors
also point out a historical variability of the column density of cold
gas, without an apparent correlation with the nuclear X-ray luminosity.
No absorption edges are reported from the SAX data; however, the lower
sensitivity and reduced observing efficiency of the LECS makes it
diffuclt to observe weak edges such as detected with {\it ASCA} in
3C~390.3.

\noindent{\bf 3C~445:} The {\it ASCA} observation of 3C~445 was
previously published by Sambruna et al. (1998), with fully consistent
results. The 90\% confidence upper limit to the K-edge of {\sc O\,
vii} is $\tau_e < 0.6$.  The 0.6--10 keV {\it ASCA} spectrum is
consistent with a dual absorber (Table 3), which also fits the
combined {\it ROSAT} plus {\it ASCA} data well. Interestingly, 3C~445
is the only source among the X-ray absorber BLRGs and QSRs of our
sample to exhibit soft X-ray emission above the extrapolation of the
hard, absorbed power law (cfr. the best-fit model in Figure 1). This
indicates the presence of reprocessed radiation at soft X-rays, of
which the dual absorber is a possible parameterization. In this
respect, the X-ray spectrum of 3C~445 is similar to the Seyfert 2
NGC~6552 (Fukazawa et al. 1994), where the ``soft X-ray excess'' was
resolved into a blend of several emission lines in a long (350 ks)
serendipitous {\it ASCA} exposure. The lines were interpreted as due
to emission from neutral elements such as O, Ne, and Mg (Reynolds et
al. 1994). The large inclinations of 3C~445 ($i \gtrsim 60$ deg;
Eracleous \& Halpern 1998) makes it conceivable that we are
intercepting the reprocessed radiation from the absorber, diluted by
the primary direct emission. We thus reanalyzed the combined {\it
ROSAT} and {\it ASCA} SIS data of 3C~445 to constrain the strength of
possible soft X-ray lines. We fitted the data with a simple continuum
model including a heavily absorbed (N$_H=(3 \pm 1) \times 10^{23}$
cm$^{-2}$) power law with $\Gamma \sim 1$ at energies $\gtrsim$ 2 keV,
plus a power law with the same slope and Galactic N$_H$ only at softer
energies. We added narrow ($\sigma=0$ keV) Gaussians with energies
fixed at the redshifted energies of neutral abundant elements in the
range 0.6--2 keV, following the model of Reynolds et al. (1994) for
NGC~6552. We find marginal ($\gtrsim$ 90\% confidence) detections in
the case of OI (0.56 keV), with EW=140$^{+100}_{-124}$ eV, and Mg(1.26
keV), with EW=45$^{+48}_{-43}$ eV. In addition, we significantly
detect ($\Delta\chi^2=10.8$ for 3 additional parameters, P$_F \gtrsim$
95\%) an emission line at 2.76 keV with EW=280 $\pm 121$ eV, which we
tentatively identify as the Si recombination continuum line. These
results support the idea that significant soft X-ray reprocessing is
present in 3C~445 from the gas responsible for the observed X-ray
absorption. High-resolution X-ray observations of 3C~445 at both soft
and medium-hard energies (e.g., with {\it XMM}) will constrain the physical
and dynamical state of the absorber(s) in this source, and its
location.

\noindent{\bf PKS 2251+113:} The {\it ASCA} data were independently
analyzed by Brandt, Laor, \& Wills (1999); our results agree with
theirs. In particular, large excess X-ray absorption of $N_{\rm H}
\sim 1 \times 10^{22}$ cm$^{-2}$ in the quasar's rest-frame is
measured with {\it ASCA}. This quasar also exhibits intrinsic UV
absorption; possible association of the UV and X-ray absorbers and the
implications for the spectral energy distribution is discussed by
Brandt et al. (1999).

\noindent{\bf PKS~0634--20:} A partial covering model describes best
the (low signal-to-noise ratio) {\it ASCA} spectrum of this NLRG. The
measured column density from the X-rays, $N_{\rm H} \sim 8 \times
10^{23}$ cm$^{-2}$, is one order of magnitude larger than the
extinction inferred from IR observations of the continuum (Simpson,
Ward, \& Wilson 1995), $N_{\rm H} \sim 7 \times 10^{22}$ cm$^{-2}$,
assuming a Galactic gas-to-dust ratio. The continuum X-ray slope,
$\Gamma \sim 1.9$, and the 2--10 keV intrinsic luminosity, $L_{\rm
2-10~keV} \sim 3 \times 10^{43}$ erg s$^{-1}$, point to the presence
of an AGN. Thus, we confirm the suggestion by Simpson et al. (1995)
that an active nucleus is present in PKS~0634--205, buried under
several magnitudes of visual extinction. The {\it ASCA} data also
suggest that $\sim$ 4\% of the primary nuclear emission is scattered
into the line of sight at soft X-rays, consistent with the presence of
dust/hot electrons around the active nucleus, as suggested by Simpson
et al. (1995).

\noindent{\bf 4C~+55.16:} The X-ray emission of this radio galaxy is
dominated by diffuse cluster emission. There is a report of a cooling
flow and lensing effects (Iwasawa et al. 1999a). 

\noindent{\bf 3C~219:} A broad component of the Pa$\alpha$ line was
detected in the IR observations of Hill et al. (1996), showing the
presence of a hidden broad-line region in this NLRG. The amount of
reddening inferred for the broad-line region is $A_{\rm V} \sim 1.8$,
implying a column density of $4 \times 10^{21}$ cm$^{-2}$ for a
Galactic gas-to-dust ratio. The {\it ASCA} data on 3C~219 confirm the
presence of a hidden AGN: the hard X-ray spectrum is described by a
heavily-absorbed power law with $\Gamma \sim 1.7-1.8$ and $N_{\rm H}
\sim 2 \times 10^{21}$ cm$^{-2}$. There is little or no flux and
spectral variability between the two {\it ASCA} observations. Our
results for the second observation are consistent with the results
reported by Brunetti et al. (1999). A thermal contribution by the host
galaxy to the observed soft X-ray spectrum is also detected in the
first observation, when the emission from the nucleus was slightly
lower (factor 1.3) than during the second exposure.

\noindent{\bf 3C~295:} This NLRG at $z$=0.51 resides in an optical
cluster of galaxies, whose X-ray emission completely dominates the
{\it ASCA} data: no point-like source is detected with {\it
ASCA}. This confirms previous X-ray observations with the {\it
Einstein} IPC and HRI (Henry \& Henriksen 1986), which indicated the
presence of a massive ($\sim$ 145 $M_{\odot}$~yr$^{-1}$) cooling flow,
with loose constraints on the X-ray temperature of the cluster ($kT
\sim$ 3--10 keV). The gas parameters measured with {\it ASCA} are a
temperature $kT \sim$ 6 keV, a metallicity $Y \sim 0.5$, and a 2--10
keV luminosity $L_{\rm 2-10~keV} \sim 6 \times 10^{44}$ erg s$^{-1}$
(Table 3). 

\noindent{\bf 3C~313:} This FR~II radio galaxy resides in a rich
optical environment (Hill \& Lilly 1991). However, the {\it ASCA}
spectrum is consistent with a power law absorbed by a Galactic column
of cold gas (Table 3). Fitting the spectrum with a Raymond-Smith model
also leads to an acceptable fit, $\chi^2_{\rm r}/dofs$=0.91/80;
however, this is worse by $\Delta\chi^2=4$ than the power-law model
which has the same number of parameters. The fitted parameters of the
thermal model are: temperature $kT=9 (> 3)$ keV and 0.1--2.4 keV
intrinsic luminosity $L_{\rm 0.1-2.4~keV} \sim 4 \times 10^{43}$ erg
s$^{-1}$ (for $Y\equiv1$), one order of magnitude lower than predicted
from the $L-T$ relationship for clusters (Figure 4; Markevitch 1998),
although the uncertainties on the temperature are large.  If the
abundance $Y$ is left free to vary, we obtain $kT \sim 10$ keV and $Y
\sim 0.06$, thus effectively mimicking a power law. We conclude that,
unless the optical cluster around 3C~313 has very unsual X-ray
properties, the X-ray emission of this radio galaxy in the {\it ASCA}
band is dominated by the non-thermal radio source.  Note, however,
that some excess flux is present in the residuals of the single
power-law model around 0.9 keV (Figure 1), which may suggest the
presence of a thermal contribution. Adding a Raymond-Smith model to
the power law decreases $\chi^2$ by $\Delta\chi^2 \sim$ 4 only, not
significant for 2 additional parameters according to the F-test. The
upper limit on the thermal component luminosity in this case is
$L_{\rm 0.5-2~keV} \lesssim 3 \times 10^{43}$ erg s$^{-1}$ (for $kT
\sim 2$ keV and $Y\equiv1$).

\noindent{\bf 3C~321:} This FRII radio galaxy has a double nucleus
(separation 4 arcsec) and large-scale diffuse structures suggestive of
a recent merger (Young et al. 1996 and references therein).  UV
spectropolarimetric observations show the presence of three distinct
emission components contributing to the UV: the host galaxy, a hidden
AGN whose emission is scattered into the line of sight (20-70\% of the
total continuum), and a young starburst (Tadhunter, Dickson, \& Shaw
1996).  The presence of a hidden AGN is also confirmed by the
detection of a broad H$\alpha$ component in polarized optical light
(Young et al. 1996).
In X-rays, 3C~321 is detected with {\it ASCA} with a poor
signal-to-noise ratio (Table 2). The SIS images show an extended
($\sim$ 2 arcmin) asymmetric structure, while a point-like source is
present at hard energies in the GIS data. Modelling of the spectrum is
hampered by the poor signal-to-noise ratio and the following results
must be taken with caution. At energies below 1 keV, the {\it ASCA}
spectrum is consistent with a Raymond-Smith plasma model with $kT
\sim$ 0.6 keV (Table 3). The intrinsic 0.5--4.5 keV luminosity of the
thermal component is $L_{\rm X,obs} \sim 6 \times 10^{41}$
erg~s$^{-1}$. This value is close to the predicted X-ray luminosity of
the starburst, $L_{\rm X,starb} \lesssim 4 \times 10^{41}$
erg~s$^{-1}$ (using equations (1) and (2) of David, Jones, \& Forman
1992, and the IRAS fluxes at 60~$\mu$m and 100~$\mu$m). While {\it
ASCA} provides initial evidence for the presence of a starburst in
3C~321, future higher resolution observations with {\it AXAF} are
needed to confirm this suggestion.  Above 3 keV, the {\it ASCA}
spectrum of 3C~321 is best described by an absorbed power law with
$N_{\rm H} \sim 10^{22}$ cm$^{-2}$ and $\Gamma \sim 1.5$. There is
some evidence in SIS0, SIS1, and GIS3 for a narrow unresolved Fe
K$\alpha$ line with EW $\sim$ 1.8 keV (Table 4). This feature persists
even when different extraction cells for the background are used.
Moreover, several ``line-like features'' are present between 0.9--3
keV at 1--2$\sigma$ level; if real, these features could be similar to
those detected in the {\it ASCA} data of a few Seyfert 2s (Ueno et
al. 1994a; Iwasawa et al. 1994; Turner et al. 1997a) and expected on
theoretical grounds in scattering-dominated AGN.
It will be important to observe 3C~321 with future high-resolution
X-ray observatories to obtain a high-quality spectrum and confirm the
marginal features in the {\it ASCA} data. 

\noindent{\bf Cygnus~A:} The X-ray emission from this galaxy is
complex, including contributions from the intracluster gas emission
and a cooling flow at lower energies (Arnaud 1996; Reynolds \& Fabian
1996). Previous observations with {\it EXOSAT} and {\it GINGA}
detected a power-law spectral component at hard X-rays (Arnaud et
al. 1987; Ueno et al. 1994b), attributed to a hidden AGN. These claims
are also confirmed by recent optical spectropolarimetry, which led to
the detection of a broad H$\alpha$ emission line in polarized light
(Ogle et al. 1997). In order to avoid the contribution of the cooling
flow and concentrate on the AGN component, we restricted our analysis
to the energy range 4--10 keV, while allowing a thermal component to
include the cluster emission. The power-law photon index and
absorption column density measured with {\it ASCA} are $\Gamma \sim
1.8$ and $N_{\rm H} \sim 1-2 \times 10^{23}$ cm$^{-2}$ (Table 3), in
good agreement with the previous {\it EXOSAT} and {\it GINGA}
results. With {\it ASCA} we also detect for the first time the Fe
K$\alpha$ line at $\sim$ 6.4 keV. Constraining the line to be narrow,
we measure an EW $\sim$ 100 eV (Table 4), consistent with an origin by
fluorescence in cold gas. However, a marginally improved fit is
obtained if the line is allowed to be broad ($\Delta\chi^2$=4.7,
significant at $P_{\rm F} \gtrsim 95$\% confidence for 1 additional
parameter), with $\sigma_{rest}=0.36^{+0.24}_{-0.11}$ keV and
EW=208$^{+140}_{-106}$ eV. Because of the spectral complexity in the
energy range 5--7 keV, where contributions from the gas cluster are
present, we regard this result as preliminary and quote only the
results for a narrow line in Table 4. While the Fe line detection
provides further unambigous proof of the presence of an AGN in the
central regions of Cygnus~A, under a heavy layer of cold gas, future
higher-resolution X-ray observations are needed to study the line
profile in more detail.

\noindent{\bf PKS~2152--69:} This NLRG has an intermediate FR~I/II
radio morphology. It is famous for exhibiting a cloud of highly
ionized gas at $\sim$ 10 arcsec from the nucleus, also emitting a blue
continuum (Tadhunter et al. 1988), which is either photoionized by a
beamed radiation from the nucleus (di Serego Alighieri et al. 1988),
or by shocks induced by jet/cloud interactions (Fosbury et al. 1998).
The {\it ASCA} data of PKS~2152--69 are consistent at hard energies
with an absorbed power law with $\Gamma \sim 1.9$, typical of the
other NLRGs in the sample, and with relatively lower luminosity,
$L_{\rm 2-10~keV} \sim 6 \times 10^{42}$ erg s$^{-1}$. The presence of
an Fe line is difficult to assess because of the thermal contribution
in the range 6--7 keV (Figure 1). At soft energies, a thermal-plasma
component is the best description of the {\it ASCA} spectrum, with
luminosity and temperature typical of a poor group of galaxies or the
corona of the elliptical host galaxy as detected in the optical
(Tadhunter et al. 1988). An archival {\it ROSAT} HRI exposure shows
that the soft X-ray emission is extended up to 1 arcmin around the
optical body of the galaxy (E. Colbert 1999, priv. comm.), confirming
the {\it ASCA} result.

\noindent{\bf 3C~109:} The {\it ASCA} spectrum of the QSR 3C~109 is
best fitted by an absorbed power law with $N_{\rm H} \sim 5 \times
10^{21}$ cm$^{-2}$ and $\Gamma \sim 1.7$ (Table 3). These values are
consistent with those previously published by Allen et al. (1997) and
with a previous {\it ROSAT} observation (Allen \& Fabian 1992). The
presence of intrinsic absorption in 3C~109 is also supported by
optical observations, which show a highly polarized continuum and
reddened Balmer lines (Goodrich \& Cohen 1992), suggestive of a dust
screen covering the active nucleus. 3C~109 also exhibits an unusually
red IR-to-optical continuum (Elvis et al. 1984). The detection of an
Fe K$\alpha$ line in the {\it ASCA} data, with EW $\sim$ 300 eV, was
previously reported by Allen et al. (1997). However, we do not confirm
this result: we find that the addition of a line to the GIS spectrum,
with EW $\sim$ 200 eV and width $\sigma \sim 0.7$ keV, improves the
fit by $\Delta\chi^2$=5, not significant for 3 additional parameters
according to the F-test.  The SIS data give only an upper limit of
EW=60 eV (for an unresolved line) and of 170 eV (for a broad
line). More sensitive observations, e.g., with {\it XMM}, are
necessary to confirm the Fe line detection in 3C~109.

\noindent{\bf 3C~234:} This classical triple radio galaxy has been
extensively studied at optical and IR wavelengths. We classify it as a
QSR following Grandi \& Osterbrock (1978) who detected a
highly-luminous broad line. Recent optical studies have shown that the
nuclear continuum and broad H$\alpha$ line are highly polarized
($\sim$ 25\%), with a position angle perpendicular to the radio axis
(Antonucci 1984; Tran, Cohen, \& Goodrich 1995; Young et
al. 1998). This was interpreted in terms of scattering into the line
of sight by a population of hot electrons or dust, the latter also
advocated to explain the steep IR-to-optical continuum (Elvis et
al. 1984).  The {\it ASCA} X-ray spectrum of 3C~234 has a low
signal-to-noise ratio, allowing us to draw only limited
conclusions. The 0.6--10 keV continuum is best described by a partial
covering model, where a large fraction ($\sim$ 67\%) of the primary
X-ray emission is scattered/transmitted into the line of sight. The
photon index is unusually low, $\Gamma \sim 0.08$, supporting a
scattering origin of the radiation in the {\it ASCA} band (as in
Seyfert 2s; Turner et al. 1997a). It is thus possible that the primary
X-rays are blocked up to 10 keV by a very large ($\gtrsim 10^{24}$
cm$^{-2}$) column density (by analogy with the Seyfert 2 galaxy
NGC~1068). At low energies, the {\it ASCA} spectrum can be described
by a Raymond-Smith thermal plasma model.  The temperature and
luminosity of this component suggest an origin in the ISM of the host
galaxy. However, another possibility is that this component consists
of a blend of soft X-ray lines around 1 keV, which would be expected
in scattering-dominated AGN on both observational and theoretical
grounds (Ueno et al. 1994a; Iwasawa et al. 1994; Netzer, Turner, \&
George 1998).

\noindent{\bf 3C~254:} Extended soft X-ray emission associated to the
optical cluster of galaxies was recently detected for this QSR in a
30~ks ROSAT HRI exposure (Crawford et al. 1999), contributing
$\lesssim$ 20\% of the total X-ray luminosity of the source. The ASCA
data are well fitted by a single power law with $\Gamma \sim 1.7$
(Table 3) absorbed by Galactic N$_H$ only; addition of a Raymond-Smith
thermal model yields a decrease of the $\chi^2$ of 4, not significant
for 3 additional parameters. The upper limit to the 0.1--2.4 keV
luminosity of the extended component is L$_{0.1-2.4~keV} \lesssim 1
\times 10^{45}$ erg s$^{-1}$, in very good agreement with the measured
luminosity from the ROSAT data (Crawford et al. 1999). 

\noindent{\bf 4C~+74.26:} This QSR is associated with a giant
double-lobed radio source. Its {\it ASCA} spectrum was previously
published by Brinkmann et al. (1998), together with optical and radio
data. These authors fitted the {\it ASCA} spectrum with a complex
model, including a power law ($\Gamma \sim 2.03$), cold reflection
(with strength $R\equiv N_{ref}/N_{pl} \sim 6$), an Fe K$\alpha$ line
($\sigma \sim 0.08$ keV, EW $\sim$ 100 eV), and a warm absorber at low
energies.  We find that the 0.6--10 keV continuum of 4C~+74.26 is
concave. When we fit it with a power law plus reflection model, we
recover a similar solution to Brinkmann et al. However, the Fe line is
no longer significantly detected in our solution, which is in striking
contrast with the strength of the reflected component. In fact, for
$R\sim6$, the expected EW of the Fe K$\alpha$ line is $\gtrsim$ 400 eV
assuming $i\lesssim 49^{\circ}$ (from the radio morphology, Brinkmann
et al. 1998) and $\Gamma=2.1$ (George \& Fabian 1991). We suggest that
the concavity of the X-ray continuum is related to the presence of a
hard tail, possibly associated with the radio jet. Fitting the
spectrum with a double power law model, we obtain $\Gamma_{\rm soft}
\sim 2$, $\Gamma_{\rm hard} \sim 0.2$. With this continuum model, the
Fe line is broad ($\sigma \sim 0.6$ keV) with EW $\sim$ 215 eV. We do
not find evidence for ionized absorption at low energies, where a cold
absorber is sufficient to fit adequately the {\it ASCA} and the
archival {\it ROSAT} data. We are only able to set an upper limit to
the optical depth of the O{\sc\,vii} edge of $\tau_e=0.8$ (at 99\%
confidence).

\noindent{\bf 3C~28:} This radio galaxy is the dominant member of the
distant ($z=0.195$) Abell cluster A115. Previous X-ray observations
with the {\it Einstein} HRI are entirely consistent with diffuse
emission centered on the optical galaxy and extending in a direction
roughly orthogonal to the radio structure (Feretti et al. 1984). The
authors interpret this component, which has a luminosity $L_{\rm X}
\sim 5 \times 10^{43}$ erg s$^{-1}$ ($\pm$ 30\%), as a cooling flow
onto the central regions of the galaxy. The {\it ASCA} spectrum is
best described by a thermal plasma model representing the emission
from A115 ($kT \sim 3.3$ keV, metallicity $Y\sim$ 0.3), plus a hard,
absorbed power-law component ($N_{\rm H} \sim 2 \times 10^{21}$
cm$^{-2}$, $\Gamma \sim 1.5$), which could be associated with a hidden
AGN. The hard tail is clearly detected in the SIS0+SIS1 spectra
($\Delta\chi^2$=12.5 for 2 additional parameters, P$_F \gtrsim$ 99\%)
but only marginally detected in the GIS2+GIS3 spectra ($\Delta\chi^2$=7,
P$_F \sim$ 95\%). Since the hard tail dominates the emission above
$\sim$ 5 keV, we examined the 5--10 keV SIS and GIS images. The
presence of a point-source was not obvious. In addition, only an upper
limit is given in the literature for the radio core emission and for
the [{\sc O\,iii}] emission line (Table 1). We thus regard the
detection of the hard tail in the {\it ASCA} data and its
interpretation as due to an active nucleus as tentative.  More
sensitive observations at hard X-rays are necessary to confirm this
result.

\noindent{\bf PKS~0131--36:} This WLRG lies in an early-type
elliptical with a dark lane crossing the main optical body of the host
galaxy (Ekers et al. 1978). Strong FIR emission is also detected
(Table 1 and references therein). In the radio, PKS~0131--36 exhibits
a compact core and two extended lobes, which are weak X-ray sources
($F_{\rm 0.5-10~keV} \sim 3 \times 10^{-13}$ erg cm$^{-2}$ s$^{-1}$)
in the X-rays (Tashiro et al. 1998), most likely via inverse Compton
scattering of the cosmic microwave background photons.  The nuclear
X-ray emission measured with {\it ASCA} is complex. At soft X-rays, a
thermal component with $kT \sim 1$ keV is detected; its intrinsic
0.5--4.5 keV luminosity, $L_{\rm 0.5-4.5~keV} \sim 9 \times 10^{40}$
erg s$^{-1}$, is compatible with the predicted X-ray luminosity of a
starburst (from the FIR fluxes at 60 and 100~$\mu$m and eqs. (1) and
(2) of David et al. 1992), $L_{\rm 0.5-4.5~keV}^{\rm starb} \sim 1
\times 10^{41}$ erg s$^{-1}$.  Thus the {\it ASCA} data provide
evidence for the presence of a starburst in PKS~0131--36, for which a
possible location is the dark lane (Brosch 1987). Alternatively, the
thermal component may originate in the hot ISM of the host galaxy. A
deep (68 ks) archival ROSAT HRI observation shows extended emission
centered on the optical body of the galaxy.  The hard X-ray spectrum
is described by a flat ($\Gamma \sim 1.3$) power law with 2--10 keV
luminosity
$L_{\rm 2-10~keV} \sim 4 \times 10^{42}$ erg s$^{-1}$, heavily
absorbed ($N_{\rm H} \sim 10^{23}$ cm$^{-2}$).  The residual flux at
low energies is consistent with 10\% of the nuclear flux being
scattered into the line of sight. An Fe line with EW $\sim$ 240 eV is
tentatively detected.

\noindent{\bf Fornax~A:} This nearby radio galaxy, located in the
outskirts of the Fornax cluster of galaxies, exhibits extended radio
lobes which also emit in the X-rays ($F_{\rm 1~keV} \sim 0.2 \mu$Jy)
via inverse Compton scattering of the cosmic microwave photons
(Feigelson et al. 1995; Kaneda et al. 1995). According to Tadhunter et
al. (1998), the nucleus of Fornax~A contains a Weak Line Radio Galaxy.
The presence of a low-luminosity AGN is also indicated by the
LINER-like optical spectrum (Baum et al. 1992). Strong FIR emission is
detected from the galaxy (Table 1). The {\it ASCA} spectrum is
described by a soft thermal-plasma model ($kT \sim 0.8$ keV, $Y \sim
0.1$), most likely due to the cluster. The FIR colors of Fornax~A,
however, suggest that a starburst could be present (Figure 5). The
0.5--4.5 keV luminosity of the thermal component is $L_{\rm
0.5-4.5~keV} \sim 9 \times 10^{40}$ erg s$^{-1}$, while the predicted
X-ray luminosity of a starburst based on the IRAS flux of Fornax~A is
$L_{\rm 0.5-4.5~keV}^{\rm starb} \sim 5 \times 10^{39}$ erg s$^{-1}$.
Thus it is possible that part of the thermal emission is due to a
starburst, which needs to be confirmed with future X-ray
observations. At high energies, an absorbed ($N_{\rm H} \sim 3.3
\times 10^{22}$ cm$^{-2}$) power law, with a poorly constrained
spectral index, is detected in the spectrum. Fixing the spectral index
at $\Gamma=1.5$ (the average value for WLRGs), we measure an intrinsic
luminosity $L_{\rm 2-10~keV} \sim 5 \times 10^{40}$ erg
s$^{-1}$. Fornax~A was previously observed with {\it ROSAT} (Kim et
al. 1998); the 0.1--2.4 keV range is dominated by the extended thermal
component. The {\it ASCA} data were previously published by Iyomoto et
al. (1998); while our results for the thermal component agree with
those of the latter authors, we can not compare directly our findings
for the power law component since Iyomoto et al. (1998) fixed the
absorption to Galactic and derive the photon index ($\Gamma=1.1 \pm
0.5$), while we left N$_H$ free and fixed the slope. However, the
power law luminosity in both cases agree well. Comparing the {\it
ASCA} results to previous observations, Iyomoto et al. (1998) conclude
that the nucleus of Fornax~A has been inactive for the last 0.1 Gyr.

\noindent{\bf IC~310:} This head-tail radio galaxy is one of the
dominant members of the Perseus cluster. The optical spectrum is
characterized by strong H$\alpha$, [N{\sc\,ii}], and [S{\sc\,ii}]
emission lines, with $L_{\sc H\alpha+[N\,ii]} \sim 1.6 \times 10^{40}$
erg s$^{-1}$ (Owen et al. 1995). While IC~310 does not meet the
selection criteria of our sample (\S2) and was not included in any of
the statistical tests, its {\it ASCA} data are presented here for the
first time.  The {\it ASCA} observations are consistent with a steep
($\Gamma \sim 2.3$) power law, confirming previous {\it ROSAT} results
(Rhee, Burns,
\& Kowalski 1994; Edge \& R\"ottgering 1995), plus a thermal component
related to the cluster emission, with $kT \sim 6$ keV, $Y \sim 0.3$.

\noindent{\bf PKS~0625--53:} This radio galaxy is classified as a WLRG
by Tadhunter et al. (1998).  The {\it ASCA} observation shows extended
emission, and the spectrum is best fitted with an absorbed
Raymond-Smith thermal model with temperature $kT \sim$ 6 keV and
metallicity $Y \sim 0.3$, typical of cluster emission. There is no
evidence for a point source; the upper limit to the 2--10~keV
luminosity of a power law component is $L_{\rm 2-10~keV}^{nucl}
\lesssim 4 \times 10^{42}$ erg s$^{-1}$ (assuming $\Gamma=1.5$).

\noindent{\bf 4C~+41.17:} This high-redshift radio galaxy is a complex
system with at least two distinct components showing up in
high-resolution {\it HST} images at rest-frame UV wavelengths (van
Breugel et al. 1998 and references therein). Star formation, possibly
induced by the interaction of the jet with the ambient gas, is
responsible for most of the unpolarized UV continuum emission (Dey et
al. 1997). The source was observed with {\it ASCA} with the SIS in
2CCD mode; a weak extended ($\sim$ 7 arcmin) source is present in the
image, with an elongated morphology and at least two distinct peaks.
However, the position of both peaks is $\gtrsim 3.5$ arcmin away from
the radio position of 4C~+41.17 (from \verb+NED+), which is much
larger than the uncertainties of $\sim$ 23 arcsec expected from the
attitude solution. Moreover, the discrepancy remains after the (small:
\verb+DETX_off+=+0.1 mm, \verb+DETY_off+=--0.66 mm) correction for
incorrect star-tracker reading which affects these data is applied
(\verb+http://heasarc.gsfc.nasa.gov/docs/asca/coord/updatecoord.html+).
We conclude that the association of the X-ray emission detected by
{\it ASCA} to 4C~+41.17 is ambiguous. However, no radio or optical
counterpart is known in the \verb+NED+ database at the {\it ASCA}
position of either X-ray peak. Integrating over the whole extended
X-ray emission yields a total of $\sim (5 \pm 0.7) \times 10^{-3}$
count s$^{-1}$ both for SIS0 and GIS2. The spectrum can be described
by a single power law with $\Gamma=1.97 \pm 0.33$ and Galactic $N_{\rm
H}=1.2 \times 10^{21}$, flux $F_{\rm 2-10~keV} \sim 2 \times 10^{-13}$
erg cm$^{-2}$ s$^{-1}$.

\noindent{\bf 3C~270:} This radio galaxy, hosted by the nearby giant
elliptical NGC~4261, is classified as a WLRG by Tadhunter et
al. (1998). Its central regions were mapped in IR and optical with HST
(Ferrarese et al. 1996). The high-resolution {\it HST} images show
that the galaxy hosts a small nuclear disk of dust which is displaced
from the nucleus; a dynamical estimate of the central mass is also
provided, $(4.9 \pm 1.0) \times 10^8~M_{\odot}$. 3C~270 was previously
studied in X-rays with the {\it ROSAT} PSPC. A diffuse thermal
component ($kT \sim 0.6$ keV for a fixed solar metallicity), possibly
associated with a cooling flow, plus an unabsorbed power law ($\Gamma
\sim 1.7$) component were found to adequately describe the {\it ROSAT}
spectrum (Worrall \& Birkinshaw 1994). The {\it ASCA} spectrum
confirms this spectral decomposition and the low X-ray luminosity
($L_{2-10~keV} \sim 9 \times 10^{40}$ erg s$^{-1}$) of the power law
component. The temperature of the soft component is found to be
slightly higher and the metallicities lower than with {\it ROSAT}. The
power-law index measured with {\it ASCA} is flatter than that measured
with {\it ROSAT}, although the latter has large uncertainties. No
excess cold absorption is detected with {\it ASCA}, in agreement with
the {\it ROSAT} findings (Worrall \& Birkinshaw 1994). An Fe line with
EW $\sim$ 260 eV is tentatively detected in the {\it ASCA} data.

\noindent{\bf Centaurus~A:} The {\it ASCA} spectrum was modelled
following the prescription of Turner et al. (1997b), who jointly
analyzed archival {\it ROSAT} HRI, PSPC, and {\it ASCA} data (see also
Sugizaki et al. 1997). The {\it ASCA} spectrum can be modelled as a
hard power law with $\Gamma \sim 1.9$ absorbed by three different
absorbers of cold gas, plus a steep ($\Gamma \sim 2.3$) power law
representing the jet synchrotron contribution, plus two Raymond-Smith
components, the softer one related to a starburst (Turner et
al. 1997b). As can be seen in Figure 1, there are line-like residuals
around 1 and 2 keV, which can be modelled as Gaussians with EW $\sim$
100 eV or less, and interpreted as the fluorescent lines of Mg, Si,
and S (Sugizaki et al. 1997). A strong Fe line is also present at 6.4
keV, with EW $\sim$ 140 eV (Table 4). Since the line is unresolved at
90\% confidence, $\sigma_{rest} \lesssim 0.12$ keV, its width was
fixed at $\sigma_{rest} \equiv 0.05$ keV in the fits to have better
constraints on the other line parameters.

\noindent{\bf NGC~6251:} This RG was recently observed with the {\it
HST} (Ferrarese \& Ford 1999). Its central regions have a complex
morphology with a warped disk of dust stretching across the optical
body of the galaxy for $\sim$ 760 pc. It appears to harbor a black
hole of mass $(4-8) \times 10^8$ $M_{\odot}$. NGC~6251 has a prominent
radio jet which emits also at X-ray wavelengths (Mack, Kerp, \& Klein
1997). In X-rays, previous observations with the {\it ROSAT} PSPC show
a diffuse thermal component ($kT \sim$ 0.5 keV), possibly associated
with a cooling flow powering the central nucleus, and a power law
component with $\Gamma \sim 2$ and little intrinsic absorption,
$N_{\rm H} \sim 5 \times 10^{20}$ (Birkinshaw \& Worrall 1992). The
{\it ASCA} spectrum is best described by a hard ($\Gamma \sim 1.8$)
power law absorbed by $N_{\rm H} \sim 7.5 \times 10^{21}$ cm$^{-2}$
(with large uncertainties), plus a thermal plasma component with $kT
\sim 1$ keV, harder than measured with the {\it ROSAT} PSPC.  The {\it
ASCA} data were previously analyzed by Turner et al. (1997a), whose
continuum model agrees with ours. We confirm the detection of an Fe
line with marginal significance ($\Delta\chi^2=6.8$, $P_{\rm F} \ge$
90\%) at a rest-frame energy E$\sim$6.6 keV, consistent with emission
from highly ionized Fe (Fe{\sc\,xxv}), and with EW $\sim$ 443~eV
(Table 4).

\noindent{\bf 3C~353:} The {\it ASCA} GIS images are dominated by the
extended emission of an elongated cluster, with the radio galaxy
3C~353 located at the edge of the structure. The nucleus of 3C~353
contains a WLRG (Tadhunter et al. 1998), and indeed the {\it ASCA}
spectrum above 2 keV is described by a power law, with $\Gamma \sim
1.9$ and heavy absorption ($N_{\rm H} \sim 4 \times 10^{22}$
cm$^{-2}$). The soft X-ray spectrum is described by a thermal
Raymond-Smith model with $kT \sim 1.4$ keV and low metallicity,
associated with the cluster. Analysis of the cluster emission at
different locations in the {\it ASCA} image shows variations of the
temperature and metallicity with position. A detailed analysis of the
cluster emission is beyond the scope of this paper and will be
presented by Iwasawa et al. (1999b).




\clearpage
\vspace*{-1in} \hspace*{-1in}
\psfig{figure=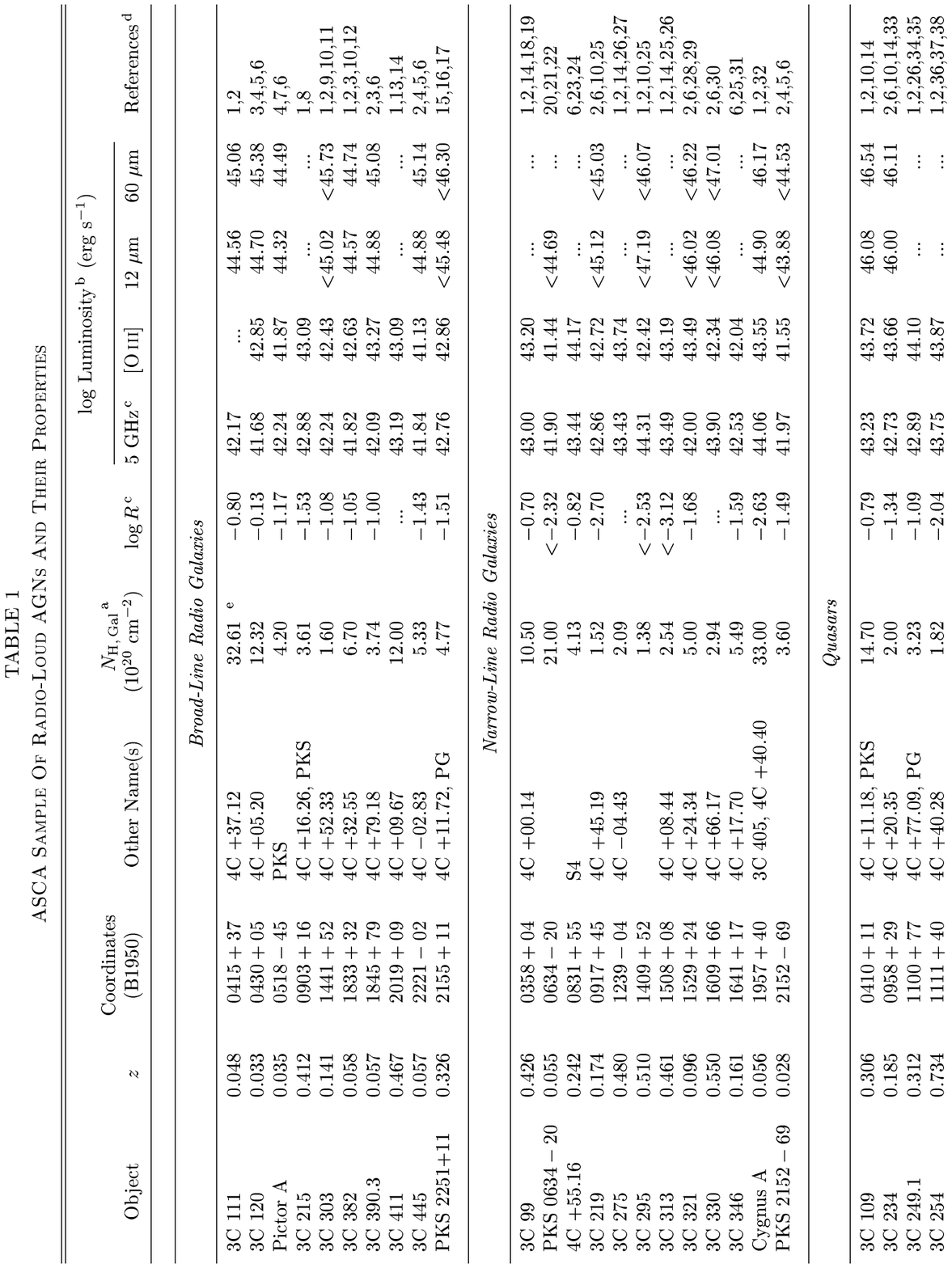,width=7.5in,rheight=8.in,angle=180}

\clearpage
\vspace*{-1in} \hspace*{-1in}
\psfig{figure=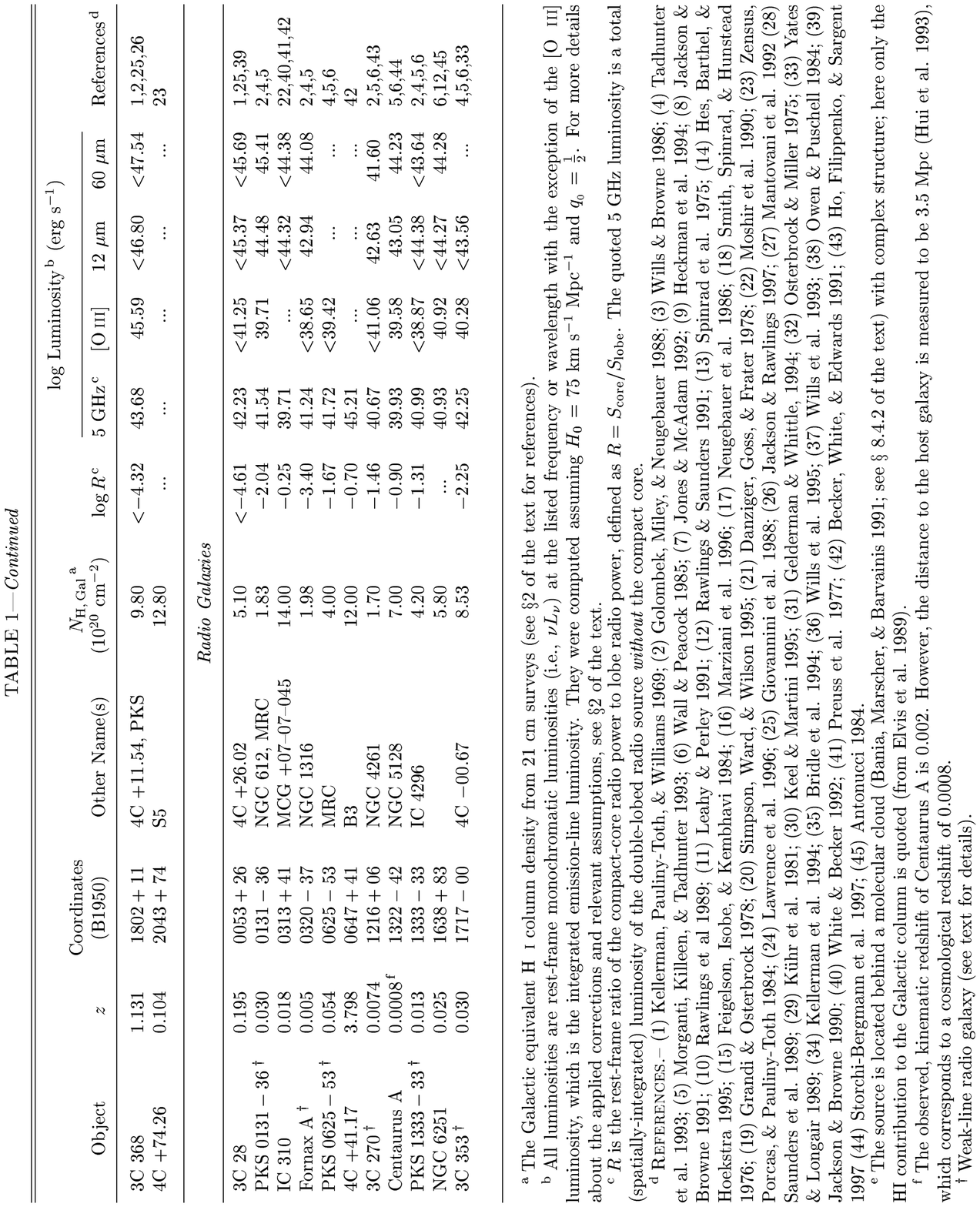,width=7.5in,rheight=8.in,angle=180}

\clearpage
\vspace*{-1in} \hspace*{-1in}
\psfig{figure=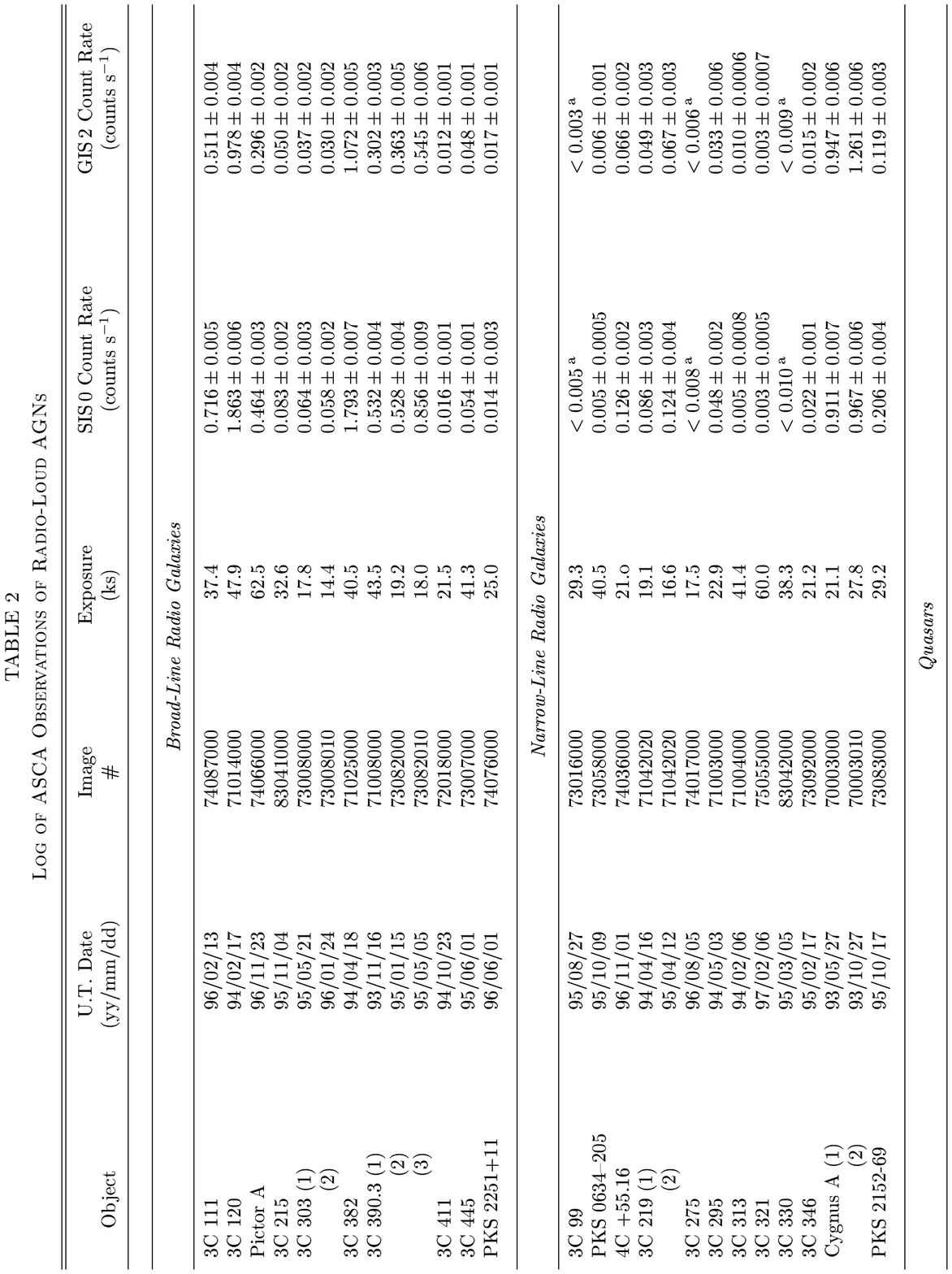,width=7.5in,rheight=8.in,angle=180}

\clearpage
\vspace*{-1in} \hspace*{-1in}
\psfig{figure=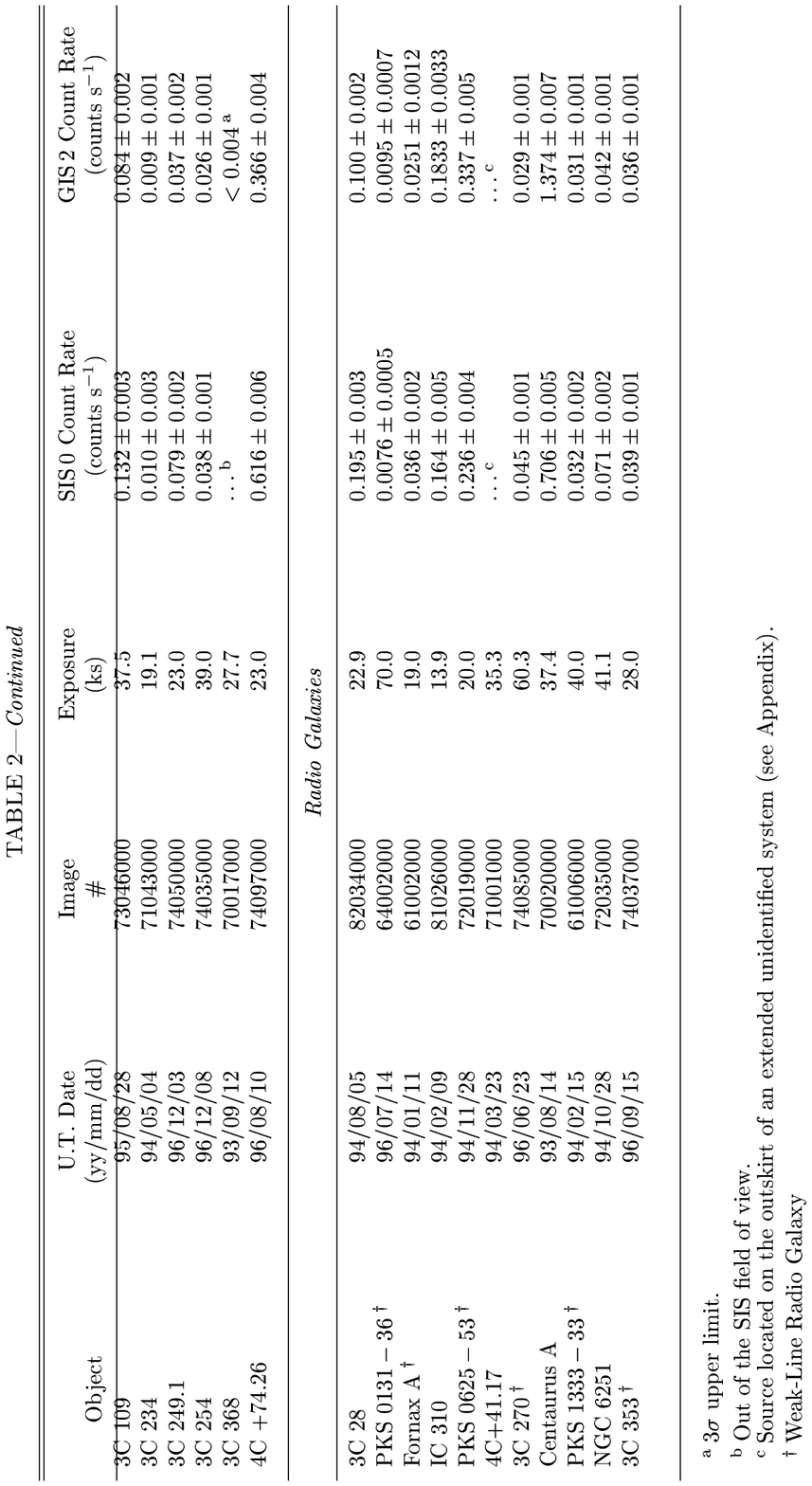,width=7.5in,rheight=8.in,angle=180}

\clearpage
\vspace*{-1in} \hspace*{-1in}
\psfig{figure=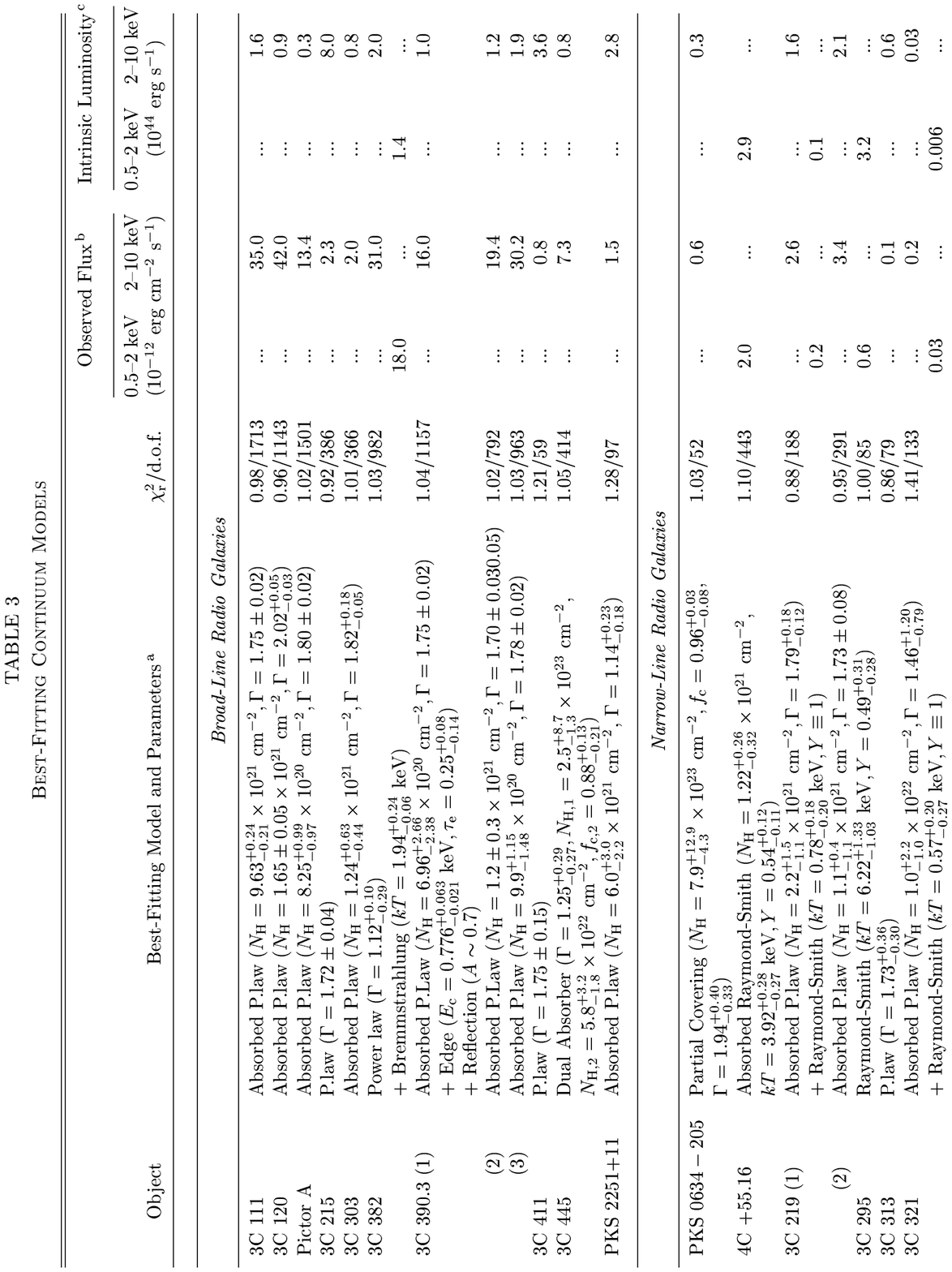,width=7.5in,rheight=8.in,angle=180}

\clearpage
\vspace*{-1in} \hspace*{-1in}
\psfig{figure=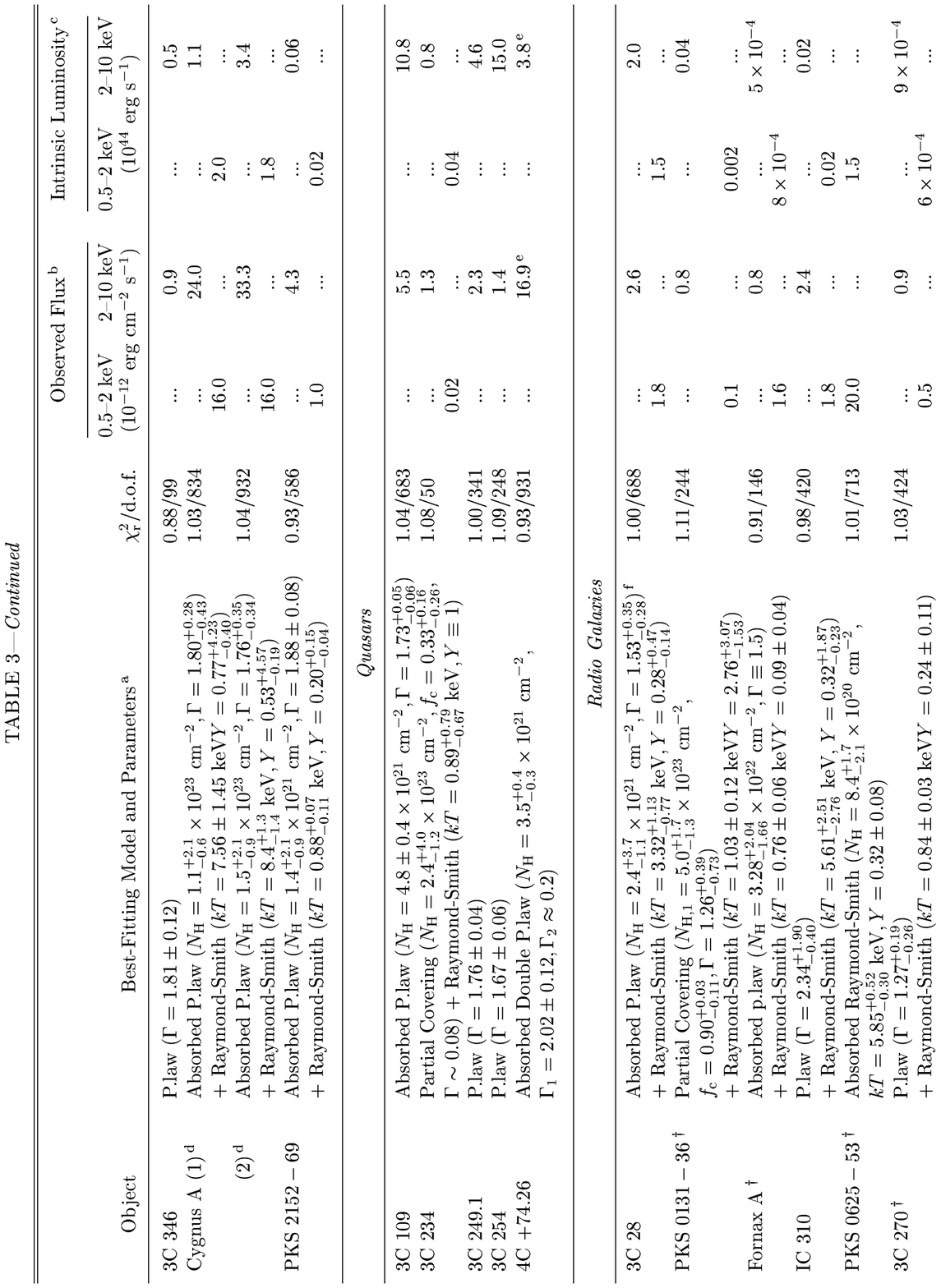,width=7.5in,rheight=8.in,angle=180}

\clearpage
\vspace*{-1in} \hspace*{-1in}
\psfig{figure=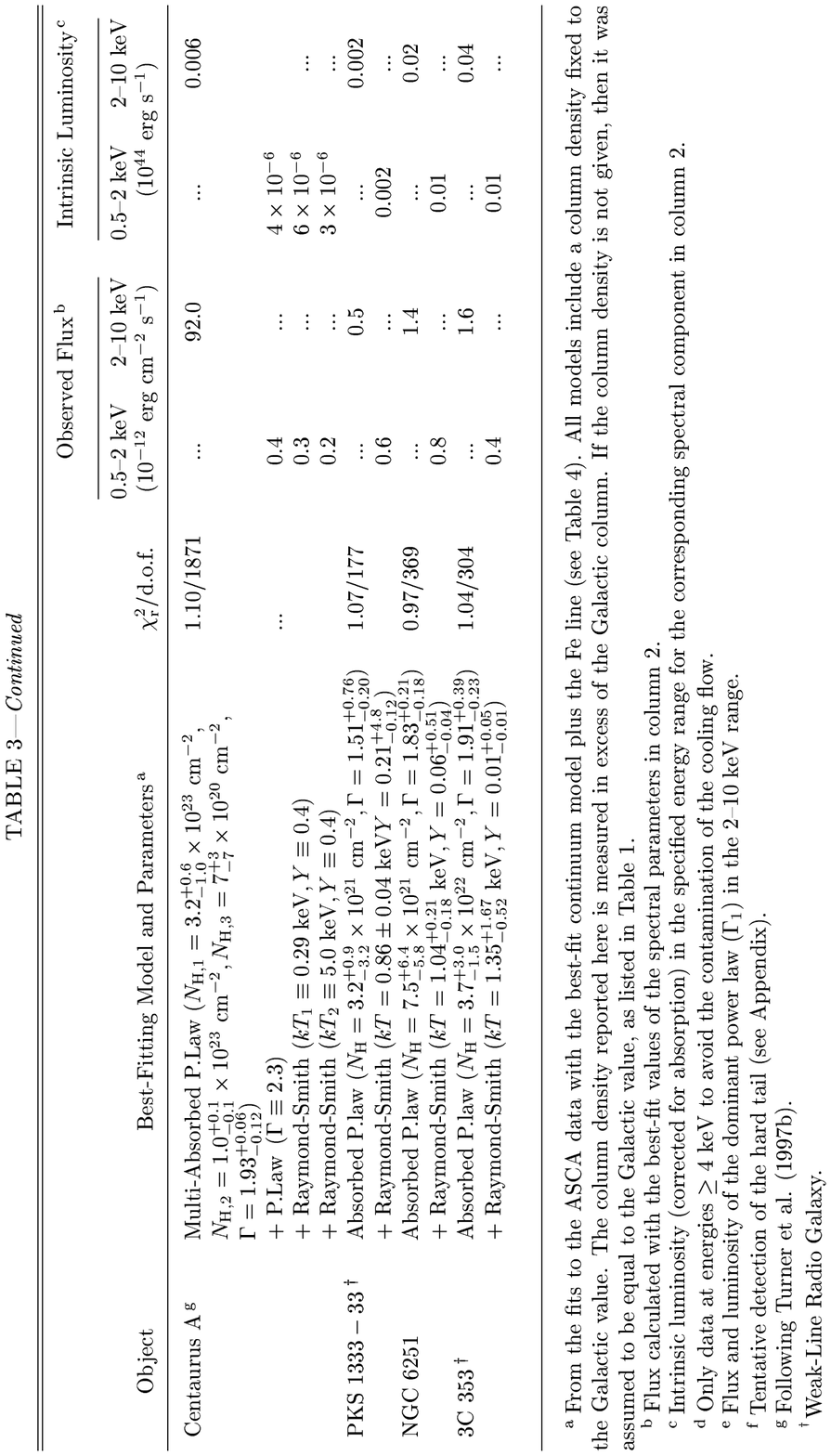,width=7.5in,rheight=8.in,angle=180}

\clearpage
\vspace*{-1in} \hspace*{-1in}
\psfig{figure=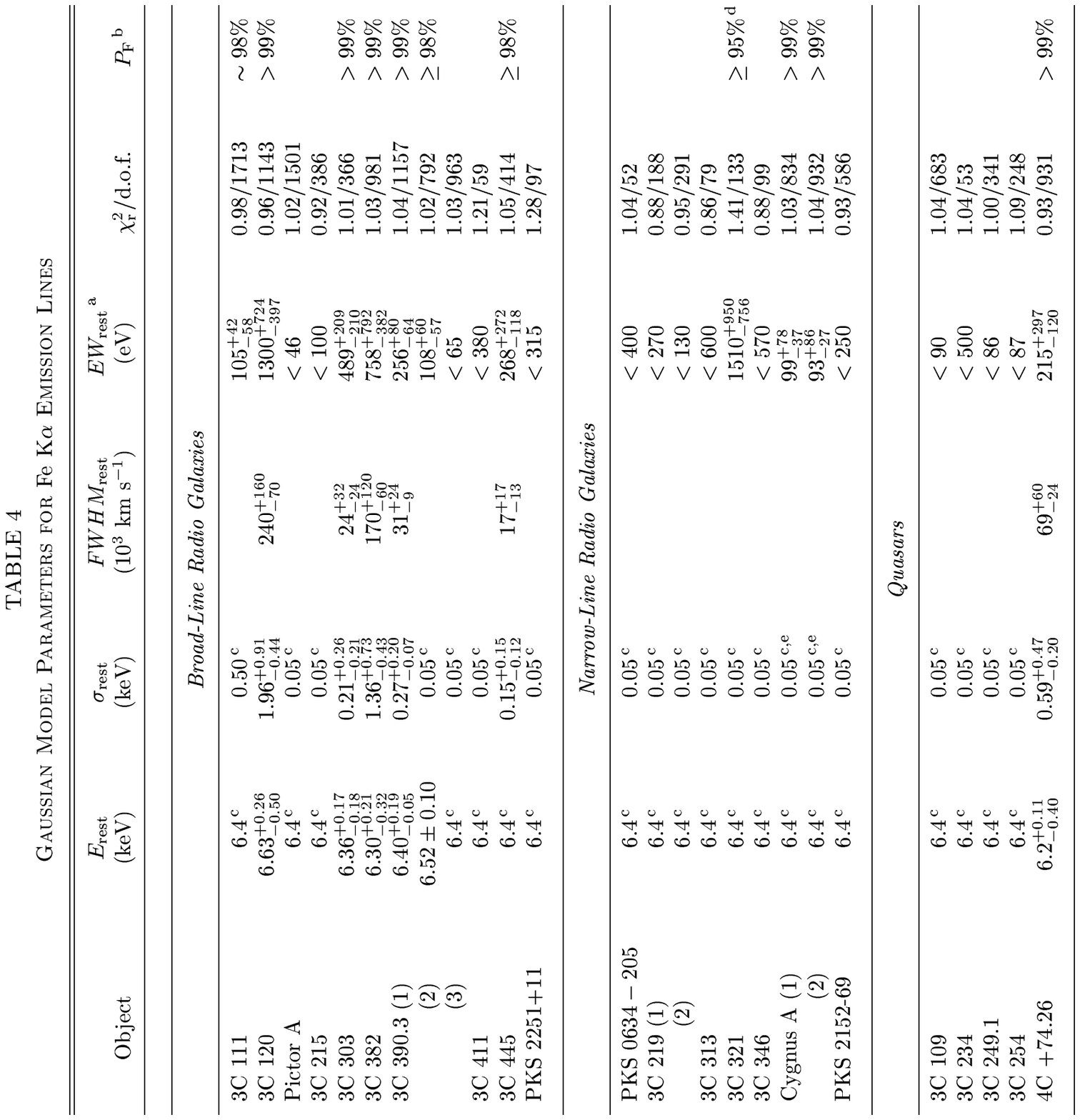,width=7.5in,rheight=8.in,angle=180}

\clearpage
\vspace*{-1in} \hspace*{-1in}
\psfig{figure=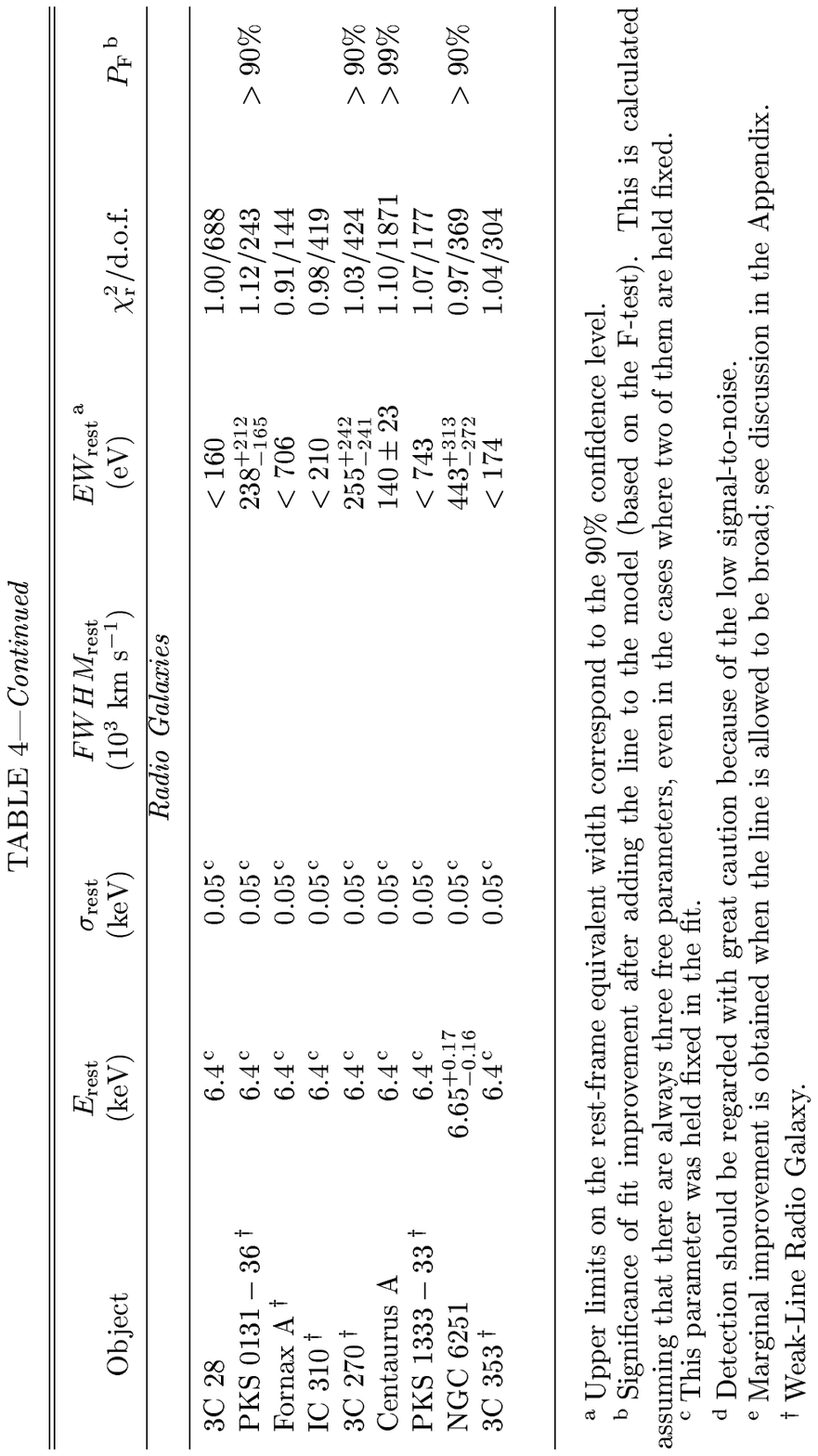,width=7.5in,rheight=8.in,angle=180}

\clearpage
\vspace*{-1in} \hspace*{-1in}
\psfig{figure=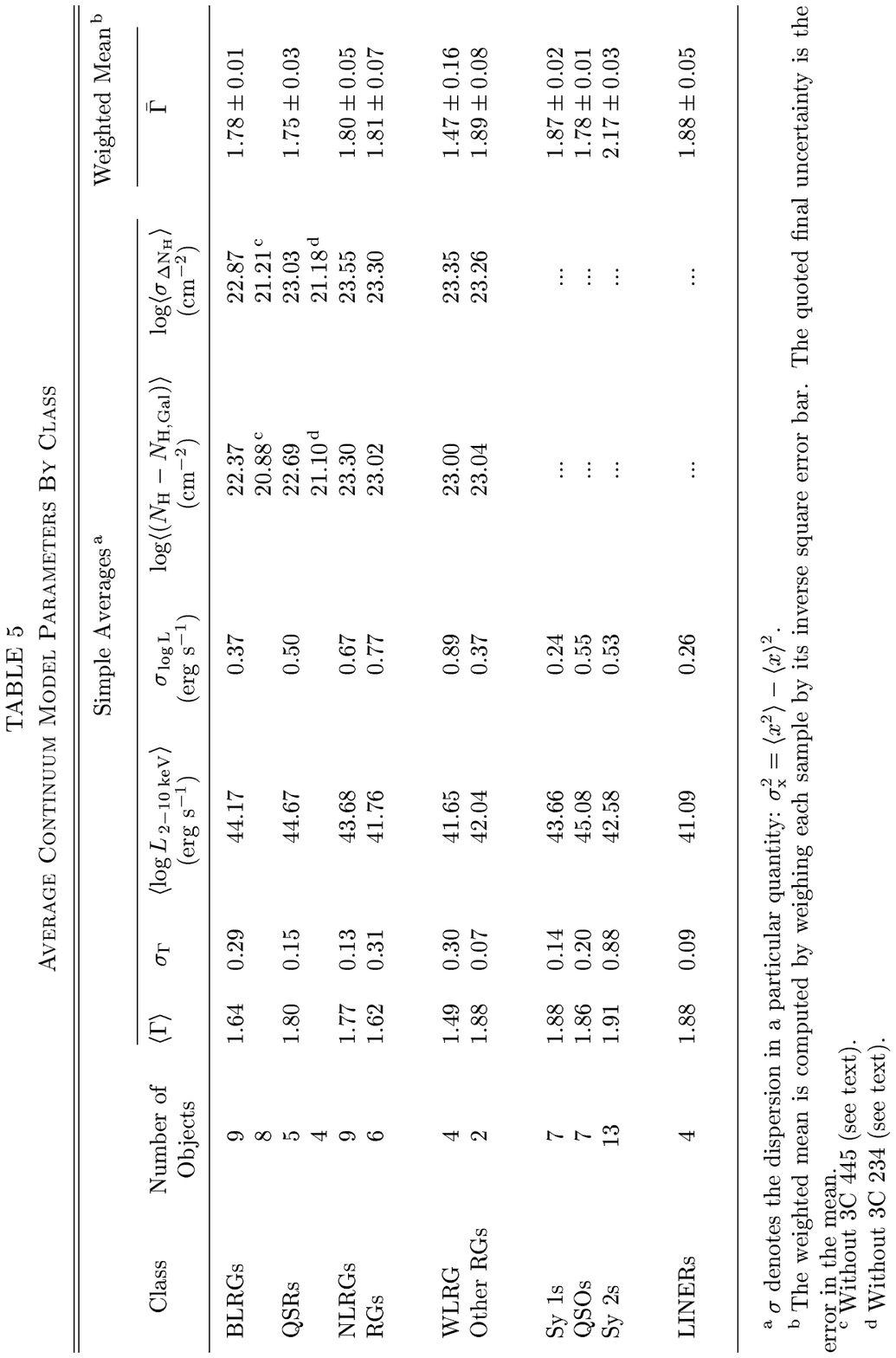,width=7.5in,rheight=8.in,angle=180}

\clearpage
\vspace*{-1in} \hspace*{-1in}
\psfig{figure=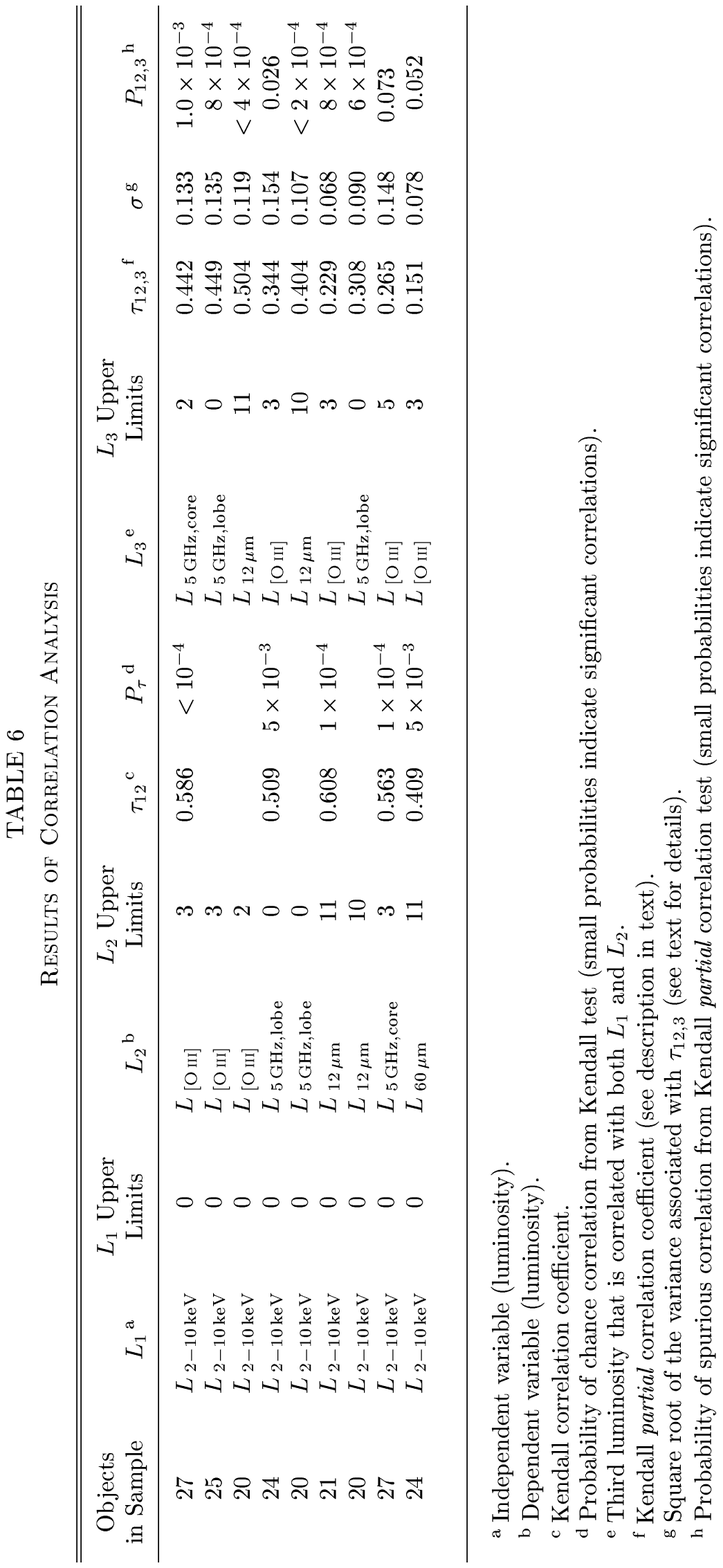,width=7.5in,rheight=8.in,angle=180}

\clearpage
\hspace*{-3in}
\psfig{figure=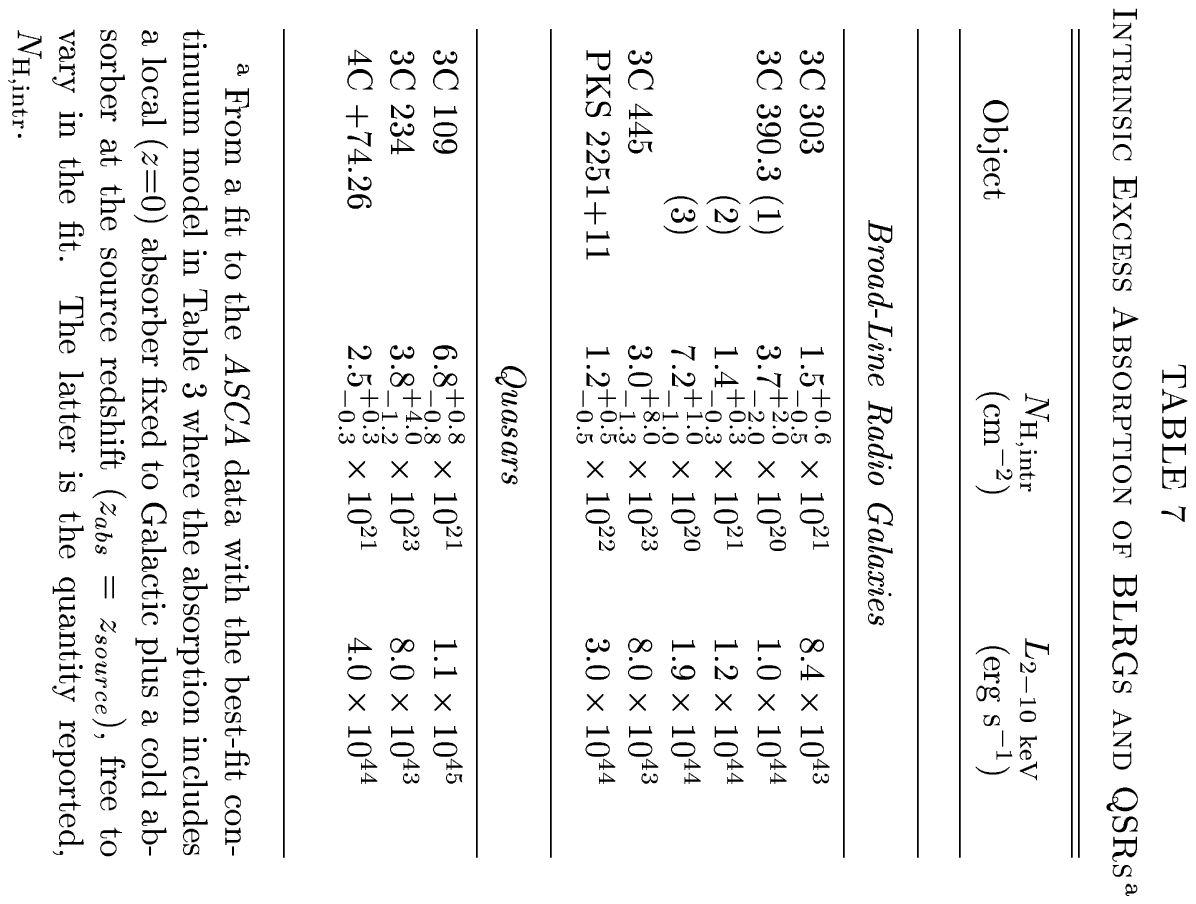,height=9in,rheight=8in,rwidth=9in,angle=90}


\clearpage
\vspace*{2in}

\noindent
{\bf Figure 1 is separate, in the form of 5 {\tt TIFF} files. It can
also be obtained in postscript form along with the complete preprint
from} 

{\tt http://www.astro.psu.edu/users/mce/preprint\_index.html}.

\vspace{2in}
\noindent {\bf Figure 1. -- } Results of spectral fits to the 0.6--10 keV
{\it ASCA} continuum of radio-loud AGN. For each source, we show the
best-fitting continuum model ({\it top panels}) and the corresponding
residuals ({\it bottom panels}).  In the cases where an Fe line was
required between 6 and 7 keV, the line profile model was excluded from
the model plotted in Figure~1 to make the line appear in the ratio
spectrum. The best-fit models are reported in Tables 3 and 4, with the
continuum parameters in Table 3 and the Fe K$\alpha$ parameters in
Table 4.  Only the SIS0 and SIS1 data are plotted here for clarity,
although the model was fitted to the data from all four detectors
jointly. 





\clearpage
\begin{center}
\begin{minipage}[t]{3in}
\centerline{\psfig{figure=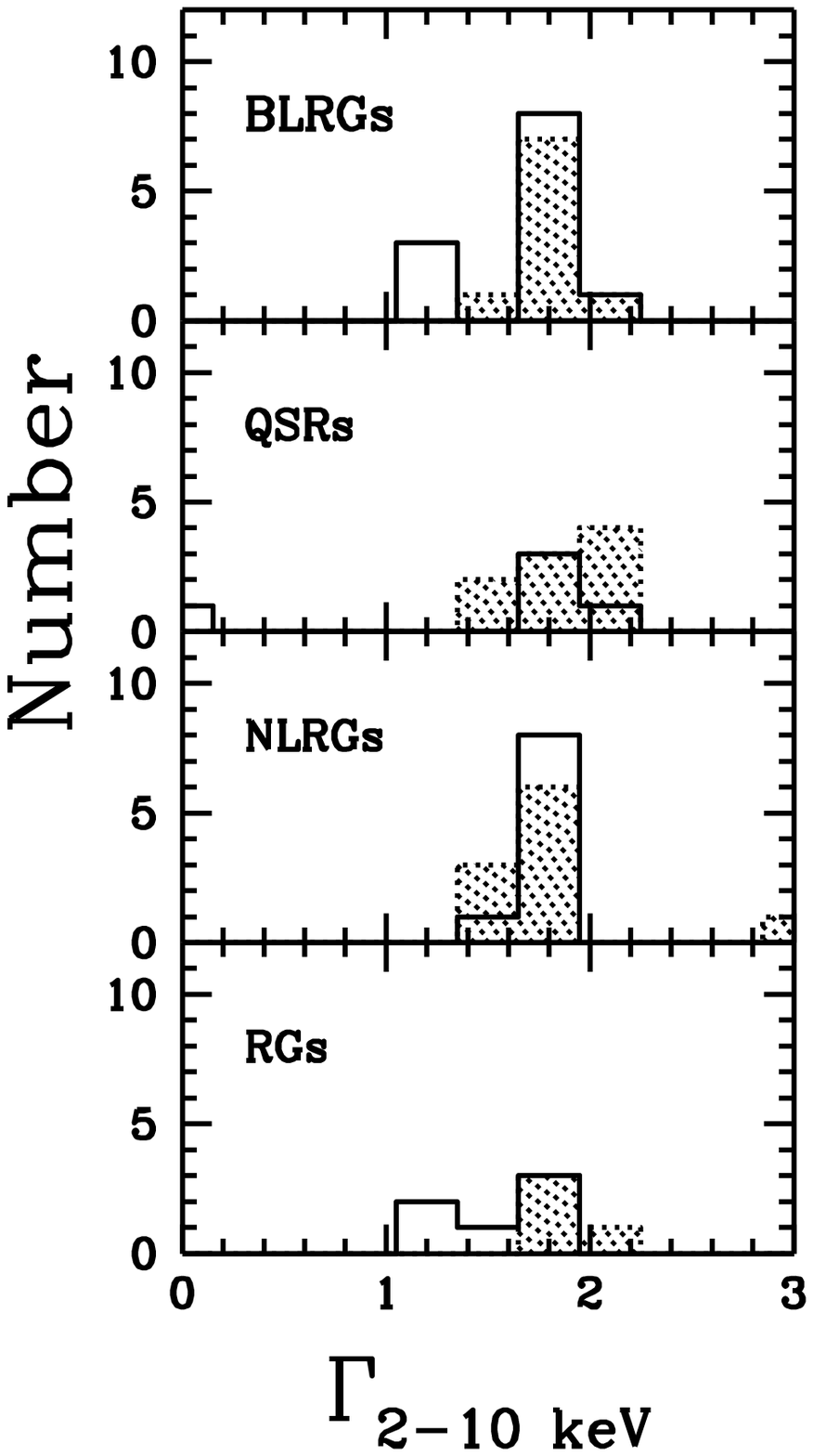,height=4in,rheight=4in,rwidth=4in}}
\medskip
\noindent{\bf Figure 2a -- } Distribution of the 2--10 keV intrinsic
photon indices for the various subclasses of radio-loud AGN {\it
(solid lines)}. The distribution spans a similar range in all
subclasses with a mean $\langle \Gamma \rangle \sim
1.7-1.8$. Superposed on the distribution for each subclass is the
distribution in the corresponding subclass of radio-quiet AGN (Seyfert
1s, Seyfert 2s, QSOs, and LINERs) in matching ranges of intrinsic
2--10 keV luminosities {\it (dotted areas)}. The distributions in
Seyfert 1s and BLRGs differ at the 94\% confidence level (based on the
Kolmogorov-Smirnov test).  Their mean values, $\langle \Gamma_{\rm
BLRG} \rangle = 1.6$ and $\langle \Gamma_{\rm Sy\,1} \rangle = 1.9$,
differ at the 93\% level (based on the t-test). The other classes of
radio-loud and radio-quiet AGN are not demonstrably different
according to the KS test.
\end{minipage}
\hfill
\begin{minipage}[t]{3in}
\centerline{\psfig{figure=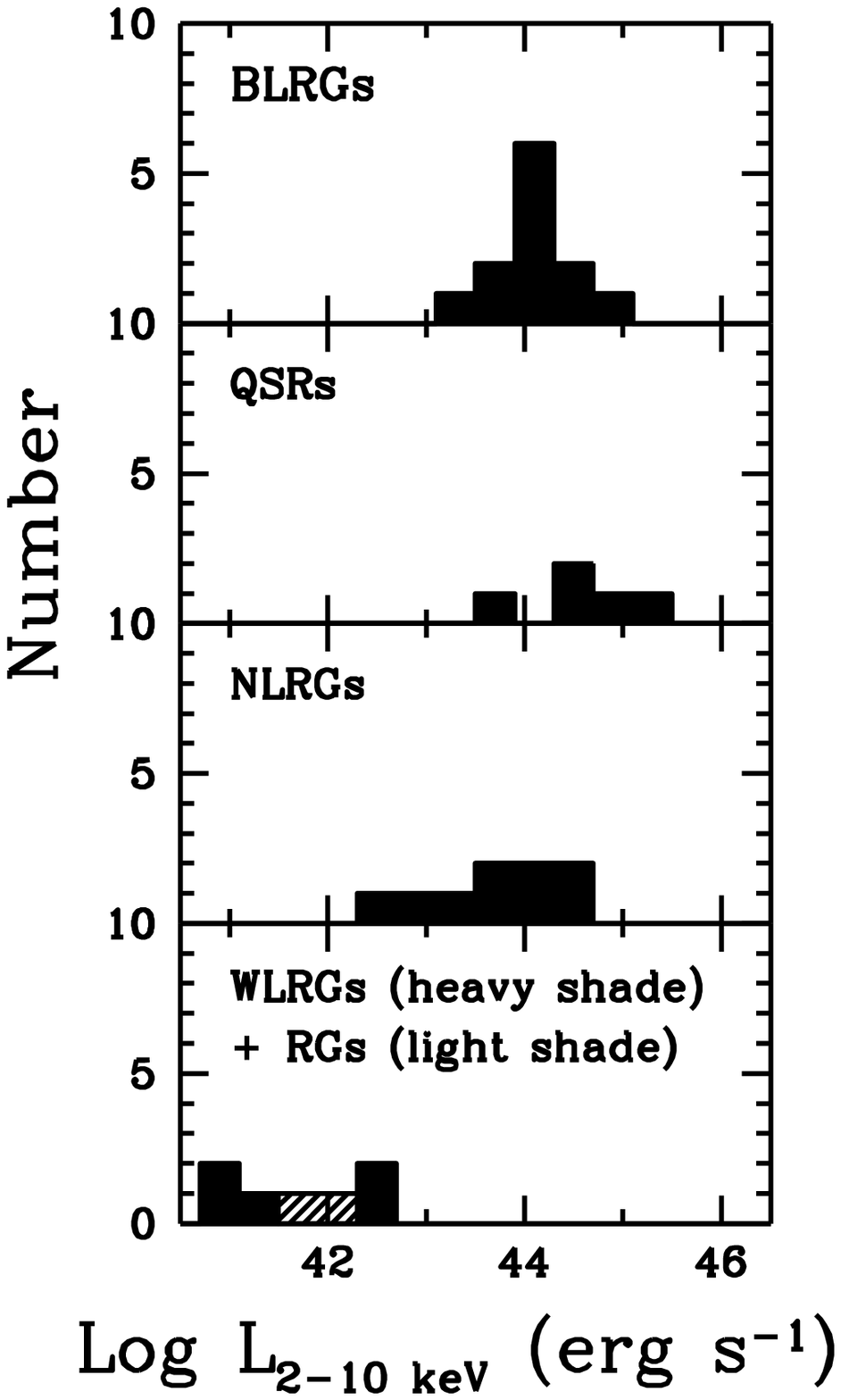,height=4in,rheight=4in,rwidth=4in}}
\medskip
\noindent{\bf Figure 2b -- } Distribution of the intrinsic
(absorption-corrected) 2--10 keV luminosity of the nuclear power-law
component in the various subclasses of radio-loud AGN. BLRGs, QSRs,
and NLRGs largely overlap, with QSRs populating the highest
luminosities and NLRGs the lowest. RGs and WLRGs are found at
luminosities lower than $10^{42}$ erg s$^{-1}$, significantly fainter
than the other subclasses.  This is at odds with the predictions of
simple orientation-based unification schemes.
\end{minipage}
\end{center}

\clearpage

\begin{center}
\begin{minipage}[t]{3in}
\centerline{\psfig{figure=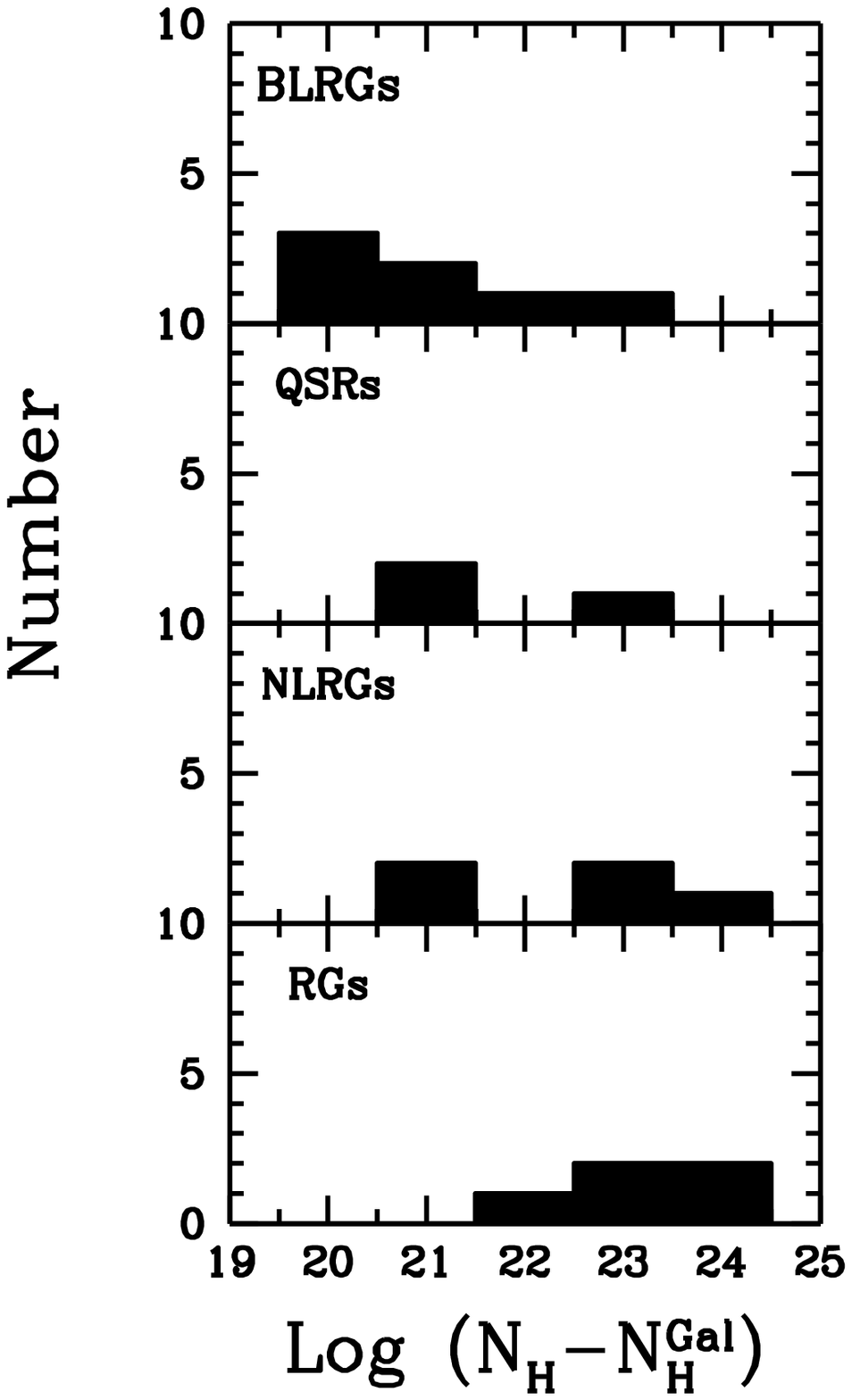,height=4in,rheight=4in,rwidth=4in}}
\medskip
\noindent{\bf Figure 2c -- } Histogram of the excess X-ray column density
for the various subclasses of radio-loud AGN from Table 3 (compensated
for the systematic overestimation by the SIS detector at low
energies).  Only sources where excess column densities were detected
with {\it ASCA} are plotted. In NLRGs and RGs the obscuring columns
are around $10^{21}$--$10^{24}$ cm$^{-2}$.  Excess X-ray absorption
with similar columns is also present in a fraction of BLRGs and QSRs.
\end{minipage}
\hfill
\begin{minipage}[t]{3in}
\centerline{\psfig{figure=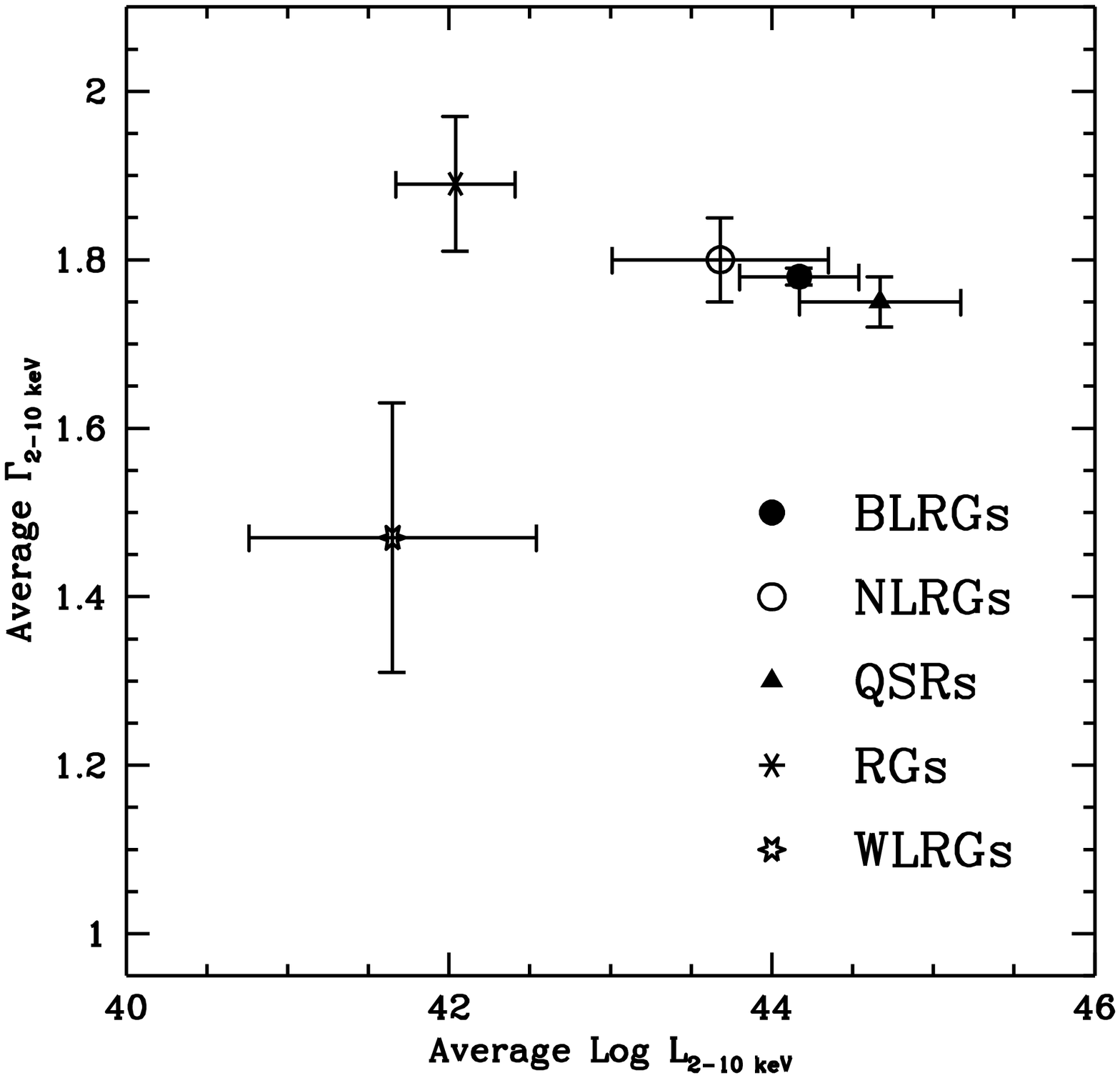,height=3in,rheight=4in,rwidth=3in}}
\medskip
\noindent{\bf Figure 2d -- } Plot of the average 2--10 keV photon
index (the weighted average in Table 5) versus the average intrinsic
2--10 keV luminosity for the various classes of radio sources. There
is an apparent trend of flatter slopes with increasing luminosity
going from RGs to QSRs, although a Kendall non-parametric correlation
test gives only a probability of $\sim$ 95\% that a correlation is
present. WLRGs stick out for having lower luminosities and flatter
slopes than the remaining sources.
\end{minipage}
\end{center}

\clearpage
\centerline{\psfig{figure=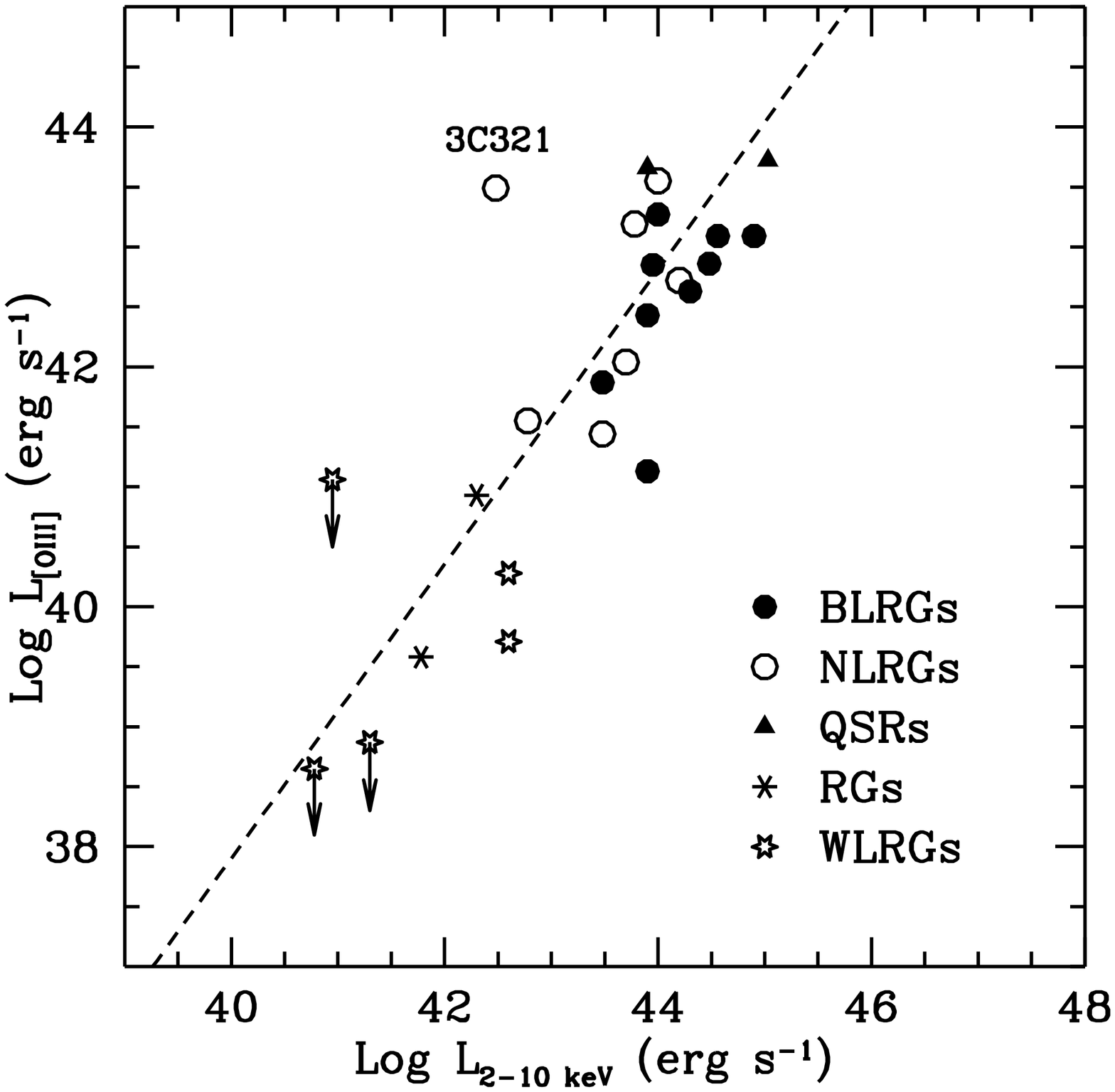,width=5in,rheight=5in}}
\noindent {\bf Figure 3a -- } Correlation between the rest-frame
luminosity of the [{\sc O\,iii}] emission lines, corrected for
Galactic extinction, and the intrinsic 2--10 keV power-law luminosity
for the various subclasses of radio-loud AGN. The correlation holds
over more than four decades in X-ray luminosity. The outlier at higher
[{\sc O\,iii}] luminosities is 3C~321, where a contribution from a
starburst could be present. Excluding this source, the upper limits,
and WLRGs, the linear regression analysis yields $\log L_{[{\rm O\,\sc
iii}]}=1.23\; \log L_{\rm 2-10~keV}-11.14$ for the sample of BLRGs,
QSRs, NLRGs, and RGs (dotted line). Note that WLRGs have abnormally
low $L_{[{\sc O\,iii}]}$ for their X-ray luminosity, which is one of
their defining properties.

\clearpage
\centerline{\psfig{figure=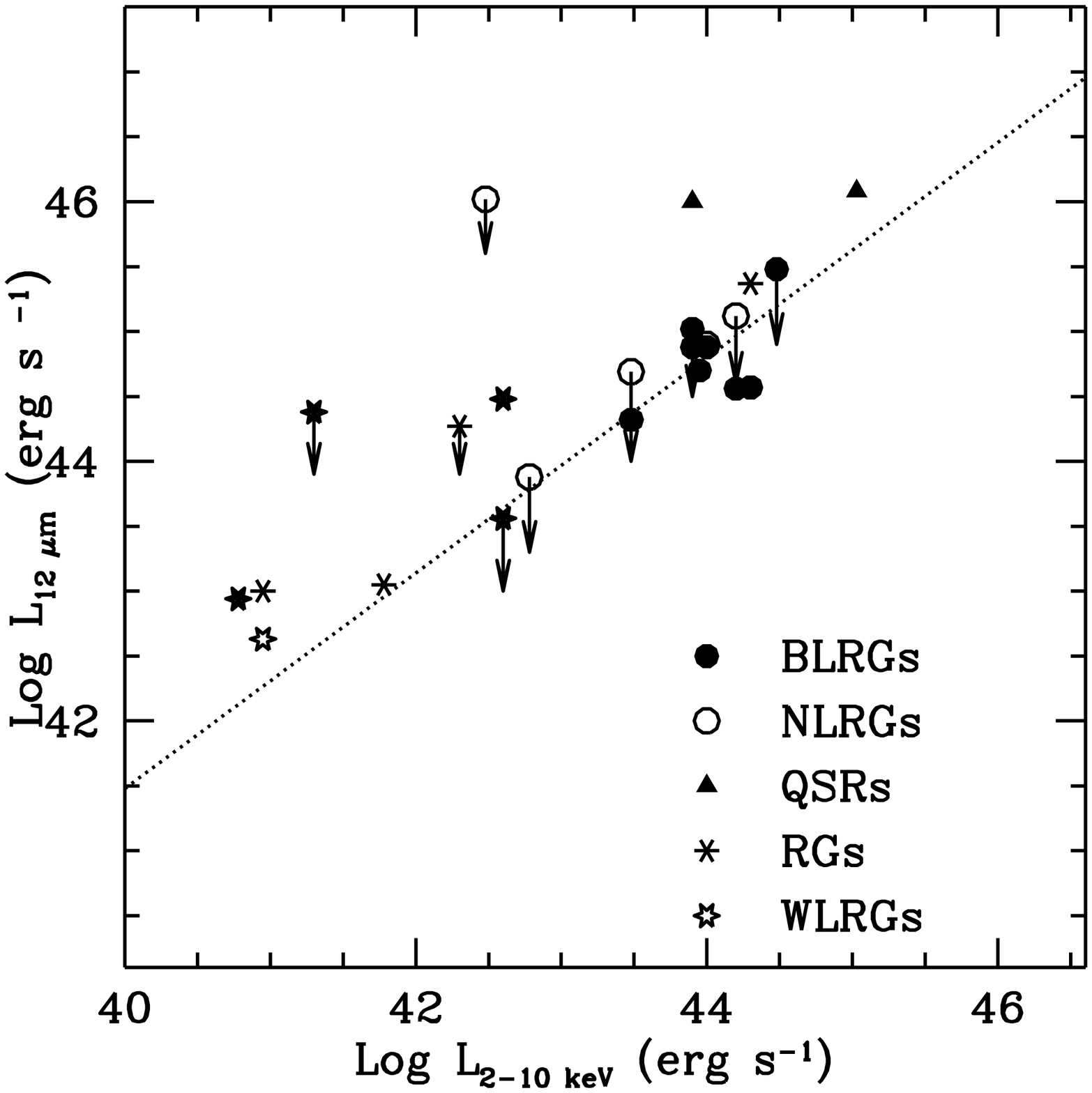,width=5in,rheight=5in}}
\noindent {\bf Figure 3b -- } Correlation between the FIR emission at
12~$\mu$m and the intrinsic 2--10 keV power-law luminosity. Excluding
the upper limits and WLRGs, a linear regression analysis gives $\log
L_{{\rm 12~\mu m}}=0.83\; \log L_{\rm 2-10~keV}+8.28$ for the
sample of BLRGs, QSRs, NLRGs, and RGs (dotted line).

\clearpage
\centerline{\psfig{figure=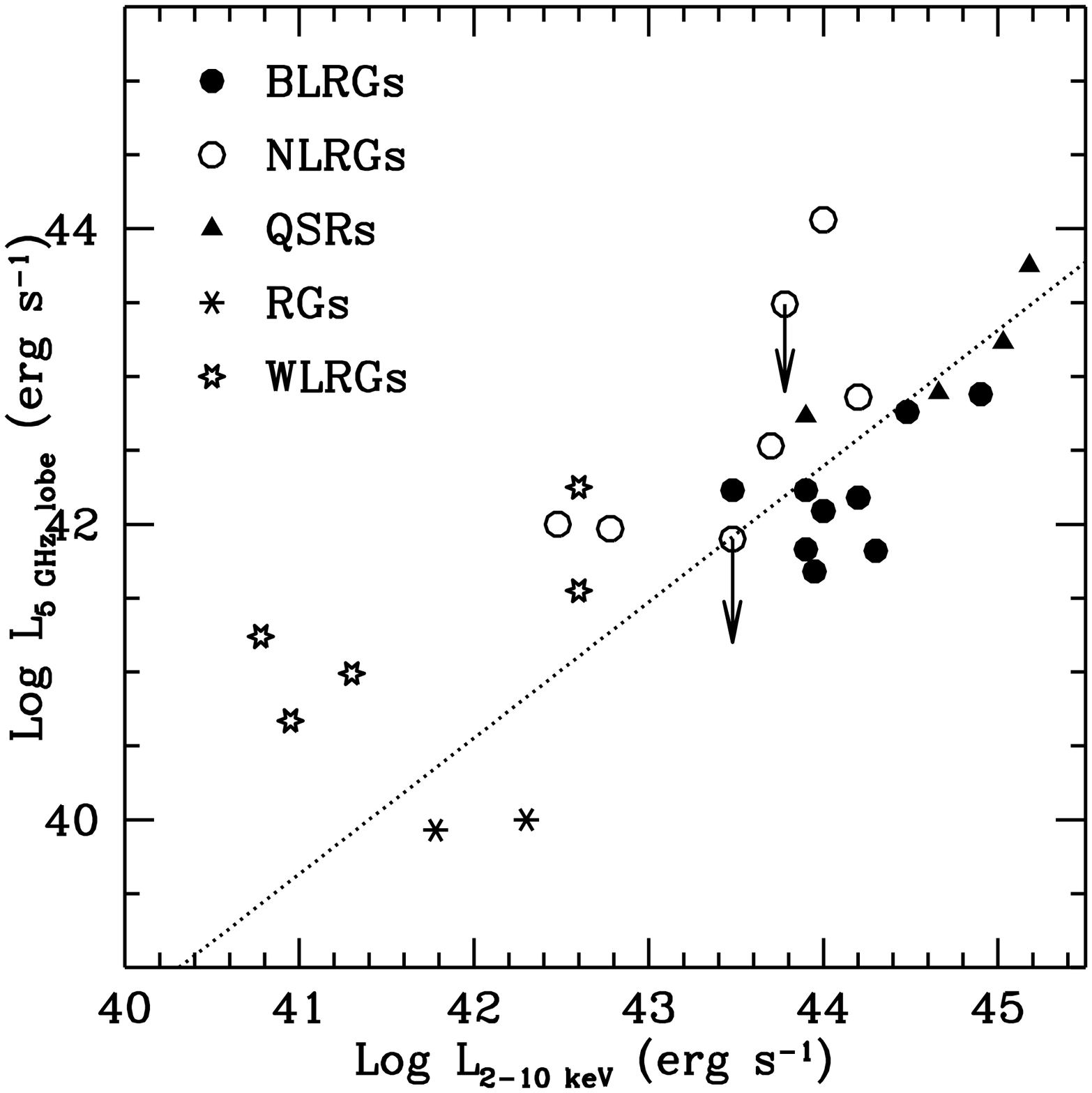,width=5in,rheight=5in}}
\noindent {\bf Figure 3c -- } Correlation between the lobe radio power at
5 GHz and the intrinsic 2--10 keV power-law luminosity. Excluding the
upper limits and WLRGs, a linear regression analysis gives $\log
L_{{\rm lobe}}=0.92 \log L_{\rm 2-10~keV}+1.91$ for the sample of
BLRGs, QSRs, NLRGs, and RGs (dotted line). WLRGs have powerful radio
lobe emission, consistent with their defining properties. 

\clearpage
\centerline{\psfig{figure=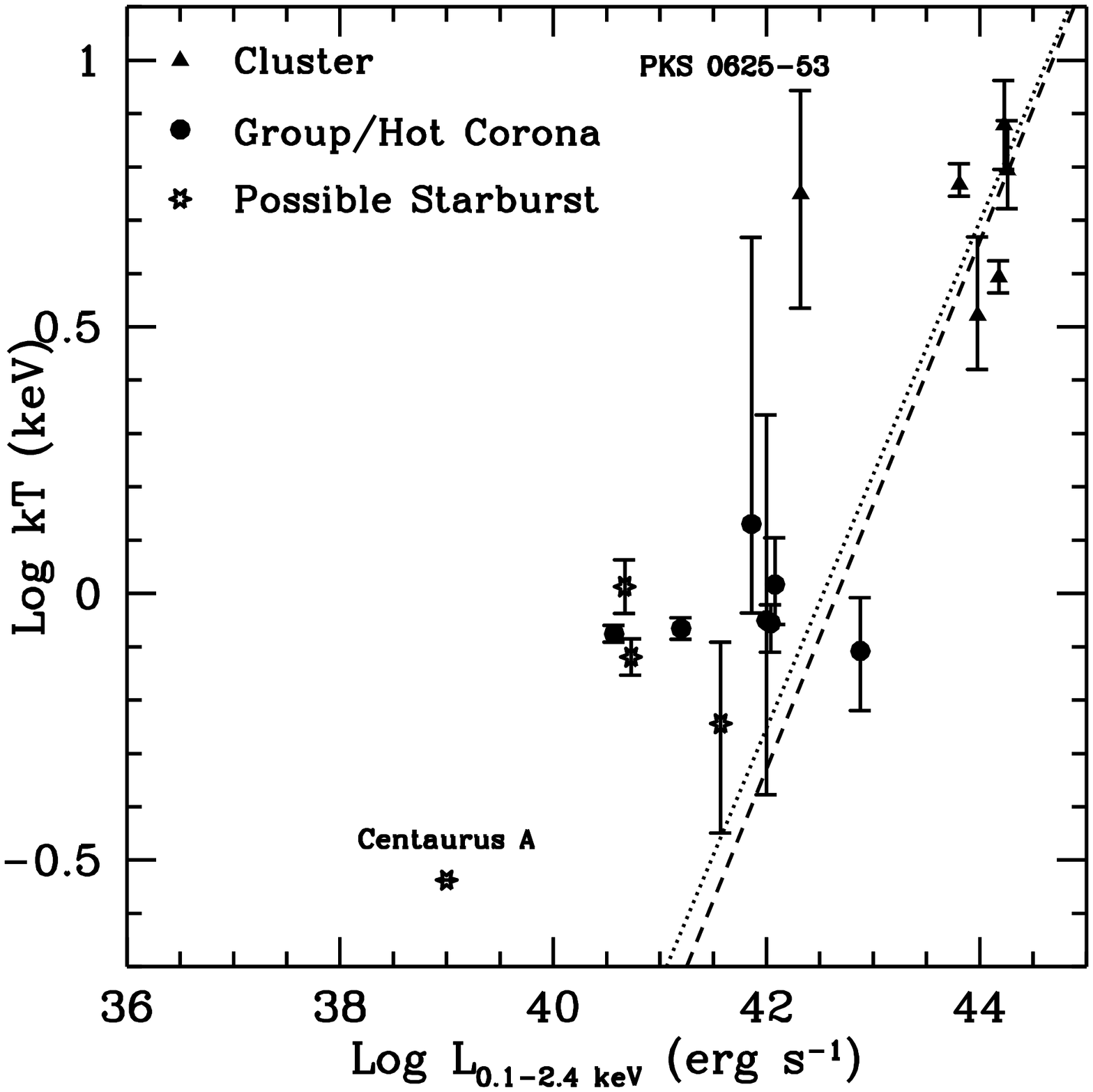,width=5in,rheight=5in}}
\noindent {\bf Figure 4 -- } Plot of the temperature versus the intrinsic
0.1--2.4 keV luminosity of the soft thermal component detected with
{\it ASCA} in several NLRGs, RGs, and WLRGs from our sample. A bimodal
distribution is apparent, with several sources grouped in the region
of high $kT$ and $L_{\rm 0.1-2.4~keV}$, typical of clusters of
galaxies {\it (filled triangles}), and others having lower $kT$ and
luminosities reminiscent of poor groups and/or hot coronae of isolated
giant ellipticals {\it (filled circles}). The sources 3C~321, 3C~234,
Centaurus~A, and PKS~0131--36, where the thermal component could have
a starburst or nuclear origin, are plotted with different symbols
({\it stars}). The solid and dotted lines represent the best-fitting
linear-regression lines to the $kT-L_x$ relationship of normal
clusters, both with and without cooling flows (from Markevitch
1998). The clusters hosting radio galaxies {\it (upper right corner)}
appear to have properties similar to normal clusters.

\clearpage
\centerline{\psfig{figure=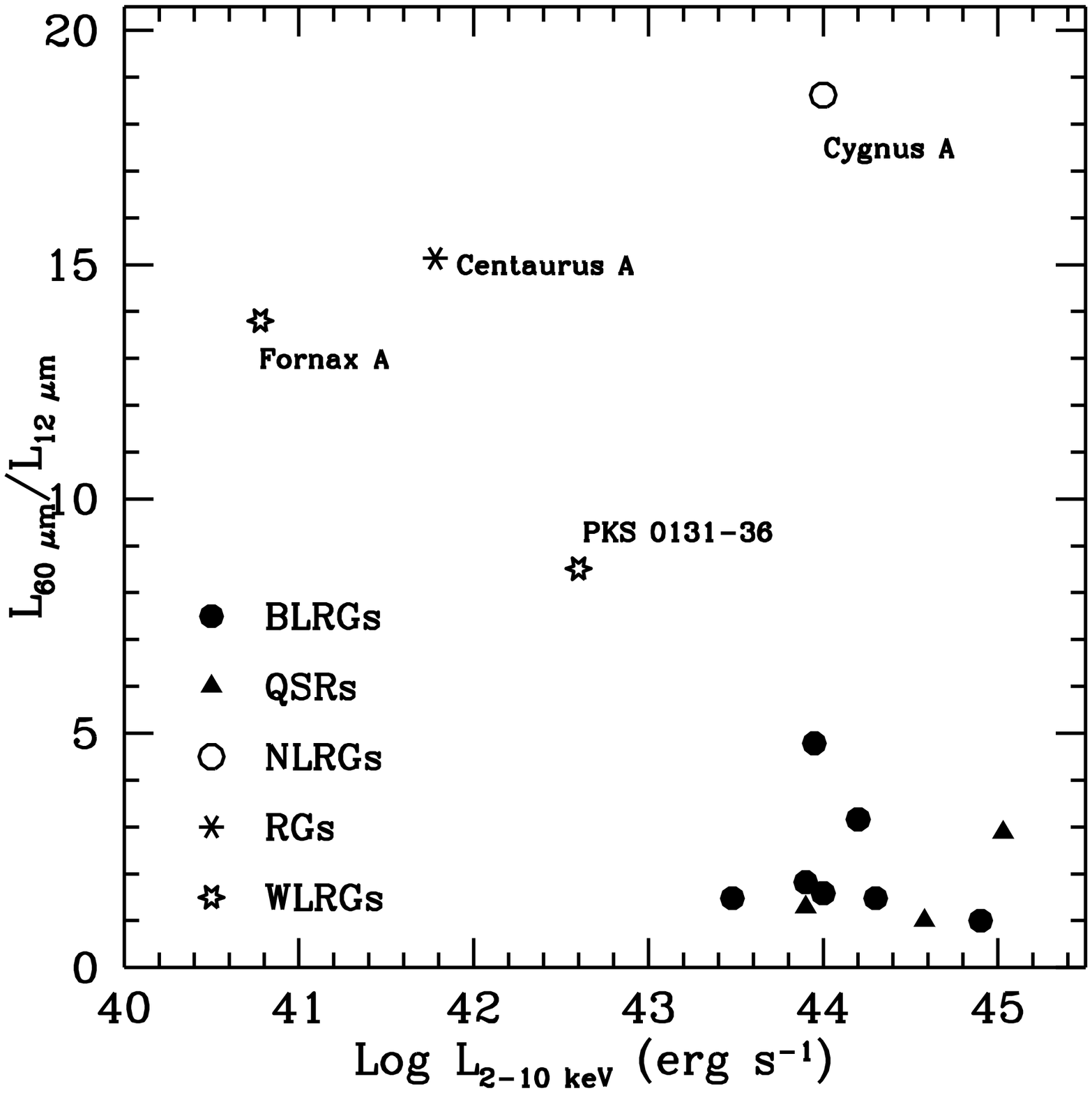,width=5in,rheight=5in}}
\noindent {\bf Figure 5 -- } Plot of the 60-to-12~$\mu$m FIR flux ratio
versus the intrinsic 2--10 keV power-law luminosity for the sources in
Table 1 with firm detections at all three wavelengths. Most BLRGs and
QSRs have ratios consistent with a non-thermal FIR emission, while
Centaurus A, Fornax A, and PKS~0131--36 have larger 60-to-12~$\mu$m
flux ratios, possibly related to an additional contribution from a
starburst. 

\clearpage
\centerline{\psfig{figure=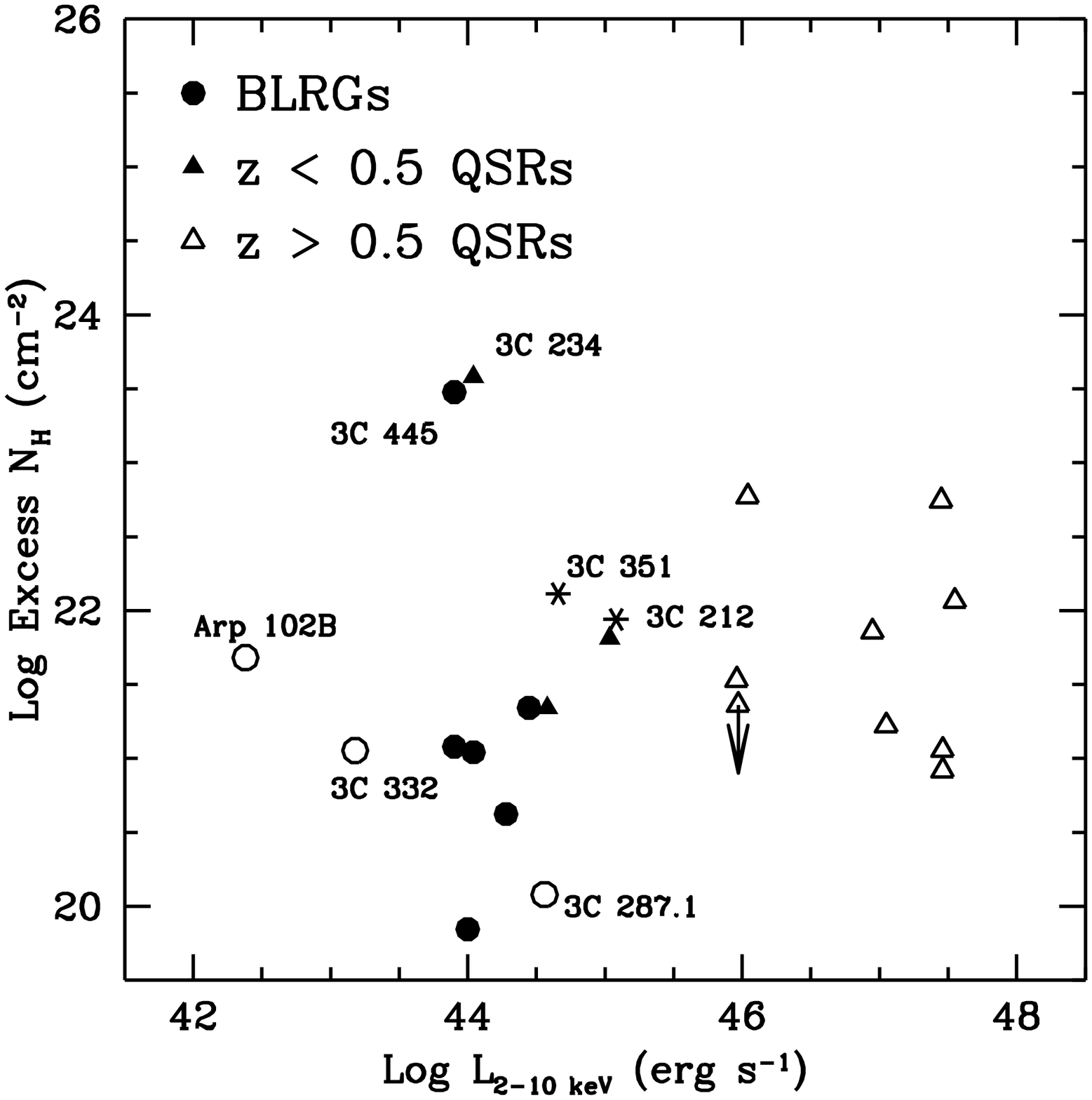,width=5in,rheight=5in}}
\noindent {\bf Figure 6 -- } Plot of the intrinsic X-ray absorption column
density versus intrinsic (absorption-corrected) nuclear X-ray
luminosity for radio-loud AGN. The filled circles and triangles are
the BLRGs and QSRs of the present study (data from Table 7). The open
triangles are the high-$z$ radio-loud QSRs studied by Cappi et
al. (1997) and Elvis et al. (1994) with {\it ROSAT} and {\it
ASCA}. Also plotted are the QSRs 3C~351 and 3C~212, where detailed
modelling of the {\it ROSAT} data indicates a highly-ionized absorber
(Fiore et al. 1993; Mathur 1994). The open circles are the BLRGs
Arp~102B, 3C~332, and 3C~287.1, where excess X-ray columns were
measured with {\it ROSAT} (Crawford \& Fabian 1995; Halpern
1997). Uncertainties on the fitted $N_{\rm H}$ are omitted for clarity
of presentation; however, they are generally large ($\sim$ 60\%),
especially for Arp~102B (80\%), 3C~332 (70\%), and 3C~287.1 (41\%),
due to the low signal-to-noise ratios of the {\it ROSAT}
spectra. Radio-loud AGN exhibit large X-ray column densities over a
wide range of X-ray nuclear luminosities; a large dispersion of
columns is observed at lower luminosities, with 3C~445 and 3C~234
having the largest column densities.

\clearpage
\centerline{\psfig{figure=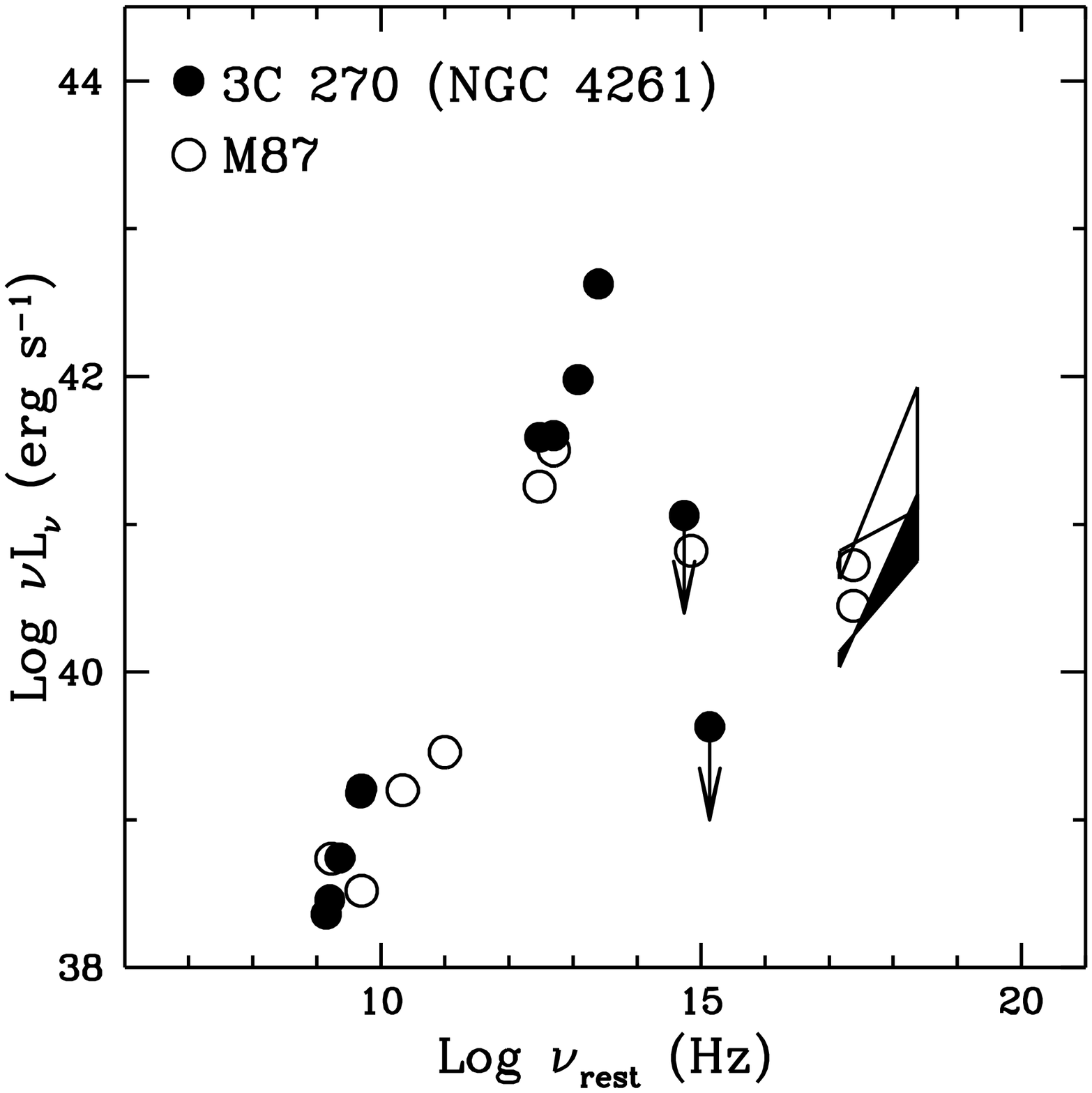,width=5in,rheight=5in}}
\noindent {\bf Figure 7 -- } Spectral energy distribution of the nuclear
emission of 3C~270 (hosted by the giant elliptical NGC~4261) from
radio to X-rays (using data in Table 1, the {\it ASCA} data presented
here, optical nuclear fluxes from Ferrarese et al. 1996, and the UV
upper limit from Zirbel \& Baum 1998). Also plotted for comparison is
the spectral energy distribution of M87 (Reynolds et al. 1996; Allen
et al. 1999), where an advection-dominated accretion flow (ADAF) is
thought to be present. The data of 3C~270 imply a ratio of the total
bolometric luminosity to the Eddington luminosity of $L_{\rm
Bol}/L_{\rm Edd} \lesssim 4 \times 10^{-4}$, placing 3C~270 in the
ADAF regime. Note the lack of a strong UV bump in 3C~270, which is
typical of other low-power radio galaxies as well, and could explain
the weak [{\sc O\,iii}] emission in these sources.

\clearpage
\centerline{\psfig{figure=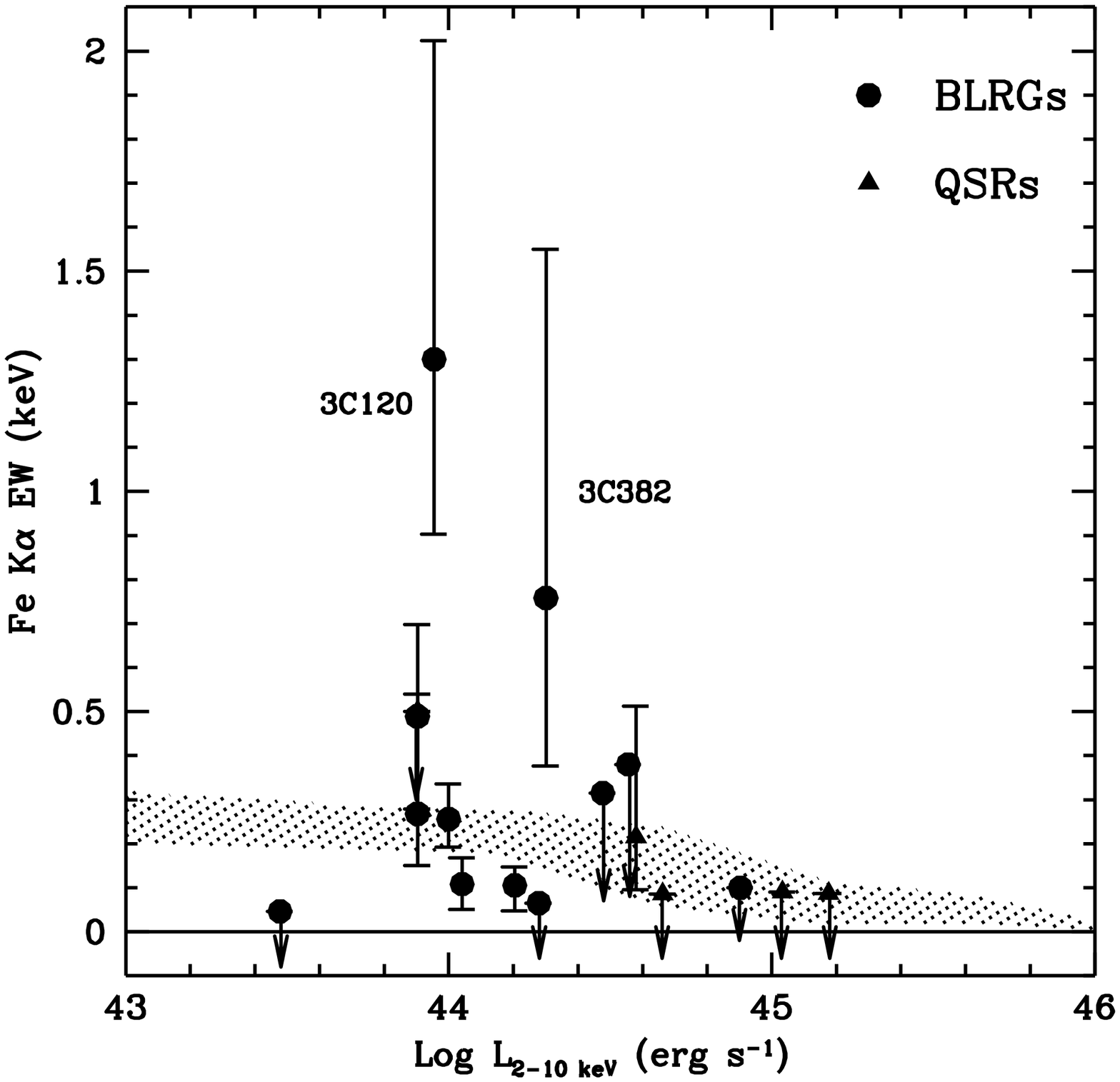,width=5in,rheight=5in}}
\noindent {\bf Figure 8 -- } Plot of the intrinsic EW of the Fe K$\alpha$
line (or upper limit at 90\% confidence) versus the
absorption-corrected power law X-ray luminosity for the sub-classes of
BLRGs and QSRs of our sample (updated from Eracleous \& Halpern
1998). The shaded strip represents the X-ray Baldwin effect at
$1\sigma$ for radio-quiet sources (Nandra et al. 1997c).  Radio-loud
AGN have a larger dispersion of intrinsic EW for a given X-ray
luminosity than radio-quiet sources.

\end{document}